\documentclass[12pt,a4paper]{article}
\pdfoutput=1

\usepackage{jheppub_modified}

\usepackage{amsfonts}

\usepackage{enumerate}
\usepackage{rotating}
\usepackage{axodraw}
\usepackage{array}
\usepackage[footnotesize]{caption}

\allowdisplaybreaks

\newcommand{\slsh}[1]{\!\!\not\! {#1}}

\title{Higgs-Pair Production via Gluon Fusion at Hadron Colliders: NLO QCD Corrections}

\author[a]{Julien Baglio,}
\author[b,c]{Francisco Campanario,}
\author[c,d]{Seraina Glaus,}
\author[c]{Margarete M\"uhlleitner,}
\author[b]{Jonathan Ronca,}
\author[e]{Michael Spira}
\author[f]{and Juraj Streicher}

\affiliation[a]{Theoretical Physics Department, CERN, CH-1211 Geneva 23, Switzerland}
\affiliation[b]{Theory Division, IFIC, University of Valencia-CSIC, E-46980 Paterna, Valencia, Spain}
\affiliation[c]{Institute for Theoretical Physics, Karlsruhe Institute of Technology, D-76131 Karlsruhe, Germany}
\affiliation[d]{Institute for Nuclear Physics, Karlsruhe Institute of Technology, D-76344 Karlsruhe, Germany}
\affiliation[e]{Paul Scherrer Institut, CH-5232 Villigen PSI, Switzerland}
\affiliation[f]{Institut f\"ur Theoretische Physik, Eberhard Karls Universit\"at T\"ubingen, Auf der Morgenstelle 14, D-72076 T\"ubingen, Germany}

\emailAdd{julien.baglio@cern.ch}
\emailAdd{Francisco.Campanario@ific.uv.es}
\emailAdd{seraina.glaus@kit.edu}
\emailAdd{milada.muehlleitner@kit.edu}
\emailAdd{Jonathan.Ronca@uv.es}
\emailAdd{Michael.Spira@psi.ch}
\emailAdd{juraj.streicher@uni-tuebingen.de}

\preprint{
\begin{flushright}
CERN--TH--2020--032 \\
IFIC/20--06 \\
FTUV--20--0301 \\
KA--TP--01--2020 \\
PSI--PR--20--03
\end{flushright}}

\abstract{Higgs-pair production via gluon fusion is the dominant production
mechanism of Higgs-boson pairs at hadron colliders. In this work, we present
details of our numerical determination of the full next-to-leading-order
(NLO) QCD corrections to the leading top-quark loops. Since gluon fusion
is a loop-induced process at leading order, the NLO calculation
requires the calculation of massive two-loop diagrams with up to four
different mass/energy scales involved.  With the current methods, this
can only be done numerically, if no approximations are used. We discuss
the setup and details of our numerical integration.
This will be followed by a phenomenological analysis of the NLO
corrections and their impact on the total cross section and the
invariant Higgs-pair mass distribution. The last part of our work will
be devoted to the determination of the residual theoretical
uncertainties with special emphasis on the uncertainties originating
from the scheme and scale dependence of the (virtual) top mass. The
impact of the trilinear Higgs-coupling variation on the total cross
section will be discussed.}

\arxivnumber{2003.03227}

\begin{document}
\maketitle
\flushbottom

\section{Introduction}
%        ============
Since the discovery of a scalar resonance \cite{Aad:2012tfa,
Chatrchyan:2012xdj} with a mass of $125.09\pm0.24$ GeV
\cite{Khachatryan:2016vau} that is compatible with the Standard Model
(SM) Higgs boson \cite{Higgs:1964ia,Higgs:1964pj,Higgs:1966ev,
Englert:1964et,Guralnik:1964eu,Kibble:1967sv}, the detailed study of the
properties of this particle has been a high priority of the analyses at
the Large Hadron Collider (LHC). Theoretical uncertainties are a
limiting factor for the accuracies reachable at the LHC. This
restriction can partly be compensated by increasing the diversity of
processes involving the Higgs boson and a broader spectrum of Higgs
couplings probed at the LHC. In order to test the nature of the Higgs
boson, its self-interactions are of particular interest. It
will be the first step towards an experimental reconstruction of the
Higgs potential. This plays a crucial role as the origin of electroweak
symmetry breaking within the SM. The initial processes that provide a
direct sensitivity to the Higgs self-couplings are Higgs-pair production
processes. They involve the trilinear Higgs
coupling at leading order (LO) \cite{Glover:1987nx, Plehn:1996wb,
Dawson:1998py, Djouadi:1999rca, Baglio:2012np}. These processes are
complementary to indirect effects induced by the Higgs self-interactions
in radiative corrections to electroweak observables and single-Higgs
processes \cite{Degrassi:2016wml, Degrassi:2017ucl} that are plagued by
unknown interference effects with other kinds of New Physics.

The Higgs self-interactions are uniquely described by the SM Higgs
potential
\begin{equation}
V = \frac{\lambda}{2} \left( \phi^\dagger \phi -
  \frac{v^2}{2}\right)^2\, ,
\end{equation}
where $\lambda$ defines the self-interaction strength of the SM Higgs
field. In unitary gauge, the Higgs doublet $\phi$ is given by
\begin{equation}
\phi = \left( \begin{array}{c} \displaystyle 0 \\ \displaystyle
\frac{v+H}{\sqrt{2}} \end{array} \right)
\end{equation}
with $v\approx 246$ GeV denoting the vacuum expectation value (vev) and
$H$ is the physical Higgs field. In the SM, the self-interaction
strength is given in terms of the Higgs mass $M_H$ by $\lambda =
M_H^2/v^2$. Expanding the Higgs field around its vev, the Higgs
self-interactions, including the corresponding permutations, are
uniquely determined as
\begin{equation}
\lambda_{H^3} = 3 \frac{M_H^2}{v}, \qquad \lambda_{H^4} = 3
\frac{M_H^2}{v^2}\, ,
\end{equation}
where $\lambda_{H^3}$ ($\lambda_{H^4}$) denotes the trilinear (quartic)
Higgs self-coupling.

\begin{figure}[hbtp]
\begin{center}
\vspace*{-1cm}

\hspace*{-2cm}
  \includegraphics[width=1.10\textwidth]{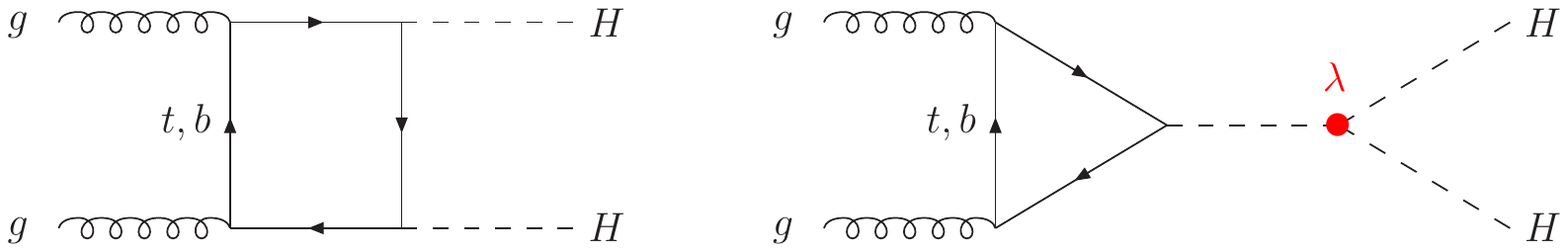} 
\vspace*{-22.5cm}

\caption{\label{fg:hhdia} \it Generic diagrams contributing to
Higgs-boson pair production via gluon fusion. The contribution of the
trilinear Higgs coupling is marked in red.}
\end{center}
\end{figure}

While the quartic Higgs coupling $\lambda_{H^4}$ cannot be probed
directly at the LHC, due to the tiny size of the triple-Higgs production
cross section \cite{Plehn:2005nk, Binoth:2006ym, Fuks:2015hna,
deFlorian:2016sit, deFlorian:2019app}\footnote{Note that Higgs pair
production will provide indirect constraints on the quartic Higgs
coupling \cite{Liu:2018peg, Bizon:2018syu, Borowka:2018pxx}.}, the
trilinear Higgs coupling can be accessed directly in Higgs-pair
production. Higgs-boson pairs are dominantly produced in the
loop-induced gluon-fusion mechanism $gg\to HH$ that is mediated by
top-quark loops supplemented by a per-cent-level contribution of
bottom-quark loops, see Fig.~\ref{fg:hhdia}. There are destructively interfering box
and triangle diagrams at LO with the latter involving
the trilinear Higgs coupling \cite{Glover:1987nx,Plehn:1996wb}. The box
diagrams provide the dominant contributions to the cross section. A
rough estimate of the dependence of the cross section on the size of the
trilinear coupling is given by the approximate relation
$\Delta\sigma/\sigma \sim -\Delta\lambda_{H^3}/\lambda_{H^3}$ in the
vicinity of the SM value of $\lambda_{H^3}$. Therefore, in order to
determine the trilinear coupling, the theoretical uncertainties of the
corresponding cross section need to be small. Thus, the inclusion of
higher-order corrections is mandatory. The QCD corrections are fully
known up to next-to-leading order (NLO) \cite{Borowka:2016ehy,
Borowka:2016ypz, Baglio:2018lrj} and at next-to-next-to-leading order
(NNLO) in the limit of heavy top quarks \cite{deFlorian:2013uza,
deFlorian:2013jea, Grigo:2014jma}. While the NLO corrections are large,
the NNLO contributions are of more moderate size. Very recently, the
next-to-next-to-next-to-leading order (N$^3$LO) QCD corrections have
been computed in the limit of heavy top quarks resulting in a small
further modification of the cross section \cite{Banerjee:2018lfq,
Chen:2019lzz, Chen:2019fhs}. This calculation uses the N$^3$LO
corrections to the effective Higgs and Higgs-pair couplings to gluons in
the heavy-top limit (HTL) \cite{Spira:2016zna}. The higher-order QCD corrections increase
the total LO cross section by about a factor of two. Recently, the full
NLO results have been matched to parton showers \cite{Heinrich:2017kxx,
Jones:2017giv} and the full NNLO results in the limit of heavy top
quarks have been merged with the NLO mass effects and supplemented by
the additional top-mass effects in the double-real corrections
\cite{Grazzini:2018bsd}.

The goal of this paper is to present in detail the calculation of
Ref.~\cite{Baglio:2018lrj} of the full NLO corrections to Higgs pair
production in gluon fusion. We rely on a direct numerical integration
of the Feynman diagrams, without any tensor reduction. We extend the
results presented in Ref.~\cite{Baglio:2018lrj}  and study not only
the LHC at center-of-mass energies of 13 and 14 TeV, but also present
numbers for a potential high-energy upgrade of the LHC (HE-LHC) at 27
TeV~\cite{Abada:2019ono} and for a provisional 100 TeV proton collider
within the Future-Circular-Collider (FCC) project
~\cite{Abada:2019lih,Benedikt:2018csr}. Special emphasis will
be given to the study of the theoretical uncertainties affecting the
results and in particular the scale and scheme uncertainty related to
the top-quark mass. We will also study the variation of the trilinear
Higgs coupling and show that the NLO mass effects shift the minimum of
the total cross section as a function of $\lambda_{H^3}$. They
vary substantially over the range of $\lambda_{H^3}$ values.

The paper is organized as follows. We present the notation of our
calculation in Section~\ref{sc:lo} and discuss the results at LO. In
Section~\ref{sc:nlo} we move to the NLO QCD corrections. We discuss
the details of the calculation of the virtual corrections in
Section~\ref{sc:virtuals}. We describe the derivation of the real
corrections in Section~\ref{sc:reals}. Our numerical analysis is
performed in Section~\ref{sc:results}. Finally, the conclusions are
given in Section~\ref{sc:conclusions}.

\section{Leading-order cross section \label{sc:lo}}
%        ===========================
At LO, Higgs-boson pair production via gluon fusion is mediated by the
generic diagrams of Fig.~\ref{fg:hhdia}, including all permutations of
the external lines. There are triangle and box diagrams with the former
involving the trilinear Higgs coupling through an $s$-channel Higgs
exchange. The LO matrix element of $g(q_1) g(q_2) \to H(p_1) H(p_2)$ can
be cast into the form
\begin{eqnarray}
{\cal M}(g^a g^b \to HH) & = & -i\,\frac{G_F\alpha_s(\mu_R) Q^2}{2\sqrt{2}\pi}
{\cal A}^{\mu\nu} \epsilon_{1\mu} \epsilon_{2\nu} \delta_{ab}
\nonumber \\[0.3cm]
\mbox{with} \qquad {\cal A}^{\mu\nu} & = & F_1 T_1^{\mu\nu} + F_2 T_2^{\mu\nu}
\, , \nonumber \\[0.3cm]
F_1 & = & C_\triangle F_\triangle + F_\Box \, , \qquad
\qquad \qquad
F_2 = G_\Box \, , \nonumber \\[0.3cm]
C_\triangle & = & \frac{\lambda_{H^3} v}{Q^2 - M_H^2 +
iM_H\Gamma_H} \nonumber \\[0.3cm]
\mbox{and} \qquad Q^2 & = & (p_1+p_2)^2 = m_{HH}^2
\label{eq:lomat}
\end{eqnarray}
with $Q=m_{HH}$ denoting the invariant Higgs-pair mass.
Here $a,b$ denote the color indices of the initial gluons,
$\epsilon_{1/2}$ their polarization vectors, $\Gamma_H$
the total Higgs width\footnote{Throughout this work, we will neglect the
total Higgs width $\Gamma_H$ in the coefficient $C_\triangle$.}, $G_F$
the Fermi constant and $\alpha_s(\mu_R)$ the strong coupling at the
renormalization scale $\mu_R$. Since in this work we neglect the small
bottom-quark contribution, the LO function of the triangle-diagram
contribution is given by the top-quark contribution,
\begin{equation}
F_\triangle (\tau_t) = \tau_t \Big[ 1 + (1-\tau_t) f(\tau_t) \Big ]
\label{eq:ftriangle}
\end{equation}
with $\tau_t = 4m_t^2/Q^2$ and the basic function
\begin{eqnarray}
f(\tau) & = & \left\{ \begin{array}{ll} \displaystyle \arcsin^2
\frac{1}{\sqrt{\tau}} & \tau \ge 1 \\ \displaystyle - \frac{1}{4} \left[
\log \frac{1+\sqrt{1-\tau}} {1-\sqrt{1-\tau}} - i\pi \right]^2 & \tau <
1 \end{array} \right. \, ,
\label{eq:ftau}
\end{eqnarray}
where $m_t$ denotes the top mass, while the more involved analytical
expressions for $F_\Box$ and $G_\Box$ can be found in
Ref.~\cite{Plehn:1996wb}. In the HTL, the LO form
factors approach the values
\begin{equation}
F_\triangle \to \frac{2}{3}, \qquad F_\Box \to -\frac{2}{3}, \qquad
G_\Box \to 0 \, .
\label{eq:ffhtl}
\end{equation}
There are two tensor structures contributing which correspond to the
total angular-momentum states with $S_z=0$ and $2$,
\begin{eqnarray}
T_1^{\mu\nu} & = & g^{\mu\nu}-\frac{q_1^\nu q_2^\mu}{(q_1q_2)}\, , \nonumber \\
T_2^{\mu\nu} & = & g^{\mu\nu}+\frac{M_H^2 q_1^\nu q_2^\mu}{p_T^2 (q_1q_2)}
-2\frac{(q_2 p_1) q_1^\nu p_1^\mu}{p_T^2 (q_1q_2)}
-2\frac{(q_1 p_1) p_1^\nu q_2^\mu}{p_T^2 (q_1q_2)}
+2\frac{p_1^\nu p_1^\mu}{p_T^2} \nonumber \\
\mbox{with} \quad p_T^2 & = & 2 \frac{(q_1 p_1)(q_2 p_1)}{(q_1 q_2)} -
M_H^2 \, ,
\end{eqnarray}
where $p_T$ is the transverse momentum of each of the final-state Higgs
bosons. Working in $n=4-2\epsilon$ dimensions, the following projectors
on the two form factors can be constructed,
\begin{equation}
P_1^{\mu\nu} = \frac{(1-\epsilon) T_1^{\mu\nu} + \epsilon
T_2^{\mu\nu}}{2(1-2\epsilon)}\, , \qquad \qquad
P_2^{\mu\nu} = \frac{\epsilon T_1^{\mu\nu} + (1-\epsilon)
T_2^{\mu\nu}}{2(1-2\epsilon)} \, ,
\end{equation}
such that
\begin{equation}
P_1^{\mu\nu} {\cal A}_{\mu\nu} = F_1 \, , \qquad \qquad
P_2^{\mu\nu} {\cal A}_{\mu\nu} = F_2 \, .
\end{equation}
Using these projectors, the explicit results of the two form factors
$F_{1,2}$ can be obtained in a straightforward manner. The analytical
expressions can be found in Refs.~\cite{Glover:1987nx, Plehn:1996wb}.
Working out the polarization and color sums of the matrix element of
Eq.~(\ref{eq:lomat}), the LO partonic cross section $\hat\sigma_{LO}$
is given by
\begin{equation}
\hat\sigma_{LO} = \frac{G_F^2\alpha_s^2(\mu_R)}{512 (2\pi)^3}
\int_{\hat t_-}^{\hat t_+} d\hat t \Big[ | F_1 |^2 + |F_2|^2
\Big]
\end{equation}
with the integration boundaries
\begin{equation}
\hat t_\pm = -\frac{1}{2} \left[ Q^2 - 2M_H^2 \mp Q^2
\sqrt{1-4\frac{M_H^2}{Q^2}} \right] \, ,
\label{eq:tbound}
\end{equation}
where the symmetry factor 1/2 for the identical Higgs bosons in the
final state is taken into account. The LO hadronic cross section
$\sigma_{LO}$ can then be derived by a convolution with the parton
densities
\begin{equation}
\sigma_{LO} = \int_{\tau_0}^1 d\tau \frac{d{\cal L}^{gg}}{d\tau}
\hat\sigma_{LO}(Q^2 = \tau s)
\end{equation}
with the gluon luminosity, given in terms of the gluon densities
$g(x,\mu_F)$,
\begin{equation}
\frac{d{\cal L}^{gg}}{d\tau} = \int_\tau^1 \frac{dx}{x} g(x,\mu_F)
g\left(\frac{\tau}{x},\mu_F\right)
\label{eq:lgg}
\end{equation}
at the factorization scale $\mu_F$ and the integration boundary
$\tau_0=4M_H^2/s$, where $s$ denotes the hadronic center-of-mass (c.m.)
energy squared.  The differential cross section with respect to the
invariant squared Higgs-pair mass $Q^2$ can be obtained as
\begin{equation}
\frac{d\sigma_{LO}}{dQ^2} = \left. \frac{d{\cal L}^{gg}}{d\tau}~
\frac{\hat\sigma_{LO}(Q^2)}{s} \right|_{\tau = \frac{Q^2}{s}} \, .
\label{eq:lodiff}
\end{equation}
As can be expected from single Higgs-boson production via gluon fusion
(see \cite{Graudenz:1992pv, Spira:1995rr, Harlander:2005rq,
Anastasiou:2009kn, Aglietti:2006tp}), the NLO QCD corrections to these LO
expressions will be large.

\section{Next-to-leading-order corrections \label{sc:nlo}}
%        =================================

The NLO QCD corrections to Higgs-pair production via gluon fusion have
been computed in the HTL, a long time ago \cite{Dawson:1998py}. 
The NLO result for the gluon-fusion cross section can be generically
expressed as \cite{Dawson:1998py} 
\begin{eqnarray}
\sigma_{NLO}(pp \rightarrow H H + X) & = &
\sigma_{LO} + \Delta
\sigma_{virt} + \Delta\sigma_{gg} + \Delta\sigma_{gq} +
\Delta\sigma_{q\bar{q}} \, , \nonumber %\\
\end{eqnarray}
\begin{eqnarray}
\sigma_{LO} & = & \int_{\tau_0}^1 d\tau~\frac{d{\cal
L}^{gg}}{d\tau}~\hat\sigma_{LO}(Q^2 = \tau s) \, , \nonumber \\ 
\Delta \sigma_{virt} & = & \frac{\alpha_s(\mu_R)}
{\pi}\int_{\tau_0}^1 d\tau~\frac{d{\cal L}^{gg}}{d\tau}~\hat
\sigma_{LO}(Q^2=\tau s)~C_{virt}(Q^2) \, , \nonumber \\ 
\Delta \sigma_{ij} & = & \frac{\alpha_{s}(\mu_R)} {\pi} \int_{\tau_0}^1
d\tau~ \frac{d{\cal L}^{ij}}{d\tau} \int_{\tau_0/\tau}^1 \frac{dz}{z}~
\hat\sigma_{LO}(Q^2 = z \tau s)\, C_{ij}(Q^2,z) \qquad (ij=gg,gq,q\bar q)
 \, ,\nonumber \\
C_{gg}(Q^2,z) & = & - z P_{gg} (z) \log \frac{\mu_F^{2}}{\tau s}
+ 6 [1+z^4+(1-z)^4] \left(\frac{\log (1-z)}{1-z} \right)_+
+ d_{gg}(Q^2,z) \, , \nonumber \\ 
C_{gq}(Q^2,z) & = & -\frac{z}{2} P_{gq}(z)
\log\frac{\mu_F^{2}}{\tau s(1-z)^2} + d_{gq}(Q^2,z) \, , \nonumber \\ 
C_{q\bar q}(Q^2,z) & = & d_{q\bar q}(Q^2,z)
\label{eq:nlocxn}
\end{eqnarray}
with $\hat\sigma_{LO}(Q^2)$ denoting the partonic cross section at LO
and the strong coupling $\alpha_s(\mu_R)$ is evaluated at the
renormalization scale $\mu_R$. The objects $d{\cal
L}^{ij}/d\tau~(i,j=g,q,\bar q)$ denote the parton-parton luminosities,
defined analogously to $d{\cal L}^{gg}/d\tau$ of Eq.~(\ref{eq:lgg}),
using the quark densities $q(x,\mu_F)$,
\begin{eqnarray}
\frac{d{\cal L}^{gq}}{d\tau} & = & \sum_{q,\bar q} \int_\tau^1 \frac{dx}{x}
\Big[ g(x,\mu_F) q\left(\frac{\tau}{x},\mu_F\right)
+ q(x,\mu_F) g\left(\frac{\tau}{x},\mu_F\right) \Big] \, , \nonumber \\
\frac{d{\cal L}^{q\bar q}}{d\tau} & = & \sum_q \int_\tau^1 \frac{dx}{x}
\Big[ q(x,\mu_F) \bar q\left(\frac{\tau}{x},\mu_F\right)
+ \bar q(x,\mu_F) q\left(\frac{\tau}{x},\mu_F\right) \Big]
\end{eqnarray}
at the factorization scale $\mu_F$ and $P_{ij}(z)~(i,j=g,q,\bar q)$ are
the specific Altarelli--Parisi splitting functions
\cite{Altarelli:1977zs}.

The quark-mass dependence is in general encoded in the LO cross section
$\hat\sigma_{LO}(Q^2)$ and the terms $C_{virt}(Q^2)$, $d_{ij}(Q^2,z)$ for
the virtual and real corrections, respectively. These expressions can
easily be converted into the differential cross section with respect to
$Q^2$,
\begin{eqnarray}
\frac{d\Delta\sigma_{virt}}{dQ^2} & = & \left.
\frac{\alpha_s\left(\mu_R\right)}{\pi}~\frac{d{\cal L}^{gg}}{d\tau}~
\frac{\hat{\sigma}_{LO}\left(Q^2 \right)}{s}~C_{virt}
\left(Q^2\right) \right|_{\tau = \frac{Q^2}{s}}, \nonumber \\[0.3cm]
\frac{d\Delta\sigma_{ij}}{dQ^2} & = & \left.
\frac{\alpha_s\left(\mu_R\right)}{\pi}\int_{\frac{Q^2}{s}}^1
\frac{dz}{z^2}~\frac{d{\cal L}^{ij}}{d\tau}~
\frac{\hat{\sigma}_{LO}\left(Q^2\right)}{s}~C_{ij}(Q^2,z) \right|_{\tau
= \frac{Q^2}{zs}} \, ,
\label{eq:nlodiff}
\end{eqnarray}
while the differential cross section at LO is given in
Eq.~(\ref{eq:lodiff}).
\begin{figure}[ht!]
\begin{center}
  \includegraphics[width=0.95\textwidth]{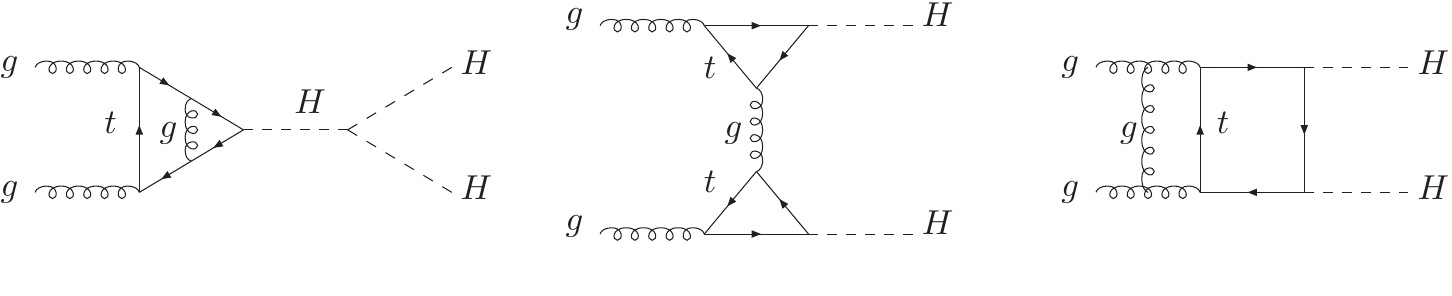} 
\vspace*{-0.5cm}

\caption{\label{fg:gghhvirt} \it Typical two-loop triangle (left),
one-particle reducible (middle) and box (right) diagrams contributing to
Higgs-pair production via gluon fusion at NLO.}
\end{center}
\end{figure}

Within the HTL, the Higgs coupling to gluons can be described by an
effective Lagrangian \cite{Ellis:1975ap, Shifman:1979eb, Inami:1982xt,
Spira:1995rr, Kniehl:1995tn}
\begin{equation}
{\cal L}_\mathrm{eff} = \frac{\alpha_s}{12\pi} G^{a\mu\nu} G^a _{\mu\nu}
\left(C_1 \frac{H}{v} - C_2 \frac{H^2}{2v^2} \right)
\label{eq:leff}
\end{equation}
involving the Wilson coefficients ($L_t = \log \mu_R^2/m_t^2$)
\cite{Chetyrkin:1997iv, Kramer:1996iq, Dawson:1998py, Schroder:2005hy,
Baikov:2016tgj, Grigo:2014jma, Spira:2016zna, Gerlach:2018hen}
\begin{eqnarray}
C_1 & = & 1 + \frac{11}{4} \frac{\alpha_s}{\pi} + \left\{
\frac{2777}{288} +
\frac{19}{16} L_t + N_F \left(\frac{L_t}{3}-\frac{67}{96} \right)
\right\} \left(\frac{\alpha_s}{\pi} \right)^2 + {\cal O}(\alpha_s^3)
\, , \nonumber \\
C_2 & = & C_1 + \left( \frac{35}{24} + \frac{2}{3} N_F \right)
\left(\frac{\alpha_s}{\pi} \right)^2 + {\cal O}(\alpha_s^3)
\label{eq:leffcoeff}
\end{eqnarray}
that are known up to N$^4$LO \cite{Schroder:2005hy, Baikov:2016tgj,
Spira:2016zna}. Since the top quark is integrated out, the number of
active flavours has been chosen as $N_F=5$.  If these effective Higgs
couplings to gluons in the calculation of the NLO QCD corrections are
used, the calculation of these is simplified to a one-loop calculation
for the virtual corrections and a tree-level one for the matrix elements
of the real corrections. The terms $C_{virt}(Q^2)$ and $d_{ij}(Q^2,z)$, for
the virtual and real corrections, approach in the HTL the simple
expressions
\begin{eqnarray}
C_{virt}(Q^2) & \to & \frac{11}{2} + \pi^2 +
C^\infty_{\triangle\triangle} +
\frac{33-2N_F}{6} \log\frac{\mu_R^2}{Q^2}, \nonumber \\
 C_{\triangle\triangle} & = &
\Re e~\frac{\int_{\hat t_-}^{\hat t_+} d\hat t \left\{ c_1 \left[
(C_\triangle F_\triangle + F_\Box) + \frac{p_T^2}{\hat t} 
G_\Box \right]^* + (\hat t \leftrightarrow \hat u) \right\}}
{\int_{\hat t_-}^{\hat t_+} d\hat t \left\{ |C_\triangle F_\triangle +
F_\Box |^2 + |G_\Box|^2 \right\}}, \nonumber \\[0.3cm]
C^\infty_{\triangle\triangle} & = & \left. C_{\triangle\triangle}
\right|_{c_1 = 2/9}, \nonumber \\[0.3cm]
d_{gg}(Q^2,z) \!\!\! & \to & \!\!\! - \frac{11}{2} (1-z)^3 \, , \
d_{gq}(Q^2,z) \to \frac{2}{3} z^2 - (1-z)^2 \, , \
d_{q\bar q}(Q^2,z) \to \frac{32}{27} (1-z)^3 \, ,
\label{eq:coeffvirt}
\end{eqnarray}
where $\hat s,\hat t, \hat u$ ($\hat s=Q^2$ at LO and for the virtual
corrections) denote the partonic Mandelstam variables and
$C_{\triangle\triangle}$ is the contribution of the one-particle
reducible diagrams, see Fig.~\ref{fg:gghhvirt}.

At NLO QCD, the full mass dependence of the LO partonic cross section has
been taken into account, while keeping the virtual corrections
$C_{virt}$ and the real corrections $d_{ij}$ in the HTL
(``Born-improved'' approach) \cite{Dawson:1998py}. This yields a
reasonable approximation for smaller invariant Higgs-pair masses and
approximates the full NLO result of the total cross section within about
15\% \cite{Borowka:2016ehy, Borowka:2016ypz, Baglio:2018lrj}. The NLO
QCD corrections in the HTL increase the cross section by $80-90\%$
\cite{Dawson:1998py}. Within the Born-improved HTL, the NNLO QCD
corrections have been obtained in Refs.~\cite{deFlorian:2013uza,
deFlorian:2013jea, Grigo:2014jma} increasing the total cross section by
a moderate amount of $20-30\%$ \cite{deFlorian:2013jea}. Beyond these
NNLO QCD corrections, the soft-gluon resummation (threshold resummation)
has been performed at next-to-next-to-leading logarithmic (NNLL)
accuracy for the total cross section and invariant mass distribution,
modifying the total cross section further by a small amount if the
central scales are chosen as $\mu_R=\mu_F=Q/2$ \cite{Shao:2013bz,
deFlorian:2015moa}.  Very recently, the N$^3$LO QCD corrections have
been computed in the Born-improved HTL resulting in a small modification
of the cross section beyond NNLO\cite{Banerjee:2018lfq, Chen:2019lzz,
Chen:2019fhs, Spira:2016zna}.  These N$^3$LO QCD corrections in the HTL
have been merged with the full top-mass effects of the NLO calculation
\cite{Chen:2019fhs}.
% \begin{figure}[hbtp]
% \begin{center}
%   \includegraphics[width=0.95\textwidth]{./plots/dia_virt.pdf} 
% \vspace*{-0.5cm}

% \caption{\label{fg:gghhvirt} \it Typical two-loop triangle (left),
% one-particle reducible (middle) and box (right) diagrams contributing to
% Higgs-pair production via gluon fusion at NLO.}
% \end{center}
% \end{figure}

The calculations in the HTL have been improved by several steps
including mass effects partially at NLO. The full mass effects in the
real correction terms $d_{ij}$ have been included by means of the full
one-loop real matrix elements for $gg\to HHg, gq \to HHq, q\bar q\to
HHg$. This improvement reduces the Born-improved HTL prediction for the
total cross section by about 10\% \cite{Frederix:2014hta,
Maltoni:2014eza} and is called the ``FTapprox'' approximation. The
calculation of the full real matrix elements has been performed by using
the {\tt MG5\_aMC@NLO} framework \cite{Alwall:2014hca, Hirschi:2015iia}.
Another improvement has been achieved by an asymptotic large-top-mass
expansion of the full NLO corrections at the level of the integral
\cite{Grigo:2013rya} and the integrand \cite{Grigo:2015dia}. This
indicated sizable mass effects in the virtual two-loop corrections
alone. In addition, the large top-mass expansion has been extended to
the virtual NNLO QCD corrections resulting in 5\% mass effects estimated
on top of the NLO result \cite{Grigo:2015dia}. The large-top-mass
expansion of the NLO QCD corrections has been used to perform a
conformal mapping of the expansion parameter and to apply Pad\'e
approximants. In this way, an approximation of the full calculation has
been achieved for $Q$ values up to about 700 GeV \cite{Grober:2017uho}.
Another approximation builds on an expansion in terms of a variable that
dominantly corresponds to the transverse momentum of the Higgs bosons.
The results of this approach show good agreement with the full
calculation for $Q$ values up to about 900 GeV \cite{Bonciani:2018omm}.
Analytical results are also available in the large-$Q$ limit
\cite{Davies:2018qvx}.  The latter have recently been combined with the
numerical results of Refs.~\cite{Borowka:2016ehy, Borowka:2016ypz} for
the full QCD corrections \cite{Davies:2019dfy}. In the following, we will
discuss the details of our NLO calculation.

\subsection{Virtual corrections \label{sc:virtuals}}
%           ===================
Typical diagrams of the two-loop virtual corrections are shown in
Fig.~\ref{fg:gghhvirt}. They can be arranged in three different classes:
(a) triangle, (b) one-particle-reducible and (c) box
diagrams\footnote{Note that we distinguish triangle and box diagrams
also at the two-loop level in terms of the number of particles attached
to the generic loop, i.e.~three particles (two gluons and an off-shell
Higgs for the triangle and two gluons and two on-shell Higgs bosons for
the box diagrams). The one-particle-reducible diagrams are a special
class.}. They contribute to the coefficient $C_{virt}(Q^2)$ of
Eq.~(\ref{eq:nlocxn}),
\begin{equation}
C_{virt}(Q^2) = 2 \Re e~\frac{\int_{\hat t_-}^{\hat t_+} d\hat t \left\{
(C_\triangle F_\triangle + F_\Box)^* [C_\triangle (\Delta
F_\triangle) + \Delta F_\Box] + G_\Box^* (\Delta G_\Box) \right\} }
{\int_{\hat t_-}^{\hat t_+} d\hat t \left\{ |C_\triangle F_\triangle +
F_\Box |^2 + |G_\Box|^2 \right\} },
\end{equation}
where $\Delta F_\triangle, \Delta F_\Box$ and $\Delta G_\Box$ denote the
virtual corrections to the corresponding LO form factors. While $\Delta
F_\triangle$ involves only virtual corrections to the triangle diagram,
$\Delta F_\Box$ and $\Delta G_\Box$ acquire contributions from the
one-particle-reducible and box diagrams.

\subsubsection{Triangle diagrams}
%              =================
\begin{figure}[hbt]
\begin{center}
  \includegraphics[width=0.9\textwidth]{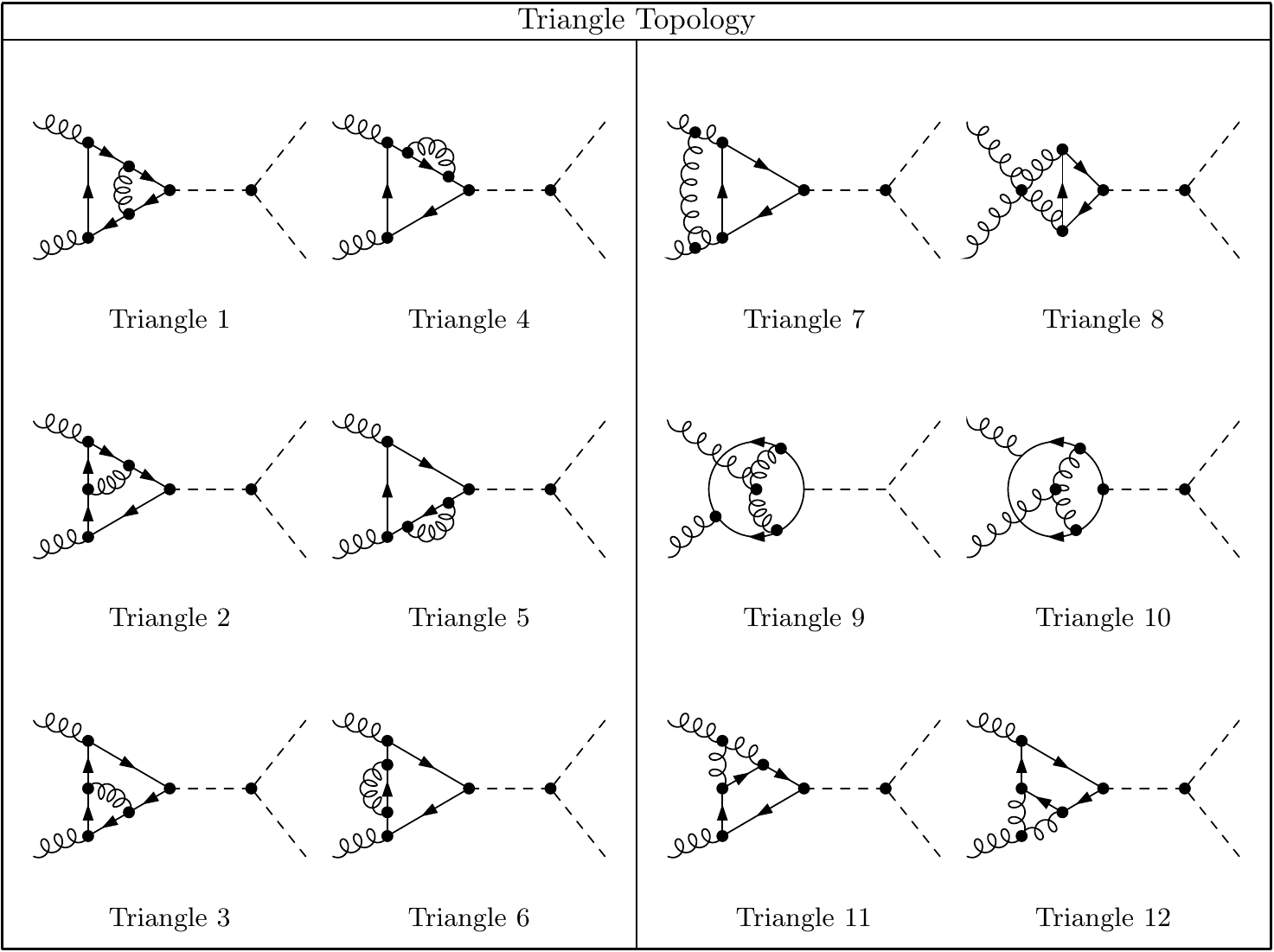} 
\caption{\label{fg:triadia} \it Two-loop triangle diagrams contributing to
Higgs-pair production via gluon fusion.}
\end{center}
\end{figure}
\noindent
The generic 2-loop triangle diagrams contributing to the virtual
coefficient $C_{virt}(Q^2)$ are shown in Fig.~\ref{fg:triadia}. They only
contribute to the spin-0 form factor $F_1$ of Eq.~(\ref{eq:lomat}) and
can be parametrized as the correction $\Delta F_\triangle$ to the
form factor $F_\triangle$,
\begin{equation}
\Delta F_\triangle = \frac{\alpha_s}{\pi}~{\cal
C}_{virt}(Q^2)~F_\triangle \, ,
\end{equation}
where ${\cal C}_{virt}(Q^2)$ denotes the {\it complex} virtual
coefficient relative to the LO form factor $F_\triangle$ of the
amplitude. This virtual coefficient is related to the
single-Higgs case so that the relative QCD corrections can be simply
obtained from the known (complex) virtual coefficient ${\cal
C}^H_{virt}(M_H^2)$ of single Higgs production \cite{Spira:1995rr,
Graudenz:1992pv, Harlander:2005rq, Anastasiou:2009kn,
Aglietti:2006tp}\footnote{The finite part of the complex virtual
coefficient ${\cal C}^H_{virt}$ has been shown in Fig.~7a of
Ref.~\cite{Spira:1995rr} after renormalization. We define the top mass
on-shell, i.e.~use the coefficient for $\mu_Q=m_Q$ of this figure for
the triangle-diagram contribution to our central prediction.},
\begin{equation}
{\cal C}_{virt}(Q^2) = \left. {\cal C}^H_{virt}(M_H^2)\right|_{M_H^2 \to
Q^2} \, .
\end{equation}
In the HTL, this virtual coefficient (before renormalization) approaches
the expression
\begin{equation}
{\cal C}_{virt} (Q^2) \to \frac{\Gamma(1-\epsilon)}{\Gamma(1-2\epsilon)}
\left(\frac{4\pi\mu_0^2 (1-i\bar\epsilon)}{-Q^2}\right)^\epsilon \left\{
-\frac{3}{2\epsilon^2} + \frac{3}{4} - \frac{\pi^2}{4} \right\}
\end{equation}
with the 't Hooft scale $\mu_0$, where the (infinitesimal) regulator
$\bar\epsilon$ defines the proper analytical continuation of this
expression. This result has to be followed by the renormalization of the
strong coupling $\alpha_s$ and the top mass $m_t$ that will be discussed
in Section \ref{sc:renorm}. In addition, we have subtracted the HTL to
obtain the pure top-mass effects at NLO (relative to the massive LO
expression $F_\triangle$) to ensure that in the end the results of the
program {\tt Hpair} \cite{hpair} can be added back. This last step will
be discussed in Section \ref{sc:renorm}, too.

\subsubsection{One-particle-reducible diagrams}
%              ===============================
The one-particle-reducible contribution is depicted in
Fig.~\ref{fg:gghhvirt} (middle diagram), where a second diagram
with the initial gluons interchanged has to be added. These will
constitute the $\hat t$- and $\hat u$-channel parts where the second is
related to the first just by the interchange $\hat t\leftrightarrow \hat
u$ [see $C_{\triangle\triangle}$ of Eq.~(\ref{eq:coeffvirt})]. The
analytical expression of the coefficient $c_1$ can be related to the top
contribution of the process $H\to Z\gamma$ \cite{Cahn:1978nz,
Bergstrom:1985hp}. The basic building block will be the one-loop
contribution of the Higgs coupling to an on-shell and an off-shell gluon
that is described, after translating all couplings and masses, by the
``effective'' Feynman rule, \\
\begin{picture}(100,100)(20,0)
\put(-120,-560){\includegraphics[width=1.00\textwidth]{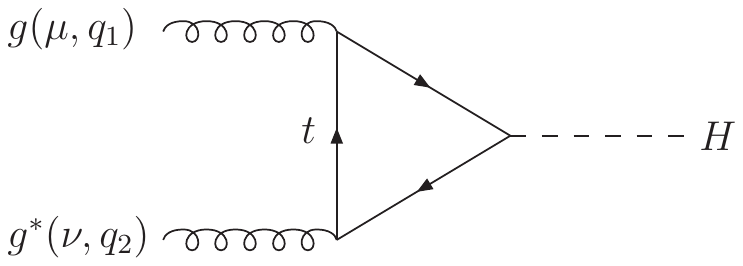}}
\put(240,58){$\displaystyle -i \frac{\alpha_s}{\pi v}
\Big[I_1(\tau,\lambda)-I_2(\tau,\lambda)\Big]
\Big[ q_2^\mu q_1^\nu - (q_1 q_2) g^{\mu\nu} \Big] \delta_{ab}\, ,$}
\end{picture} \\[-0.5cm]
where the functions $I_{1,2}$ are defined as \cite{Gunion:1989we}
\begin{eqnarray}
I_1(\tau,\lambda) & = & \frac{\tau\lambda}{2(\tau-\lambda)} +
\frac{\tau^2\lambda^2}{2(\tau-\lambda)^2} \left[ f(\tau) - f(\lambda)
\right]
+ \frac{\tau^2\lambda}{(\tau-\lambda)^2} \left[ g(\tau)
- g(\lambda) \right], \nonumber \\
I_2(\tau,\lambda) & = & - \frac{\tau\lambda}{2(\tau-\lambda)}\left[
f(\tau) - f(\lambda) \right],
\end{eqnarray}
with $\tau = 4m_t^2/m_H^2$, $\lambda = 4 m_t^2/q_2^2$ and the basic
functions
\begin{eqnarray}
g(\tau) & = & \left\{ \begin{array}{ll} \displaystyle \sqrt{\tau-1}
\arcsin \frac{1}{\sqrt{\tau}} & \tau \ge 1 \\ \displaystyle
\frac{\sqrt{1-\tau}}{2} \left[ \log \frac{1+\sqrt{1-\tau}}
{1-\sqrt{1-\tau}} - i\pi \right] & \tau < 1 \end{array} \right.
\end{eqnarray}
and $f(\tau)$ defined in Eq.~(\ref{eq:ftau}). Implementing this
building block for the two top loops of the one-particle-reducible
diagrams, one arrives at the final coefficient $c_1$ of
Eq.~(\ref{eq:coeffvirt}),
\begin{eqnarray}
c_1 & = & 2 \Big[ I_1(\tau,\lambda_{\hat t}) -I_2(\tau,\lambda_{\hat
t}) \Big]^2
\label{eq:1prc1}
\end{eqnarray}
with $\lambda_{\hat t} = 4 m_t^2/\hat t$ (and $\lambda_{\hat u} = 4
m_t^2/\hat u$ for the $\hat t\leftrightarrow \hat u$ interchanged
contribution accordingly).  This expression, inserted in the coefficient
$C_{\triangle\triangle}$ of Eq.~(\ref{eq:coeffvirt}), determines the
contribution of the one-particle-reducible diagrams analytically and
agrees with the previous calculation of Ref.~\cite{Degrassi:2016vss}. In
the HTL, this coefficient approaches the value $c_1\to 2/9$ in accordance
with Eq.~(\ref{eq:coeffvirt}). We have subtracted the HTL with $c_1=2/9$
from the coefficient $C_{\triangle\triangle}$ in order to account for
the NLO top-mass effects only so that eventually the results of the program
{\tt Hpair} \cite{hpair} can be added back. While the total effect of
the one-particle-reducible contributions on the total cross section
ranges below the per-cent level, the finite mass effects at NLO
contribute less than one per mille.

\begin{figure}[hbt]
\begin{center}
 \includegraphics[width=0.80\textwidth]{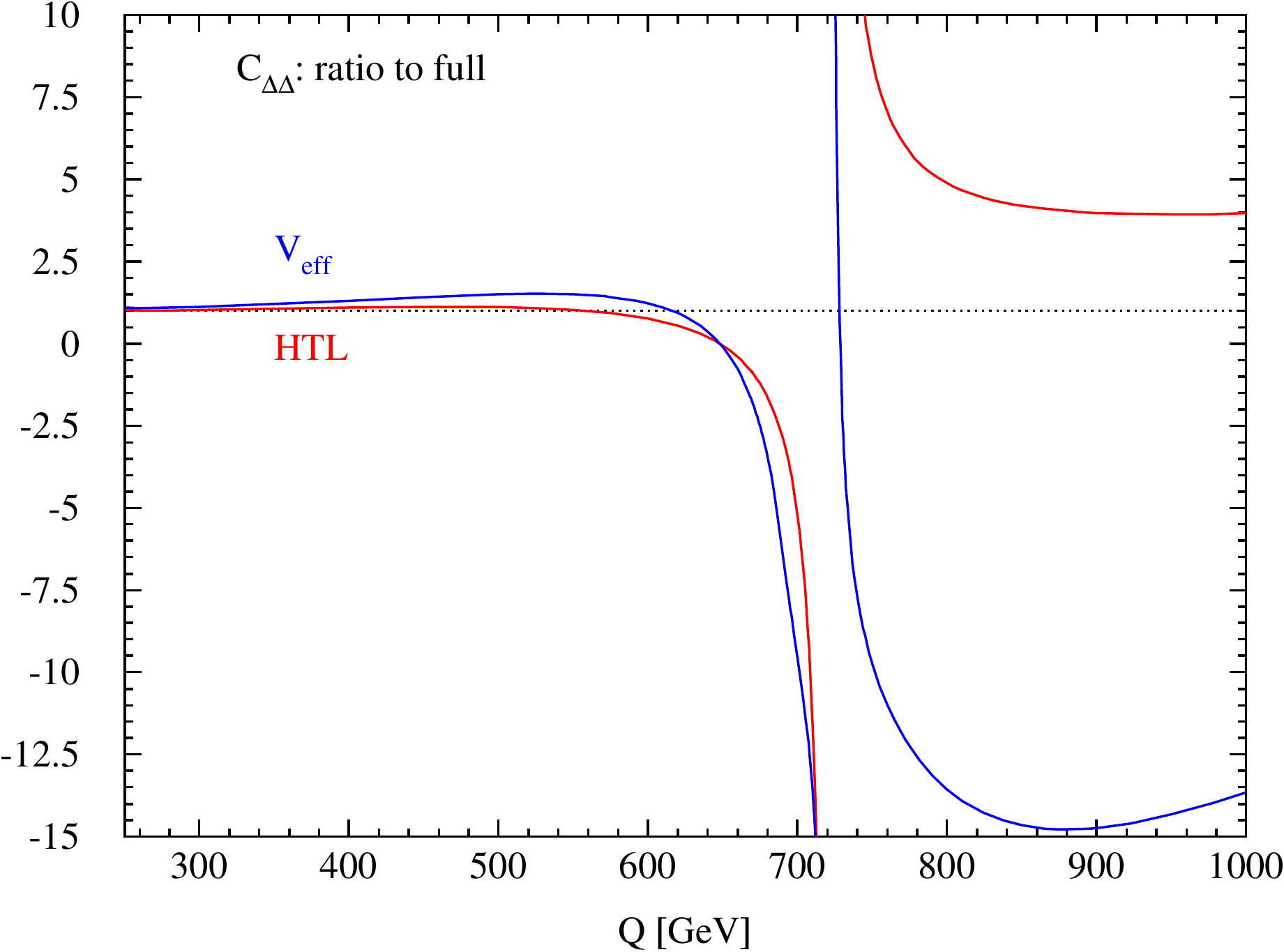} 
\caption{\label{fg:1prcomp} \it Comparison of the approximation of
Ref.~\cite{deFlorian:2017qfk} (blue) for the one-particle-reducible
contributions and the HTL (red), both normalized to the full analytical
expression.  The singularity at about 720 GeV is due to a sign change of
the exact expression.}
\end{center}
\end{figure}

Reference \cite{deFlorian:2017qfk} has proposed an approximation of this
one-particle-reducible contribution in terms of the triangle form factor
of two on-shell external gluons,
\begin{eqnarray}
C_{\triangle\triangle} & = &
\Re e~\frac{\int_{\hat t_-}^{\hat t_+} d\hat t \left[
(C_\triangle F_\triangle + F_\Box)^* V_{eff}^2\right]}
{\int_{\hat t_-}^{\hat t_+} d\hat t \left\{ |C_\triangle F_\triangle +
F_\Box |^2 + |G_\Box|^2 \right\}}, \nonumber \\[0.3cm]
V_{eff} & = & F_\triangle (\bar\tau_t)
\end{eqnarray}
with $\bar\tau_t = 16 m_t^2/Q^2$ [i.e.~$\tau_t$ of
Eq.~(\ref{eq:ftriangle}) evaluated at half the invariant Higgs-pair mass
$Q/2$ instead of $Q$], where the function $F_\triangle$ can be found in
Eq.~(\ref{eq:ftriangle}). Since Ref.~\cite{deFlorian:2017qfk} works in
the HTL, the contribution of the second form factor $F_2$ vanishes,
i.e.~$G_\Box\to 0$, and the approximation $V_{eff}^2/2$ is in fact
treated as an approximation for the coefficient $c_1$ of the exact
expression of $C_{\triangle\triangle}$ as given in
Eq.~(\ref{eq:coeffvirt})\footnote{Since $V_{eff}$ is symmetric with
respect to $\hat t \leftrightarrow \hat u$ the additional factor 2
emerges from the second term in the numerator of
$C_{\triangle\triangle}$ in Eq.~(\ref{eq:coeffvirt}).}. Thus, the
approximate expression involving the coefficient $c_1$ has to be
compared to the corresponding expression involving the exact coefficient
$c_1$ of Eq.~(\ref{eq:1prc1}). This comparison is presented, normalized
to the exact expression, in Fig.~\ref{fg:1prcomp} and shows that the
approximation of Ref.~\cite{deFlorian:2017qfk} is not better than the
HTL.
% \begin{figure}[hbt]
% \begin{center}
% \vspace*{-5.5cm}

% \hspace*{-2cm}
%  \includegraphics[width=0.90\textwidth]{./plots/1pr_comp.pdf} 
% \vspace*{-5.5cm}

% \caption{\label{fg:1prcomp} \it Comparison of the approximation of
% Ref.~\cite{deFlorian:2017qfk} (blue) for the one-particle-reducible
% contributions and the HTL (red), both normalized to the full analytical
% expression.  The singularity at about 720 GeV is due to a sign change of
% the exact expression.}
% \end{center}
% \end{figure}

\subsubsection{Box diagrams}
%              ============
The third class of two-loop contributions to the virtual corrections is
given by the box diagrams. The generic box diagrams are shown in
Figs.~\ref{fg:boxdia1}--\ref{fg:boxdia4} in the Appendix. The
simultaneous exchange of the gluons and Higgs bosons has to be added to
complete the set of diagrams. The only exception is diagram 44 that is
already totally symmetric so that in the final end there are 93 two-loop
box diagrams.  The generic 47 diagrams are grouped into 6 topology
classes. The first 5 topologies contain only a virtual threshold for
$Q^2 > 4m_t^2$. The diagrams of topology 6 on the other hand develop a
second threshold for $Q^2>0$, because two virtual gluon lines next to
the external gluons can be cut. This implies that the form factors are
complex in the entire $Q^2$ range. Therefore, a dedicated treatment of
this last topology in terms of a suitably constructed infrared
subtraction term to isolate the associated infrared singularities is
required.

In the following, we will exemplify our method for the boxes
39 of topology 5 and 45 of topology 6. The diagrams of topologies 1--5
are treated analogously to box 39 and those of topology 6 analogously to
box 45. The algebraic manipulation of the traces and projections
onto the form factors have been performed with the help of the
symbolic tools {\tt FORM}~\cite{Vermaseren:2000nd,Kuipers:2012rf},
{\tt Reduce}~\cite{Hearn:1971zza}, and {\tt Mathematica}~\cite{Mathematica}.
Our method of Feynman parametrization and end-point subtraction
to isolate the ultraviolet singularities for the numerical integration
has first been applied to the NLO two-loop QCD corrections to
$H\to\gamma\gamma, Z\gamma$ in Refs.~\cite{Djouadi:1990aj, Spira:1991tj}
and later to the squark-loop contributions to $h,H \leftrightarrow
gg,\gamma\gamma$ within the minimal supersymmetric extension of the SM
\cite{Muhlleitner:2006wx}. The method of the infrared subtraction as
applied to topology 6 originates from numerical cross checks of the full
NLO QCD corrections to single Higgs production in
Refs.~\cite{Spira:1995es, Spira:1995rr, Graudenz:1992pv,
Muhlleitner:2006wx}. The stabilization of virtual thresholds by
integration by parts of the integrand has first been applied to the
SUSY--QCD corrections to single Higgs production in
Refs.~\cite{Muhlleitner:2010nm, Muhlleitner:2010zz}.  The basic idea
behind the integration by parts is to reduce the power of the
threshold-singular denominator and in this way to stabilize the
numerical integration. The treatment of the thresholds in our approach
is performed by replacing the squared top mass $m_t^2$ by a complex
counter part
\begin{equation}
m_t^2 \to m_t^2 (1-i\bar\epsilon)
\label{eq:imaginary}
\end{equation}
with a positive regulator $\bar\epsilon > 0$ to ensure proper
micro-causality. This defines the analytical continuation of our
two-loop box integrals. In the following, the parameter $\bar\epsilon$
will be kept finite in our numerical analysis, while the narrow-width
limit $\bar\epsilon\to 0$ is achieved by a Richardson extrapolation
\cite{Richardson}. This will be discussed in more detail in the
following paragraphs.

\paragraph{Box 39}~\\
%          ======
\begin{figure}[hbt]
\begin{center}
\vspace*{-1.0cm}

\hspace*{1.5cm}
 \includegraphics[width=1.15\textwidth]{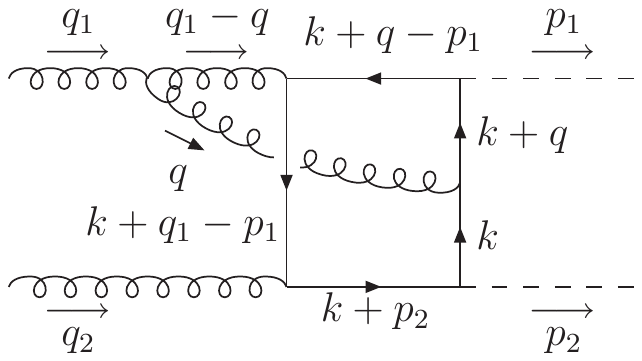} 
\vspace*{-24.0cm}

\caption{\label{fg:box39} \it Explicit definitions of the virtual
momenta in box 39.}
\end{center}
\end{figure}
\noindent
Using the definition of real and virtual momenta as in
Fig.~\ref{fg:box39}, the contribution to the tensor $A^{\mu\nu}$ [see
Eq.~(\ref{eq:lomat})] of the virtual two-loop corrections is given by
\begin{eqnarray}
%A_{39}^{\mu\nu} \!\! & \!\!\!\! = \!\!\!\! & \!\! \frac{3}{16}\,
A_{39}^{\mu\nu} & = & \frac{3}{16}\,
\frac{\alpha_s}{\pi}\, (4\pi)^4\, B_{39}^{\mu\nu} \, , \nonumber \\
B_{39}^{\mu\nu} \!\! & \!\!\!\! = \!\!\!\! & \!\! \int \frac{d^n k d^n
q}{(2\pi)^{2n}} \frac{Tr\Big\{ (\,\slsh{k}+\slsh{q}-\slsh{p}_1+m_t)
(\,\slsh{k}+\slsh{q}+m_t) \gamma^\sigma (\,\slsh{k}+m_t)
(\slsh{k}+\slsh{p}_2+m_t) \gamma^\nu
(\,\slsh{k}+\slsh{q}_1-\slsh{p}_1+m_t) \gamma^\rho \Big\}}
{[(k+q)^2-m_t^2] [(k+q-p_1)^2-m_t^2] [(k+p_2)^2-m_t^2]
[(k+q_1-p_1)^2-m_t^2]} \nonumber \\[0.3cm]
& & \qquad \times~\frac{g_{\rho\sigma} (2q-q_1)^\mu - g^\mu_\rho
(q-2q_1)_\sigma - g^\mu_\sigma (q+q_1)_\rho}{(k^2-m_t^2)(q-q_1)^2 q^2}
\, ,
\label{eq:mat39}
\end{eqnarray}
where $k,q$ are the loop momenta that are integrated over. The Feynman
parametrization is first performed for the integration over $k$.
We provide Feynman parameters $x_1,\ldots,x_4$ for the first four
propagators in the denominator and $1-\sum_i x_i$ for the last one
($k^2-m_t^2$). Performing the substitutions
\begin{equation}
x_1 = (1-x)(1-y)\, , \quad x_2 = (1-x)y\, , \quad x_3 = xzr \, ,
\quad x_4 = xz(1-r) \, ,
\end{equation}
we arrive at a four-dimensional integral over $x,y,z,r$ with integration
boundaries from 0 to 1. To symmetrize the $n$-dimensional
$k$-integration, we have to perform the shift
\begin{eqnarray}
k & \to & k - Q_1 \, , \nonumber \\
Q_1 & = & (1-x)q + xzq_1 + xzr q_2 - [(1-x)y+xz] p_1 \, ,
\end{eqnarray}
in both the numerator and denominator. The residual (properly
normalized) denominator after the $k$-integration is treated as a
propagator for the second loop integration over $q$. We attribute
additional Feynman parameters $x_5, x_6$ to this residual propagator and
the next one [$(q-q_1)^2$] and $1-x_5-x_6$ for the last one ($q^2$) in
Eq.~(\ref{eq:mat39}). Performing the substitution\footnote{Note that
$s$ denotes a Feynman parameter here and not the squared hadronic
c.m.~energy. The same holds for $z$.}
\begin{equation}
x_5 = s\, , \quad x_6 = (1-s)t\, ,
\end{equation}
we again arrive at integrals over $s,t$ from 0 to 1. This latter
parametrization requires the shift
\begin{eqnarray}
q & \to & q - Q_2 \, , \nonumber \\
Q_2 & = & - [zs + (1-s)t] q_1 - zrs q_2 - (y-z)s p_1
\end{eqnarray}
in the numerator and denominator to be able to perform the loop
integration over $q$ symmetrically. After projecting on the two
form factors, we finally arrive at integrals of the type
\begin{equation}
\Delta F_i = \frac{\alpha_s}{\pi}~\Gamma(1+2\epsilon)
\left(\frac{4\pi\mu_0^2}{m_t^2}\right)^{2\epsilon} \int_0^1 d^6 x~
\frac{x^\epsilon(1-x)^\epsilon s^{-1-\epsilon} H_i(\vec
x)}{N^{3+2\epsilon}(\vec x)}
\label{eq:fi39}
\end{equation}
with $\vec x = (x,y,z,r,s,t)$ and $d^6x = dx\,dy\,dz\,dr\,ds\,dt$.
$H_i(\vec x)$ denotes the full numerator, including regular factors of
the Jacobians due to the Feynman parametrization and substitutions, and
singular as well as higher powers of the dimensional regulator
$\epsilon$, and $N(\vec x)$ the final denominator,
\begin{eqnarray}
N(\vec x) & = & 1 + \rho_s xzr \Big\{ xz + (1-x) [zs+(1-s)t] \Big\}
\nonumber \\
& & \quad - \rho_t x \Big\{ z(1-y-r) + (y-z)[z+(1-x)(1-s)(t-z)] \Big\}
\nonumber \\
& & \quad + \rho_u xzr \Big\{ xz + (1-x) [zs+(1-s)y] \Big\} \nonumber \\
& & \quad - \rho_H \Big\{ [xz+(1-x)y][1-xz-(1-x)y] - x(1-x)s(y-z)^2
\Big\} \, ,
\end{eqnarray}
where we define $\rho_s = \hat s/m_t^2 = Q^2/m_t^2$, $\rho_t = (\hat
t-M_H^2)/m_t^2$, $\rho_u = (\hat u-M_H^2)/m_t^2$ and $\rho_H =
M_H^2/m_t^2$.  The singular powers in $\epsilon$ of $H_i(\vec x)$ arise
from powers of $k^2$ and $q^2$ in the numerators of the final
integrations of the loop momenta $k$ and $q$. It is important that the
final denominator develops the form of $1+O(1/m_t^2)$ to ensure that no
further ultraviolet nor infrared singularities arise from this part of
the integrand.

The integral for $\Delta F_i$ of Eq.~(\ref{eq:fi39}) is singular for
$s\to 0$. To separate this singularity from the integral, we perform an
endpoint subtraction,
\begin{eqnarray}
\Delta F_i & = & \frac{\alpha_s}{\pi}~\Gamma(1+\epsilon)
\frac{\Gamma(1-\epsilon)}{\Gamma(1-2\epsilon)}
\left(\frac{4\pi\mu_0^2}{m_t^2}\right)^{2\epsilon}  \left[ \Delta
F_{i,1} + \Delta F_{i,2} \right] \, , \nonumber \\
\Delta F_{i,1} & = & \int_0^1 \frac{d^6 x}{s} \left\{ \frac{H_i(\vec
x)}{N^3(\vec x)} (1+\epsilon L) - \left. \frac{H_i(\vec x)}{N^3(\vec x)}
\right|_{s=0} (1+\epsilon L_0) \right\} \, , \nonumber \\
\Delta F_{i,2} & = & -\frac{1}{\epsilon} \int_0^1 d^5 x \left.
\frac{H_i(\vec x)}{N^3(\vec x)} \right|_{s=0} \left[1+\epsilon L_1 +
\epsilon^2 \left( \frac{L_1^2}{2} + 3\zeta_2 \right) \right] \nonumber \\
\mbox{with} \qquad L & = & \log\frac{x(1-x)}{s} - 2 \log N(\vec x) \, ,
\nonumber \\
L_0 & = & \log\frac{x(1-x)}{s} - 2 \log N(\vec x)|_{s=0} \, , \nonumber \\
L_1 & = & \log [x(1-x)] - 2 \log N(\vec x)|_{s=0} \, ,
\end{eqnarray}
where in the second term $\Delta F_{i,2}$ the integration over $s$ has
been performed analytically and the integration measure is given by $d^5
x = dx\,dy\,dz\,dr\,dt$. It should be noted that in the terms $L,
L_0, L_1$ the logarithms of the denominator $N$ need to be linear in $N$
to be consistent with the analytical continuation along the proper
Riemann sheet. We have checked numerically that the first (subtracted)
part $\Delta F_{i,1}$ is finite for each order in the dimensional
regulator $\epsilon$ by introducing cuts in the integration boundaries,
i.e.~integrating from $\tilde\epsilon$ to $1-\tilde\epsilon$, varying
$\tilde\epsilon$ down to $10^{-10}$ and checking that the integrals
become independent of $\tilde\epsilon$.

These integrals are numerically stable below the virtual $t\bar
t$-threshold, i.e.~for $Q^2 < 4 m_t^2$ or $\rho_s < 4$. However, above
this threshold, the integrals have to be stabilized. We have achieved
this stabilization by means of integration by parts with respect to the
Feynman parameter $z$. The denominator is a quadratic polynomial in $z$,
\begin{eqnarray}
N(\vec x) & = & a z^2 + b z + c \nonumber \\
\mbox{with} \qquad a & = & x [\rho_s r + \rho_t + \rho_u r + \rho_H]
[1 - (1-x)(1-s)] \, , \nonumber \\
b & = & \rho_s x(1-x)r(1-s)t - \rho_t x [1-r - (1-x)(1-s)(y+t)]
\nonumber \\
& + & \rho_u x(1-x)y r (1-s) - \rho_H x [1-2(1-x)y(1-s)]
\, , \nonumber \\
c & = & 1 - \rho_t x(1-x)y(1-s)t - \rho_H (1-x) y [1-y + xy(1-s)] \, .
\end{eqnarray}
To simplify the integration by parts, we insert a unit factor
$\Delta/\Delta$ with $\Delta=4ac-b^2$ in the integrand and replace
$\Delta$ in the numerator by the expression
\begin{equation}
\Delta = 4 a N - (\partial_z N)^2 = 4 a N - (2az+b)^2 \, .
\end{equation}
Then the following manipulation can be performed,
\begin{eqnarray}
\int_0^1 dz~\frac{H_i(\vec x)}{N^3} & = & \frac{1}{\Delta} \left\{
\left. \left[\frac{2a+b}{2N^2} H_i(\vec x) + \frac{\partial_z H_i(\vec
x)}{2N}\right]\right|_{z=1} - \left. \left[\frac{b}{2N^2} H_i(\vec x) +
\frac{\partial_z H_i(\vec x)}{2N}\right]\right|_{z=0} \right. \nonumber \\
& & \left. + \int_0^1 dz~\left[ \frac{3a}{N^2} H_i(\vec x) -
\frac{\partial_z^2 H_i(\vec x)}{2N} \right] \right\}
\end{eqnarray}
and analogously for integrals involving additional powers of $\log N$
factors in the numerator of the integrand. The progress achieved with
these integrations by parts is that the maximal power of the denominator
in the new integral is reduced by one compared to the original integral.
One could perform additional integrations by parts with respect to
another Feynman parameter. However, we did not investigate this further,
since the stability we achieved at this point has been sufficient for
the numerical integrations for the top loops\footnote{For the bottom
loops, additional stabilization of the numerical integration is required.
This is left for future work.}.

After performing the integrations by parts, the integral is stable for
regulators $\bar \epsilon$ [see Eq.~(\ref{eq:imaginary})] down to $0.05$
for the relevant Higgs mass, top mass and $Q^2$ range. Since this is
still apart from the plateau of the narrow-width limit, we performed a
Richardson extrapolation \cite{Richardson} from finite values of $\bar
\epsilon$ down to zero. Richardson extrapolation is possible since the
$\bar \epsilon$-dependence of the integral is polynomial for small
values of $\bar \epsilon$. The basic principle behind this extrapolation
method is very simple: let a function $f(\bar\epsilon)$ behave for small
$\bar\epsilon$ as
\begin{equation}
f(\bar\epsilon) = f(0) + {\cal O}(\bar\epsilon^n) \, .
\end{equation}
If we know $f(\bar\epsilon)$ for two different values $\bar\epsilon$
and $t\bar\epsilon$, we can construct the new function
\begin{equation}
R_1(\bar\epsilon,t) = \frac{t^n f(\bar\epsilon)-f(t
\bar\epsilon)}{t^n-1} \, .
\label{eq:richardson}
\end{equation}
This function shows a better convergence towards the value at
$\bar\epsilon=0$,
\begin{equation}
R_1(\bar\epsilon,t) = f(0) + {\cal O}(\bar\epsilon^{n+1}) \, .
\end{equation}
Our integrals $I(\bar\epsilon)$ behave for small values of $\bar\epsilon$ as
\begin{equation}
I(\bar\epsilon) = I(0) + {\cal O}(\bar\epsilon)
\end{equation}
so that the first new extrapolation function in our case is given by
\begin{equation}
R_1(\bar\epsilon,t) = \frac{t I(\bar\epsilon)-I(t \bar\epsilon)}{t-1}
= I(0) + {\cal O}(\bar\epsilon^2) \, .
\end{equation}
Using an additional value of $\bar\epsilon$, this method can be
repeated iteratively for the new function obtained by applying
Eq.~(\ref{eq:richardson}),
\begin{equation}
R_2(\bar\epsilon,t) = \frac{t^2 R_1(\bar\epsilon)-R_1(t
\bar\epsilon)}{t^2-1} = I(0) + {\cal O}(\bar\epsilon^3) \, .
\end{equation}
In this way, the estimated error is reduced by each additional iteration.
We have used this method for a set of $\bar\epsilon$ separated by
factors of $t=2$. Then, we obtain the following extrapolation
polynomials,
\begin{eqnarray}
R_1(\bar\epsilon) & = & 2 I(\bar\epsilon) -I(2\bar\epsilon) = I(0) +
{\cal O}(\bar\epsilon^2) \, , \nonumber \\
R_2(\bar\epsilon) & = & \frac{1}{3} \Big[ 8 I(\bar\epsilon) - 6
I(2\bar\epsilon) + I(4\bar\epsilon) \Big] = I(0) + {\cal
O}(\bar\epsilon^3) \, , \nonumber \\
R_3(\bar\epsilon) & = & \frac{1}{21} \Big[ 64 I(\bar\epsilon) - 56
I(2\bar\epsilon) + 14 I(4\bar\epsilon) - I(8\bar\epsilon) \Big] = I(0)
+ {\cal O}(\bar\epsilon^4) \, , \nonumber \\
R_4(\bar\epsilon) & = & \frac{1}{315} \Big[ 1024 I(\bar\epsilon) - 960
I(2\bar\epsilon) + 280 I(4\bar\epsilon) - 30 I(8\bar\epsilon) +
I(16\bar\epsilon) \Big] = I(0) + {\cal O}(\bar\epsilon^5)
\end{eqnarray}
and so on. We have used extrapolation polynomials up to
$R_9(\bar\epsilon)$. To determine the extrapolation error, we have
chosen different sets of $\bar\epsilon$ values and derived the spread of
the extrapolated values appropriately (see Section \ref{sc:results} for
more details).

\paragraph{Box 45}~\\
%          ======
\begin{figure}[hbt]
\begin{center}
\vspace*{-1.0cm}

\hspace*{1.5cm}
 \includegraphics[width=1.15\textwidth]{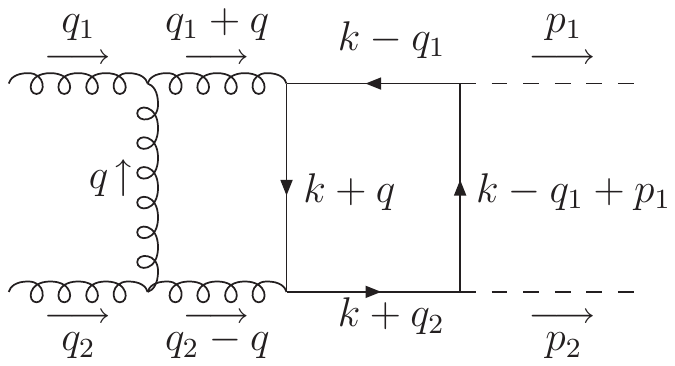} 
\vspace*{-24.0cm}

\caption{\label{fg:box45} \it Explicit definitions of the virtual
momenta in box 45.}
\end{center}
\end{figure}
\noindent
Based on the distribution of the loop and external momenta of
Fig.~\ref{fg:box45}, the contribution to the two-loop matrix element is
given by
\begin{eqnarray}
A_{45}^{\mu\nu} & = & \frac{3}{8}\, \frac{\alpha_s}{\pi}\, (4\pi)^4\,
B_{45}^{\mu\nu} \, , \nonumber \\
B_{45}^{\mu\nu} & \!\!\! = \!\!\! & \int \frac{d^n k d^n q}{(2\pi)^{2n}}
\frac{Tr\Big\{ (\,\slsh{k}-\slsh{q}_1+m_t)
(\,\slsh{k}-\slsh{q}_1+\slsh{p}_1+m_t) (\,\slsh{k}+\slsh{q}_2+m_t)
\gamma^\sigma (\,\slsh{k}+\slsh{q}+m_t) \gamma^\rho \Big\}}
{[(k+q)^2-m_t^2] [(k+q_2)^2-m_t^2] [(k+p_1-q_1)^2-m_t^2]
[(k-q_1)^2-m_t^2]} \nonumber \\[0.3cm]
& & \qquad \times~\frac{\Big\{ g_{\rho\tau} (2q+q_1)^\mu - g^\mu_\rho
(q+2q_1)_\tau - g^\mu_\tau (q-q_1)_\rho \Big\}}{(q+q_1)^2 (q-q_2)^2 q^2}
\nonumber \\[0.3cm]
& & \qquad \times~\Big\{ g^{\nu\tau} (q+q_2)_\sigma + g^\nu_\sigma
(q-2q_2)^\tau -g_\sigma^\tau (2q-q_2)^\nu \Big\} \, .
\label{eq:mat45}
\end{eqnarray}
Following the same procedure as for box 39 for the Feynman
parametrization, we have first performed the parametrization of the
$k$-integration following the ordering of the denominator of
Eq.~(\ref{eq:mat45}). The shift in the loop momentum $k$ and the
corresponding substitutions of the Feynman parameters are given by
\begin{eqnarray}
k & \to & k - Q_1 \, , \nonumber \\
Q_1 & = & (1-x)q - xyq_1 + x(1-y) q_2 + xyz p_1 \, , \nonumber \\
x_1 & = & (1-x)\, , \quad x_2 = x(1-y)\, , \quad x_3 = xyz \, .
\end{eqnarray}
Performing the second loop integration over $q$ with the residual
(normalized) denominator of the $k$ integration as the first propagator
of the $q$ integration, attributing the additional Feynman parameters
$x_4, x_5, x_6$ to the remaining propagators in Eq.~(\ref{eq:mat45}) and
applying the substitutions\footnote{Again $z,s$ denote Feynman
parameters here.}
\begin{equation}
x_4 = rs\, , \quad x_5 = 1-s\, , \quad x_6 = (1-r)st \, ,
\end{equation}
we arrive at the final expressions for the shift of $q$ and the
denominator that contribute to the two form factors,
\begin{eqnarray}
q & \to & q - Q_2 \, , \nonumber \\
Q_2 & = & [yrs + 1-s] q_1 - [(1-y)rs+(1-r)st] q_2 - yzrs p_1 \, , \nonumber \\
N(\vec x) & = & r - \rho_s x \Big\{ xy(1-y)r + (1-x)
[1-s+yrs][(1-r)t+(1-y)r] \Big\} \nonumber \\
& & \quad - \rho_t xyzr \Big\{ 1-xy - (1-x)[yrs+1-s] \Big\}
- \rho_H xyzr \Big\{ 1-xyz-(1-x)yzrs \Big\}
\nonumber \\
& & \quad - \rho_u xyzr \Big\{ x(1-y) + (1-x)s [(1-r)t+(1-y)r] \Big\}
\end{eqnarray}
and the final integrals of the two form factors ($i=1,2$) can be cast into
the form
\begin{equation}
\Delta F_i = \Gamma(1+2\epsilon)
\left(\frac{4\pi\mu_0^2}{m_t^2}\right)^{2\epsilon} \int_0^1 d^6 x~
\frac{x^{1+\epsilon} (1-x)^\epsilon r^{1+\epsilon} s^{-\epsilon}
H_i(\vec x)}{N^{3+2\epsilon}(\vec x)} \, ,
\end{equation}
where $H_i(\vec x)$ contains all additional regular Feynman-parameter
factors from Jacobians and the normalization of the denominator of the
first loop-integration over $k$. It develops a singular Laurent-expansion in
$\epsilon$. The final denominator exhibits the basic form of
$r+O(1/m_t^2)$, so that the additional singular behavior is entirely
controlled by the limit of small $r$. Since the denominator is of the
form
\begin{eqnarray}
N(\vec x) & = & a r^2 + b r + c \, , \nonumber \\
\mbox{where} \qquad a & = & x(1-x)ys \Big[-\rho_s (1-y-t) + \rho_t yz -
\rho_u z(1-y-t) + \rho_H y z^2\Big] \, , \nonumber \\
b & = & 1 - \rho_s x \Big\{ xy(1-y) +(1-x)[(1-s)(1-y-t)+yst] \Big\}
- \rho_H xyz (1-xyz) \nonumber \\
&  & -\rho_t xyz [1-xy - (1-x)(1-s)] - \rho_u xyz [x(1-y)+(1-x)st]
\, , \nonumber \\
c & = & - \rho_s x(1-x)(1-s)t
\label{eq:abc}
\end{eqnarray}
with $a,c = {\cal O}(1/m_t^2)$ and $b=1+{\cal O}(1/m_t^2)$ and the
infrared singularities are universal (relative to the LO expressions)
the coefficient $a$ does not contribute to the infrared singularity
structure, because $a$ is subleading relative to $b$ in the limit $r\to
0$. Thus, we can construct infrared subtraction terms that turn the
contributions to the form factors into
\begin{eqnarray}
\Delta F_i & = & \frac{\alpha_s}{\pi}~\Gamma(1+2\epsilon)
\left(\frac{4\pi\mu_0^2}{m_t^2}\right)^{2\epsilon} (G_1 + G_2) \, , \nonumber \\
G_1 & = & \int_0^1 d^6 x~x^{1+\epsilon} (1-x)^\epsilon r^{1+\epsilon}
s^{-\epsilon} \left\{ \frac{H_i(\vec x)}{N^{3+2\epsilon}(\vec x)}
-\frac{H_i(\vec x)|_{r=0}}{N_0^{3+2\epsilon}(\vec x)} \right\} \, , \nonumber \\
G_2 & = & \int_0^1 d^6 x~x^{1+\epsilon} (1-x)^\epsilon r^{1+\epsilon}
s^{-\epsilon} \frac{H_i(\vec x)|_{r=0}}{N_0^{3+2\epsilon}(\vec
x)} \nonumber \\
\mbox{with} \qquad N_0(\vec x) & = & br + c \, .
\end{eqnarray}
Numerically, we have tested that the subtracted integral $G_1$ (after
expansion in the dimensional regulator $\epsilon$) is finite for each
coefficient of the expansion in $\epsilon$ individually by integrating
the Feynman-parameter integrals from $\tilde\epsilon$ to
$1-\tilde\epsilon$ with $\tilde\epsilon$ varied down to $10^{-10}$. The
second integral $G_2$ can be integrated over the Feynman parameter $r$
analytically giving rise to hypergeometric functions,
\begin{equation}
G_2 = \frac{1}{2+\epsilon} \int_0^1 d^5x~\frac{x^{1+\epsilon}
(1-x)^\epsilon s^{-\epsilon}}{c^{3+2\epsilon}}
~_2F_1\left(3+2\epsilon, 2+\epsilon; 3+\epsilon; -\frac{b}{c}\right)
\left. H_i(\vec x) \right|_{r=0}
\end{equation}
with $d^5x = dx\, dy\, dz\, ds\, dt$. Since this integral is singular
for $c\to 0$, we have to invert the last argument of the hypergeometric
function. Using the transformation relation
\begin{eqnarray}
~_2F_1 (a,b;c;z) & = &
\frac{\Gamma(c)\Gamma(b-a)}{\Gamma(b)\Gamma(c-a)}(-z)^{-a}~_2F_1\left(
a,1-c+a; 1-b+a; \frac{1}{z}\right) \nonumber \\
& & \qquad + \frac{\Gamma(c)\Gamma(a-b)}{\Gamma(a)\Gamma(c-b)}
(-z)^{-b}~_2F_1\left(b,1-c+b; 1-a+b; \frac{1}{z}\right) \, ,
\end{eqnarray}
the special property
\begin{equation}
~_2F_1 (a,0;c;z) = 1
\end{equation}
and suitable end-point subtractions of the residual singular integrals
analogous to box 39, we arrive at the final decomposition of the initial
Feynman-parameter integral
\begin{eqnarray}
%\begin{eqnarray*}
\Delta F_i & = & \frac{\alpha_s}{\pi}~\Gamma(1+\epsilon)
\frac{\Gamma(1-\epsilon)}{\Gamma(1-2\epsilon)}
\left(\frac{4\pi\mu_0^2}{m_t^2}\right)^{2\epsilon} \sum_{j=1}^6 S_j
\, , \nonumber \\[0.0cm]
S_1 & = & \int_0^1 d^6x~xr \left\{ \frac{H_i(\vec x)}{N^3(\vec x)} \left[
1+\epsilon L + \epsilon^2 \left( \frac{L^2}{2} + 3\zeta_2\right) \right]
\right. \nonumber \\
& & \qquad\quad - \left. \frac{H_i(\vec x)|_{r=0}}{(c+br)^3} \left[
1+\epsilon L_0 + \epsilon^2 \left( \frac{L_0^2}{2} + 3\zeta_2\right)
\right] \right\} \, , \nonumber \\[0.0cm]
S_2 & = & -\int_0^1 d^6x~x \frac{H_i(\vec x)|_{r=0}}{(b+cr)^3} \left\{
1+\epsilon L_1 + \epsilon^2 \left( \frac{L_1^2}{2} + 3\zeta_2\right)
+ \epsilon^3 \left( \frac{L_1^3}{6} + 3\zeta_2 L_1 \right) \right\}
\, , \nonumber \\[0.0cm]
S_3 & = & -\int_0^1 \frac{d^5x}{2\rho_s (1-x)(1-s)t} \left\{
\frac{H_i(\vec x)|_{r=0}}{b^2} \left[
1-\epsilon (L_2+2) + \epsilon^2 \left( \frac{L_2^2}{2} + 2L_2 + 2\zeta_2
+ 4 \right) \right] \right. \nonumber \\
& & \qquad\quad + \frac{H_i(\vec x)|_{r,t=0,s=1}}{b_0^2} \left[
1-\epsilon (L_3+2) + \epsilon^2 \left( \frac{L_3^2}{2} + 2L_3 + 2\zeta_2
+ 4 \right) \right] \nonumber \\
& & \qquad\quad - \frac{H_i(\vec x)|_{r=0,s=1}}{b_1^2} \left[
1-\epsilon (L_4+2) + \epsilon^2 \left( \frac{L_4^2}{2} + 2L_4 + 2\zeta_2
+ 4 \right) \right] \nonumber \\
& & \qquad\quad \left. - \frac{H_i(\vec x)|_{r,t=0}}{b_2^2} \left[
1-\epsilon (L_5+2) + \epsilon^2 \left( \frac{L_5^2}{2} + 2L_5 + 2\zeta_2
+ 4 \right) \right] \right\} \, , \nonumber \\[0.0cm]
S_4 & = & -\int_0^1 \frac{dx\, dy\, dz\, ds}{2\rho_s (1-x)(1-s)} \left\{
\frac{H_i(\vec x)|_{r,t=0}}{b_2^2} \left[
-\frac{1}{\epsilon}+ L_6+2 - \epsilon \left( \frac{L_6^2}{2} + 2L_6 + 2\zeta_2
+ 4 \right) \right. \right. \nonumber \\
& & \qquad\qquad\qquad\qquad\qquad \left. + \epsilon^2 \left( \frac{L_6^3}{6}
+ L_6^2 + 2(\zeta_2+2) L_6 - 2\zeta_3+4\zeta_2+8 \right) \right] \nonumber \\
& & \qquad\quad - \frac{H_i(\vec x)|_{r,t=0, s=1}}{b_0^2} \left[
-\frac{1}{\epsilon}+ L_7+2 - \epsilon \left( \frac{L_7^2}{2} + 2L_7 + 2\zeta_2
+ 4 \right) \right. \nonumber \\
& & \qquad\qquad\qquad\qquad\qquad \left. \left. + \epsilon^2 \left(
\frac{L_7^3}{6} + L_7^2 + 2(\zeta_2+2) L_7 - 2\zeta_3+4\zeta_2+8 \right)
\right] \right\} \, , \nonumber \\[0.0cm]
%\right] \right\}
%\end{eqnarray*}
%\begin{eqnarray}
S_5 & = & -\int_0^1 \frac{dx\, dy\, dz\, dt}{2\rho_s (1-x)t} \left\{
\frac{H_i(\vec x)|_{r=0,s=1}}{b_1^2} \left[
-\frac{1}{\epsilon}+ L_8+2 - \epsilon \left( \frac{L_8^2}{2} + 2L_8 + \zeta_2
+ 4 \right) \right. \right. \nonumber \\
& & \qquad\qquad\qquad\qquad\qquad \left. + \epsilon^2 \left( \frac{L_8^3}{6}
+ L_8^2 + (\zeta_2+4) L_8 +2\zeta_2+8 \right) \right] \nonumber \\
& & \qquad\quad - \frac{H_i(\vec x)|_{r,t=0, s=1}}{b_0^2} \left[
-\frac{1}{\epsilon}+ L_9+2 - \epsilon \left( \frac{L_9^2}{2} + 2L_9 + \zeta_2
+ 4 \right) \right. \nonumber \\
& & \qquad\qquad\qquad\qquad\qquad \left. \left. + \epsilon^2 \left(
\frac{L_9^3}{6} + L_9^2 + (\zeta_2+4) L_9 +2\zeta_2+8 \right)
\right] \right\} \, , \nonumber \\[0.0cm]
S_6 & = & -\int_0^1 dx\, dy\, dz~\frac{H_i(\vec
x)|_{r,t=0,s=1}}{2\rho_s (1-x)b_0^2} \left\{ \frac{1}{\epsilon^2} -
\frac{1}{\epsilon} (L_{10}+2) + \frac{L_{10}^2}{2} + 2L_{10} + \zeta_2 +
4 \right. \nonumber \\
& & \qquad\qquad\qquad\qquad\qquad \left. - \epsilon \left( \frac{L_{10}^3}{6}
+ L_{10}^2 + (\zeta_2+4) L_{10} +2\zeta_2+8 \right) \right\} \, .
\end{eqnarray}
The logarithms used in the expressions above are defined as
\begin{eqnarray}
L & = & \log\left(\frac{x(1-x)r}{s}\right) - 2 \log N \, , \qquad\qquad\qquad\,
L_0 = \log\left(\frac{x(1-x)r}{s}\right) - 2 \log (c+br) \, , \nonumber \\
L_1 & = & \log\left(\frac{x(1-x)r}{s}\right) - 2 \log (b+cr) \, , \qquad\qquad
L_2 = \log\left[-\rho_s s(1-s)t\right] + \log b \, , \nonumber \\
L_3 & = & \log\left[-\rho_s s(1-s)t\right] + \log b_0 \, , \qquad\qquad\qquad\ \ \,
L_4 = \log\left[-\rho_s s(1-s)t\right] + \log b_1 \, , \nonumber \\
L_5 & = & \log\left[-\rho_s s(1-s)t\right] + \log b_2 \, , \qquad\qquad\qquad\ \ \,
L_6 = \log\left[-\rho_s s(1-s)\right] + \log b_2 \, , \nonumber \\
L_7 & = & \log\left[-\rho_s s(1-s)\right] + \log b_0 \, , \qquad\qquad\qquad\quad\,
L_8 = \log\left(-\rho_s t\right) + \log b_1 \, , \nonumber \\
L_9 & = & \log\left(-\rho_s t\right) + \log b_0
\, , \qquad\qquad\qquad\qquad\qquad\!\!
L_{10} = \log\left(-\rho_s\right) + \log b_0
\end{eqnarray}
and the remaining objects $b_0, b_1, b_2$ as
\begin{eqnarray}
b_0 & = & b|_{t=0,s=1} \, , \qquad\qquad b_1 = b|_{s=1} \, , \qquad\qquad
b_2 = b|_{t=0}
\end{eqnarray}
with $b$ from Eq.~(\ref{eq:abc}).

Box 45 contains a second threshold for $Q^2>0$ so that even below the
$t\bar t$-threshold, integrations by parts are required to stabilize the
integrand numerically. These integrations by parts are performed for the
Feynman parameter $r$ in the contributions $S_{1,2}$ along the same
lines as for box 39, while the integrals $S_{3-6}$ are stable without
integrations by parts.

\subsubsection{Renormalization \label{sc:renorm}}
\begin{figure}[hbt]
\begin{center}
\vspace*{-1.0cm}

\hspace*{-1.5cm}
 \includegraphics[width=1.15\textwidth]{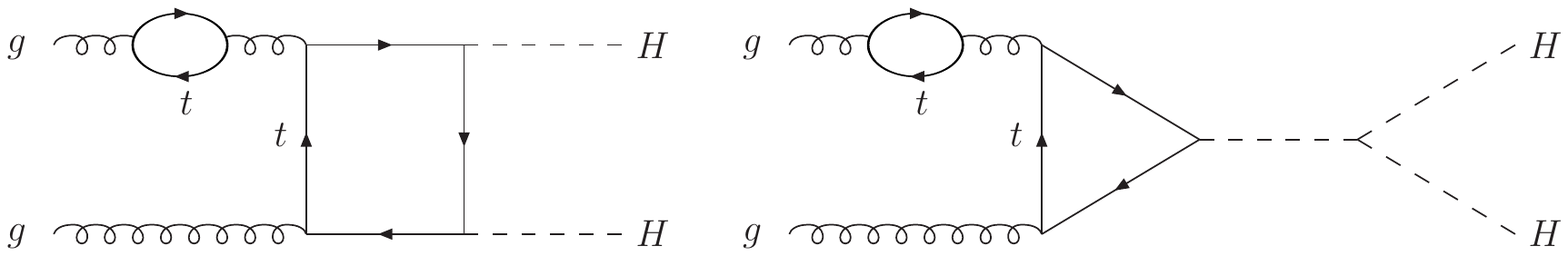} 
\vspace*{-24.0cm}

\caption{\label{fg:toploop} \it Typical diagrams with external top loops.}
\end{center}
\end{figure}

%              ===============
The strong coupling $\alpha_s$ has been renormalized in the
$\overline{\rm MS}$ scheme with the top quark decoupled, i.e.~the
renormalization constant is given by
\begin{eqnarray}
\alpha_{s,0} & = & \alpha_s(\mu_R) + \delta\alpha_s \, , \nonumber \\
\frac{\delta\alpha_s}{\alpha_s} & = & \frac{\alpha_s}{\pi}
\Gamma(1+\epsilon) \left(\frac{4\pi\mu_0^2}{\mu_R^2}\right)^\epsilon
\left\{ -\frac{33-2(N_F+1)}{12\epsilon} + \frac{1}{6} \log
\frac{\mu_R^2}{m_t^2} \right\}
\end{eqnarray}
with $N_F=5$. This choice ensures that there are no artificial large
logarithms of the top mass for the available energy range of the LHC in
the final result, since we do not introduce top densities inside the
proton, i.e.~work in a five-flavour scheme. The additional logarithm of
the top mass cancels against the diagrams with a top loop within the
external gluon lines, see Fig.~\ref{fg:toploop}. This leads to the total
contribution related to the renormalization of the strong coupling
\begin{equation}
\delta_{\alpha_s} F_i = \frac{\alpha_s}{\pi} \Gamma(1+\epsilon)
\left(\frac{4\pi\mu_0^2}{\mu_R^2}\right)^\epsilon \left\{
- \frac{33-2N_F}{12\epsilon} \right\} F_{i,LO} \, ,
\end{equation}
where the LO form factors $F_i$ have to be used in $n$ dimensions,
i.e.~including higher orders in the dimensional regulator $\epsilon$.

For our default prediction, we have renormalized the top mass on-shell so
that the renormalization constant is given by
\begin{eqnarray}
m_{t,0} & = & m_t - \delta m_t \, , \nonumber \\
\frac{\delta m_t}{m_t} & = & \frac{\alpha_s}{\pi}
\Gamma(1+\epsilon) \left(\frac{4\pi\mu_0^2}{m_t^2}\right)^\epsilon
\left\{ \frac{1}{\epsilon} + \frac{4}{3} \right\} \, .
\end{eqnarray}
The explicit contribution of the mass counterterm can either be obtained
by calculating the corresponding counterterm diagrams or, in much more
elegant manner, by differentiating the LO form factors with respect to
the top mass,
\begin{equation}
\delta_{m_t} F_i = - \delta m_t \frac{\partial F_{i,LO}}{\partial m_t}
\, ,
\end{equation}
where we followed the second option. For the renormalization of the top
mass in terms of the $\overline{\rm MS}$ mass, a counterterm
\begin{eqnarray}
m_{t,0} & = & \overline{m}_t(\mu_t) - \delta \overline{m}_t \, , \nonumber \\
\frac{\delta \overline{m}_t}{\overline{m}_t(\mu_t)} & = & \frac{\alpha_s}{\pi}
\Gamma(1+\epsilon) \left(\frac{4\pi\mu_0^2}{\mu_t^2}\right)^\epsilon
\frac{1}{\epsilon}
\end{eqnarray}
has to be used with the LO and NLO expressions of the form factors
expressed in terms of the $\overline{\rm MS}$ top mass
$\overline{m}_t(\mu_t)$. For the evaluation of the $\overline{\rm MS}$
top mass, we use the N$^3$LO relation between the pole and $\overline{\rm
MS}$ mass \cite{Gray:1990yh, Chetyrkin:1999ys, Chetyrkin:1999qi,
Melnikov:2000qh},
\begin{eqnarray}
{\overline{m}}_{t}(m_{t}) & = & \frac{m_{t}}{\displaystyle 1+\frac{4}{3}
\frac{\alpha_{s}(m_t)}{\pi} + K_2
\left(\frac{\alpha_s(m_t)}{\pi}\right)^2 + K_3
\left(\frac{\alpha_s(m_t)}{\pi}\right)^3}
\label{eq:mspole}
\end{eqnarray}
with $K_2\approx 10.9$ and $K_3 \approx 107.11$. The scale dependence of
the $\overline{\rm MS}$ mass is treated at N$^3$LL,
\begin{eqnarray}
{\overline{m}}_{t}\,(\mu_t)&=&{\overline{m}}_{t}\,(m_{t})
\,\frac{c\,[\alpha_{s}\,(\mu_t)/\pi ]}{c\, [\alpha_{s}\,(m_{t})/\pi ]}
\label{eq:msbarev}
\end{eqnarray}
with the coefficient function \cite{Tarasov:1982gk, Chetyrkin:1997dh}
\begin{eqnarray}
c(x)=\left(\frac{7}{2}\,x\right)^{\frac{4}{7}} \, [1+1.398x+1.793\,x^{2}
- 0.6834\, x^3]\, .
\end{eqnarray}
Since we are interested in the finite top-mass effects on top of the LO
ones, we have subtracted in addition the Born-improved HTL of the virtual
corrections involving the full top-mass dependence at LO
\cite{Dawson:1998py}. This yields the additional subtraction term
\begin{equation}
\delta_{HTL} F_i = \frac{\alpha_s}{\pi}
\frac{\Gamma(1-\epsilon)}{\Gamma(1-2\epsilon)}
\left(\frac{4\pi\mu_0^2}{-m_t^2 \rho_s}\right)^\epsilon \left\{
\frac{3}{2\epsilon^2} + \frac{33-2N_F}{12\epsilon}
\left(\frac{\mu_R^2}{-m_t^2 \rho_s}\right)^{-\epsilon}
- \frac{11}{4} + \frac{\pi^2}{4} \right\} F_{i,LO} \, .
\end{equation}
After adding this subtraction term, the result of {\tt Hpair} can simply
be added back to the NLO top-mass effects obtained in this way for the
virtual corrections. Thus, the total counterterm plus HTL-subtraction is
given by
\begin{equation}
\delta F_i = \delta_{\alpha_s} F_i + \delta_{m_t} F_i + \delta_{HTL} F_i
\, .
\end{equation}
The addition of this term results in an infrared and ultraviolet finite
result for the virtual corrections as we have explicitly checked
numerically. It should be noted that we have defined this total
subtraction term with the imaginary part $\bar\epsilon$ for the top mass
to be consistent with our treatment of the two-loop diagrams. For the
two-loop triangle diagrams, this total subtraction term is included in
the narrow-width approximation according to the known result for the
single-Higgs case.

\subsubsection{Differential cross section}
%              ==========================
The final numerical integrations have been performed by {\tt Vegas}
\cite{Lepage:1980dq} for the differential cross sections $d\sigma/dQ^2$
of Eq.~(\ref{eq:nlodiff}), i.e.~the integration over $\hat t$ is
included. Each individual box diagram is divergent in $\hat t$ at the
lower and upper bound of the $\hat t$-integration in general. To
stabilize the $\hat t$-integration, we have performed a suitable
substitution to smoothen the integrand,
\begin{equation}
\hat t_1 = m_t^2 e^y + t_{1-}
\end{equation}
with $\hat t_1 = \hat t-M_H^2, \hat u_1 = \hat u-M_H^2$ and $\hat
t_{1\pm} = \hat t_{\pm}-M_H^2$, where the integration boundaries $\hat
t_{\pm}$ are given in Eq.~(\ref{eq:tbound}). By means of this
substitution, we can rewrite the integration over $\hat t_1$ generically
as\footnote{The symmetrization of the integrand $f(\hat t_1, \hat u_1)$
for the $y$ integration is a straightforward result of this
substitution.}
\begin{equation}
\int_{\hat t_{1-}}^{\hat t_{1+}} \frac{d\hat t_1}{\hat t_1\hat u_1 -
\hat s M_H^2} f(\hat t_1, \hat u_1) = \int_{y_-}^{y_+}
\frac{dy}{t_+-t_-} \Big[ f(\hat t_1,\hat u_1) + f(\hat u_1,\hat t_1)
\Big] \, ,
\end{equation}
where $f(\hat t_1, \hat u_1)$ denotes the corresponding virtual matrix
element with the (singular) denominator $\hat t_1\hat u_1 - \hat s
M_H^2$ extracted and the integration boundaries read
\begin{eqnarray}
y_+ & = & \log \frac{(t_+-t_-)(1-\tilde\epsilon)}{m_t^2} \, , \nonumber \\
y_- & = & \log \frac{(t_+-t_-)\tilde\epsilon}{m_t^2} \, ,
\end{eqnarray}
where we have introduced a cut $\tilde\epsilon$ for the upper and lower
bound of the $\hat t_1$-integration (after rewriting this into an
integral from 0 to 1 and replacing these integration boundaries by
$\tilde\epsilon$ and $1-\tilde\epsilon$). We have checked that the total
sum of all box diagrams becomes independent of this cut by varying
$\tilde\epsilon$ down to $10^{-10}$, i.e.~that the total sum is again
finite\footnote{Note that also the individual LO box diagrams are not
finite with respect to the $\hat t$ integration, but the sum of all
three LO boxes is.}. 

\subsection{Real corrections \label{sc:reals}}
%           ================

We are left with the evaluation of the real contributions to complete
the picture of the NLO QCD corrections. As we are interested in the
calculation of the top-mass effects on top of the HTL calculation
that is provided by {\tt Hpair}, we use the universality of the infrared
divergent pieces to subtract the Born-improved HTL contributions
$d\sigma_{ij}^{\text{HTL}}$ in such a way that our integration of the
real contributions $d\Delta\sigma_{ij}^{\text{mass}} = d\sigma_{ij} -
d\sigma_{ij}^{\text{HTL}}$ is finite. We construct a local subtraction
term for the partonic channels $d\hat{\sigma}_{ij}$,
\begin{equation}
d\Delta\hat{\sigma}^{\text{mass}}_{ij}(p_k)  =
d\hat{\sigma}_{ij}(p_k) - d\hat{\sigma}_{\text{LO}}(\tilde{p}_k)
\frac{d\hat{\sigma}_{ij}^{\text{HTL}}(p_k)}{d\hat{\sigma}_{\text{LO}}^{\text{HTL}}(\tilde{p}_k)},
\end{equation}
where $p_k$ denote the four-momenta from the full $2\to 3$
phase-space and $\tilde{p}_k$ stand for the mapping of the momenta
$p_k$ on a $2\to 2$ sub-phase-space. As the results in the HTL limit are
given in the Born-improved approximation in which the pure HTL is
rescaled with the full LO matrix elements, we need to map the full
$2\to 3$ phase-space onto a projected $2\to 2$ phase-space to
construct the subtraction term involving this rescaling to the full LO
contribution $ d\hat{\sigma}_{\text{LO}}$.

\begin{figure}[hbtp]
  \begin{center}
    \includegraphics[width=0.75\textwidth]{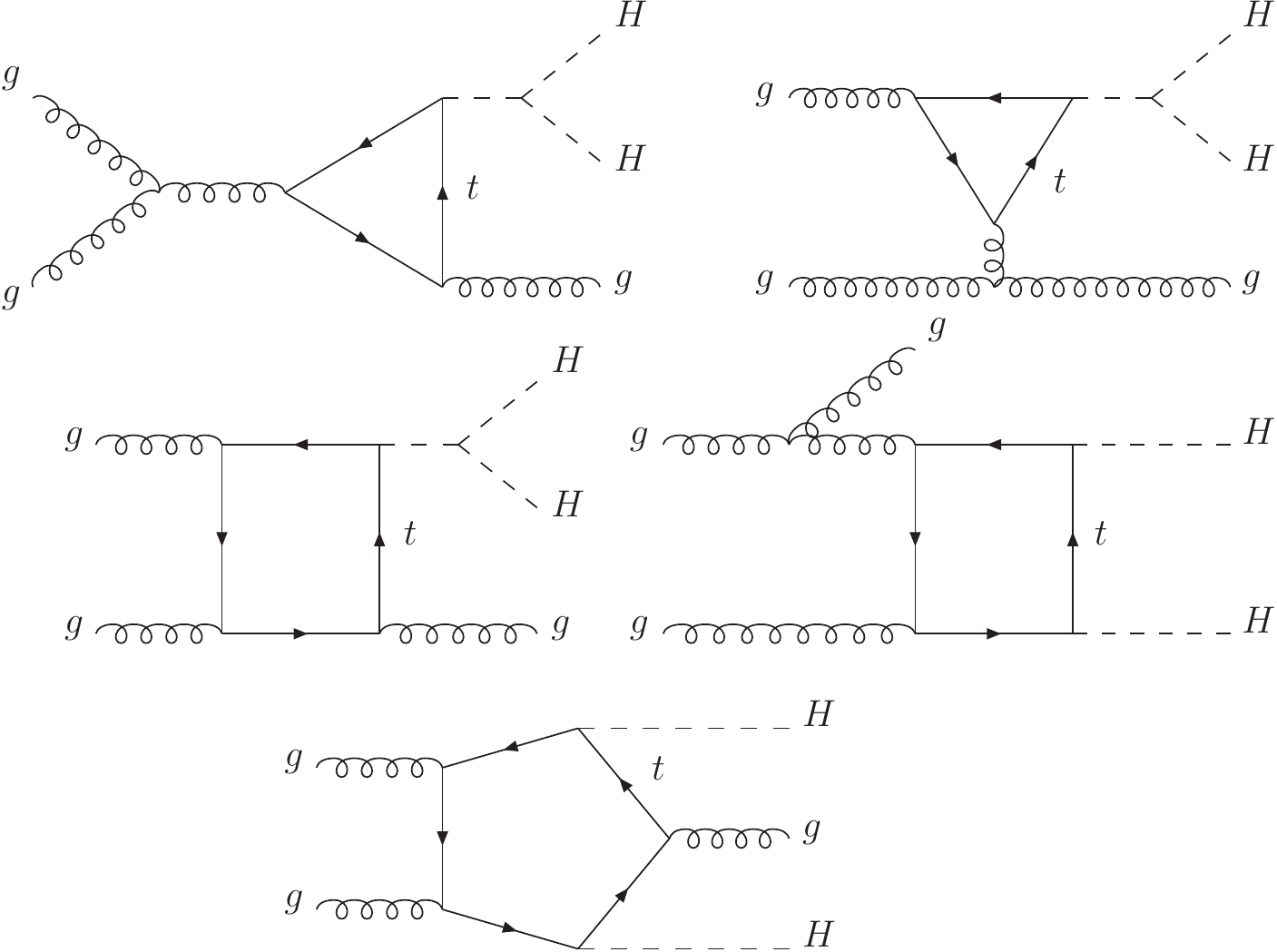} 
    \caption{\label{fg:gghhreal} \it Typical one-loop triangle (upper row),
      box (middle row), and pentagon (lower row) diagrams for the
      partonic channel $g g\to H H g$ contributing to the real
      corrections of Higgs-pair production via gluon fusion at NLO in
      QCD.}
  \end{center}
\end{figure}

\begin{figure}[hbtp]
  \begin{center}
    \includegraphics[width=0.75\textwidth]{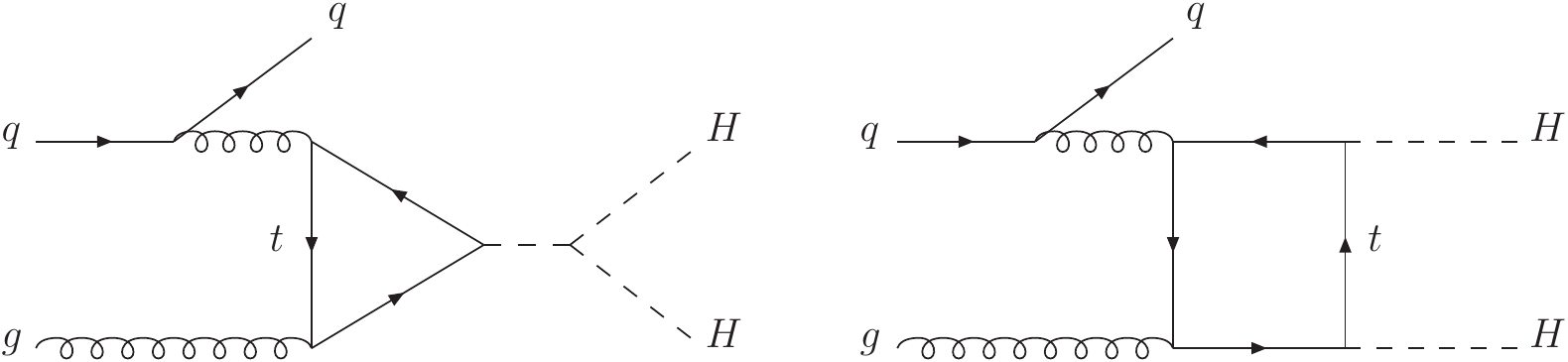}\vspace*{0.5cm}
    \includegraphics[width=0.75\textwidth]{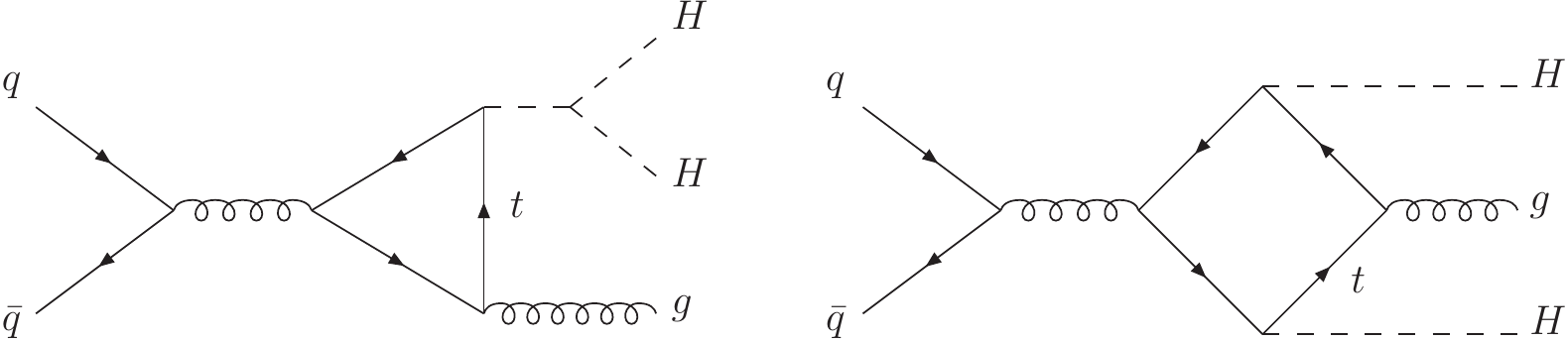}
    \caption{\label{fg:gghhreal2} \it Typical one-loop triangle and box
      diagrams for the partonic channels $q g\to H H q$ (upper row)
      and $q\bar{q}\to H H g$ (lower row), contributing to the real
      corrections of Higgs-pair production via gluon fusion at NLO in
      QCD.}
  \end{center}
\end{figure}

The mapping is done by using the transformation formulae for initial-state
emitter and initial-state spectator in the construction of dipole
subtraction terms, i.e.~using Eqs.~(5.137-5.139) of
Ref.~\cite{Catani:1996vz}. The (mapped) momenta of the initial-state
partons are $p_{1/2}$ ($\tilde{p}_{1/2}$), the (mapped) momenta of the
final-state Higgs bosons are $p_{3/4}$ ($\tilde{p}_{3/4}$), and the
momentum of the radiated parton is $p_5$. For the initial-state partons,
we use the following mapping,
\begin{equation}
\tilde{p}_1 = p_1,\qquad \tilde{p}_2 = p_2 \left(1-\frac{(p_5
    p_1) + (p_5 p_2)}{(p_1 p_2)}\right).
\end{equation}
In order to transform the Higgs momenta, we introduce the variables $K$ and
$\tilde{K}$,
\begin{equation}
K = p_1+p_2-p_5,\qquad \tilde{K} = \tilde{p}_1+\tilde{p}_2
\end{equation}
allowing us to define
\begin{align}
\tilde{p}_3 
  & = p_3 - 2\,\frac{p_3 (K+\tilde{K})}{(K+\tilde{K})^2}
    \left(K+\tilde{K}\right) + 2\,\frac{(p_3 K)}{K^2}
    \tilde{K},\nonumber\\
\tilde{p}_4 
  & = p_4 - 2\,\frac{p_4 (K+\tilde{K})}{(K+\tilde{K})^2}
    \left(K+\tilde{K}\right) + 2\,\frac{(p_4 K)}{K^2} \tilde{K}.
\end{align}
The HTL matrix elements are calculated analytically. We introduce the
partonic center-of-mass energy $\hat{s}$, and the Mandelstam variables
$\hat{t}=(p_1-p_5)^2$ and $\hat{u}=(p_2-p_5)^2$. The invariant squared
Higgs-pair mass is $Q^2=\hat{s}+\hat{t}+\hat{u}$. The real spin- and
colour-averaged matrix elements are
\begin{align}
\overline{\Big|\mathcal{M}_{gg\to HHg}^{\text{HTL}}\Big|^2}
& = \frac{\alpha_s^3(\mu_R)G_F^2}{12\pi}\,
  \frac{\hat{s}^4+\hat{t}^4+\hat{u}^4+Q^8}{\hat{s}\hat{t}\hat{u}}\left(1-\frac{3
  M_H^2}{Q^2-M_H^2}\right)^2,\nonumber\\
\overline{\Big|\mathcal{M}_{qg\to HHq}^{\text{HTL}}\Big|^2}
& = \frac{\alpha_s^3(\mu_R)G_F^2}{27\pi}\,
  \frac{\hat{s}^2+\hat{u}^2}{-\hat{t}}\left(1-\frac{3
  M_H^2}{Q^2-M_H^2}\right)^2,\nonumber\\
\overline{\Big|\mathcal{M}_{q\bar{q}\to HHg}^{\text{HTL}}\Big|^2}
& = \frac{8\alpha_s^3(\mu_R) G_F^2}{81\pi}\,
  \frac{\hat{t}^2+\hat{u}^2}{\hat{s}}\left(1-\frac{3
  M_H^2}{Q^2-M_H^2}\right)^2,
\end{align}
and the LO matrix element in the HTL reads
\begin{equation}
\overline{\Big|\mathcal{M}_{\text{LO}}^{\text{HTL}}\Big|^2} =
\frac{\alpha_s^2(\mu_R) G_F^2}{288\pi^2}\,
Q^4 \left(1-\frac{3 M_H^2}{Q^2-M_H^2}\right)^2.
\end{equation}
The full one-loop matrix elements have been generated with {\tt
FeynArts}~\cite{Hahn:2000kx} and {\tt FormCalc}~\cite{Hahn:1998yk}. They
contain triangle, box, and pentagons diagrams. Generic diagrams for the
contribution $g g\to H H g$ are given in Fig.~\ref{fg:gghhreal}, generic
diagrams for the contributions $q g\to H H q$ and $q\bar{q}\to H H g$
are displayed in Fig.~\ref{fg:gghhreal2}. The numerical
evaluation of the scalar integrals~\cite{tHooft:1978jhc} as well as the
tensor reduction has been performed using the techniques developed in
Refs.~\cite{vanOldenborgh:1990yc,Denner:2002ii,Denner:2005nn,Denner:2010tr}
and implemented in the library {\tt Collier 1.2}~\cite{Denner:2016kdg}.
The latter has been interfaced to the analytic expressions generated by
{\tt FormCalc} with an in-house routine. In order to improve our
numerical stability, we have implemented a technical collinear cut in
the phase-space parametrization. The integration of the scattering angle
$\theta$ of the radiated parton in the c.m.~system is restricted to the range
$|\!\cos\theta|< 1-\delta$ with $\delta=10^{-4}$. We have checked that our
results are stable against a variation of $\delta$ from $10^{-4}$ to
$10^{-6}$ and therefore they are not affected by our choice for this
technical cut.%
We have cross-checked the final mass-effects of the real corrections
against the results presented in the literature~\cite{Frederix:2014hta,
Maltoni:2014eza, Borowka:2016ehy, Borowka:2016ypz} and we have obtained
agreement.

\section{Results \label{sc:results}}
%           =======

Our numerical results will be presented for the invariant
Higgs-pair-mass distributions for different
c.m.~energies, i.e.~14 TeV for the LHC, 27 TeV for a potential
high-energy LHC (HE-LHC) and 100 TeV for a provisional proton collider
within the Future-Circular-Collider (FCC) project. The Higgs mass has
been chosen as $M_H=125$ GeV and the top pole mass as $m_t=172.5$ GeV.
The results for the full NLO cross sections have been obtained with two
different PDF sets, {\tt MMHT2014} \cite{Harland-Lang:2014zoa} and {\tt
PDF4LHC15} \cite{Butterworth:2015oua}, that are taken from the {\tt
LHAPDF-6} library \cite{Buckley:2014ana}. The central scale choices for
the renormalization and factorization scales are $\mu_F=\mu_R=Q/2$ and
the input value $\alpha_s(M_Z)$ is chosen according to the PDF set used.
Since {\tt MMHT2014} contains a LO set, these PDFs are used for the
evaluation of the consistent K-factors with the NLO (LO) cross section
calculated with NLO (LO) $\alpha_s$ and PDFs. The whole calculation of
the virtual and real corrections has been performed at least twice independently adopting also
different Feynman parametrizations of the virtual two-loop diagrams. The
real corrections have been derived with different parametrizations of
the real phase-space. Both calculations agree within the numerical
errors. We work in the narrow-width approximation of the top quark so
that the Richardson extrapolation has to be applied to reach this
limit for the two-loop box diagrams.\footnote{Finite top-width effects
  have been estimated to amount to $\sim -2\%$
  \cite{Maltoni:2014eza}. The effects are slightly larger in the
  vicinity of the virtual $t\bar t$ threshold, $Q^2\sim 4m_t^2$.}.

\subsection{Differential cross section\label{sec:diffxs}}
%              ==========================
For the differential cross section, we have computed a grid of
$Q$-values from 250~GeV to 1.5~TeV. In order to get a reliable result
for the total cross section later on, we have used steps of 5~GeV
between $Q=250$~GeV and $Q=300$~GeV, steps of 25~GeV between
$Q=300$~GeV and $Q=700$~GeV, and steps of 50~GeV for
$Q>700$~GeV. After applying the integrations by parts to each
individual virtual diagram, we reached reliable results of 
our numerical integrations for $\bar\epsilon$ values [see
Eq.~(\ref{eq:imaginary})] down to about 0.05.  In order to obtain the
result in the narrow-width approximation ($\bar\epsilon \to 0$), we have
performed a Richardson extrapolation applied to the results for
different values\footnote{Note that a Richardson
extrapolation of the integrand {\it before} integration provides an
alternative to stabilize the numerical integration.} of $\bar\epsilon$. We adopt
$\bar\epsilon$ values $\bar\epsilon_n= 0.025\times 2^n$ ($n=0\dots 10$).
For bins close to threshold, $Q=300,325,350$~GeV, we use the set
$n=0\dots 8$. For $Q \in [375,475]$~GeV, we use $n=1\dots 9$ while we
use $n=2\dots 10$ for Q values in the range $Q \in [500,700]$~GeV. For
$Q$ values starting at 750~GeV, we restrict the extrapolation to
$n=2\dots 6$. In this way, we obtain a series of extrapolated results up
to the ninth order in the dominant region and up to the fifth order in
the tails for large $Q$. We define an estimate of the theoretical error
due to the Richardson extrapolation as the difference of the
extrapolated results at fifth and fourth order. In addition, we multiply
this error by a factor of two close to the virtual $t\bar t$ threshold
in order to be conservative. The total estimated
Richardson-extrapolation error ranges below the per-cent level and is
added in quadrature to the statistical integration error.

\begin{figure}[htb!]
  \centering
  \includegraphics[width=0.45\textwidth]{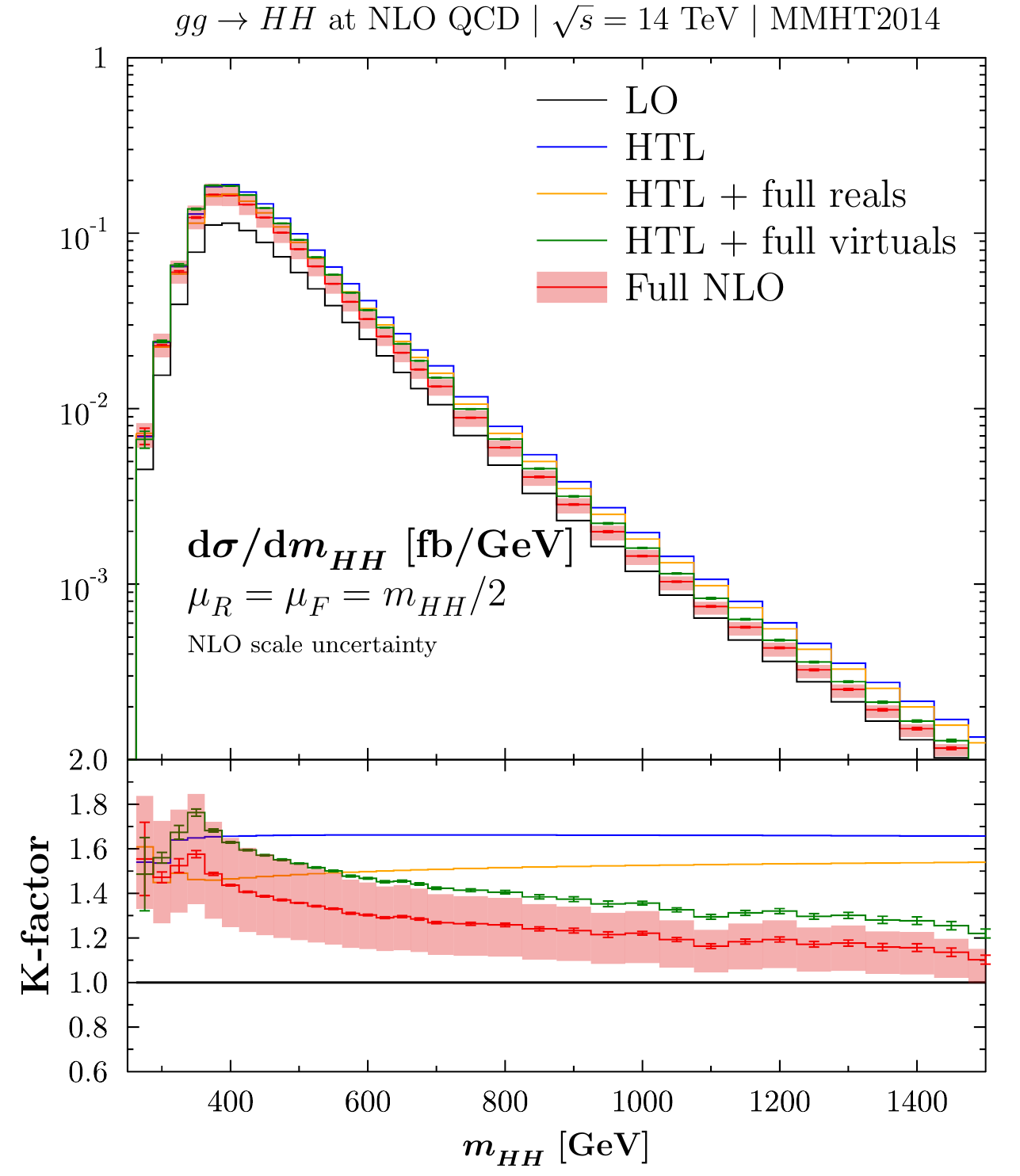} 
  \hspace{3mm}
  \includegraphics[width=0.45\textwidth]{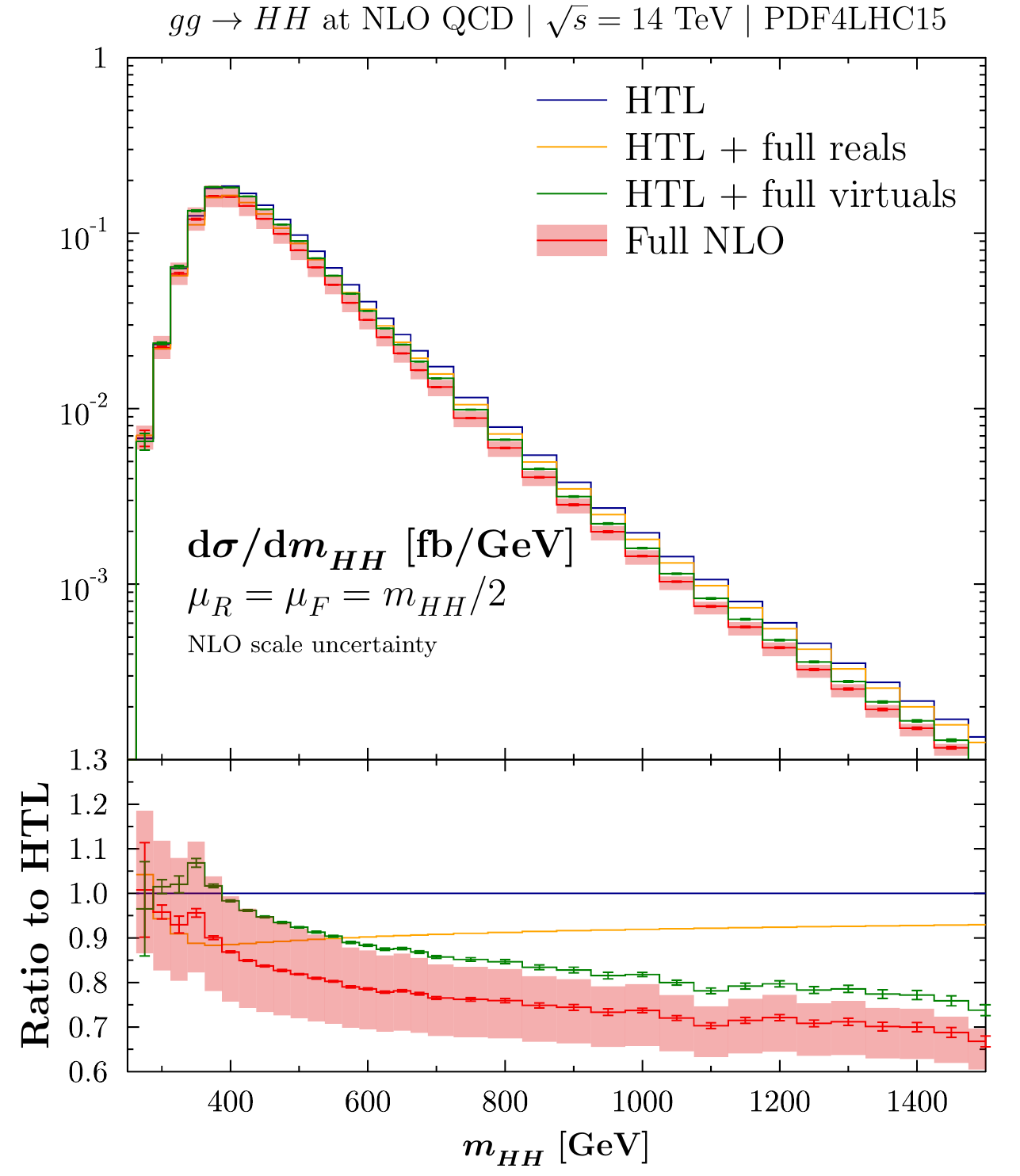}
%  \hspace{3mm}
  \caption[]{Invariant Higgs-pair-mass distributions for Higgs boson
pair production via gluon fusion at the 14 TeV LHC as a function of
$Q=m_{HH}$. LO results (in black), HTL results (in blue), HTL results
including the full real corrections (in yellow), HTL results including
the full virtual corrections (in green, including the numerical errors),
and the full NLO QCD results (in red, including the numerical errors).
Left: Results with the {\tt MMHT2014} PDF set, the panel below displays
the K-factors for the different results. Right: Results with the {\tt
PDF4LHC15} PDF set, the panel below displays the ratio to the NLO
Born-improved HTL result for the different calculations. The red band
indicates the renormalization and factorization scale uncertainties for
results including the full NLO QCD corrections.}
  \label{fig:distrib_14}
\end{figure}

\begin{figure}[htb!]
  \centering
  \includegraphics[width=0.45\textwidth]{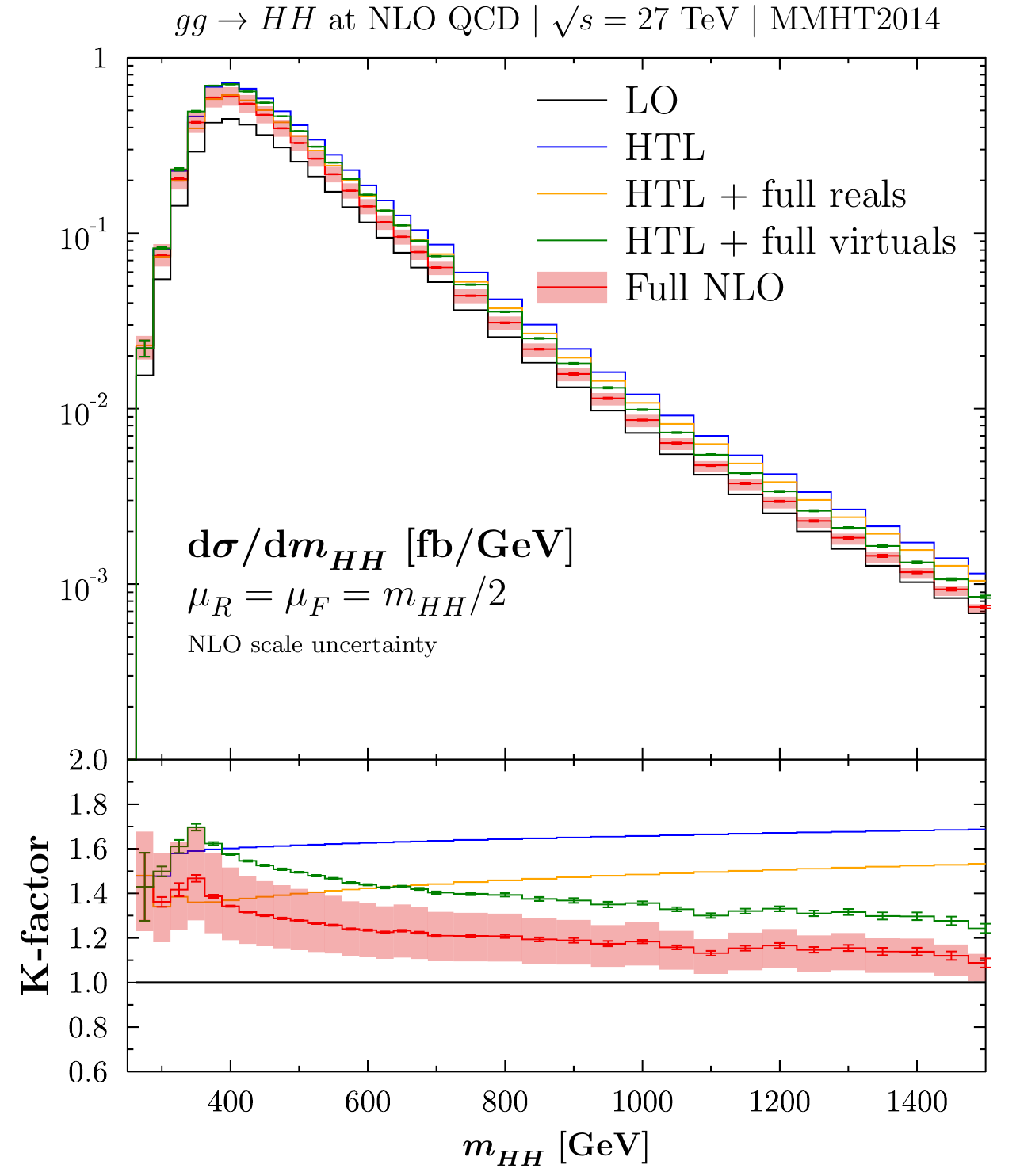}
  \hspace{3mm}
  \includegraphics[width=0.45\textwidth]{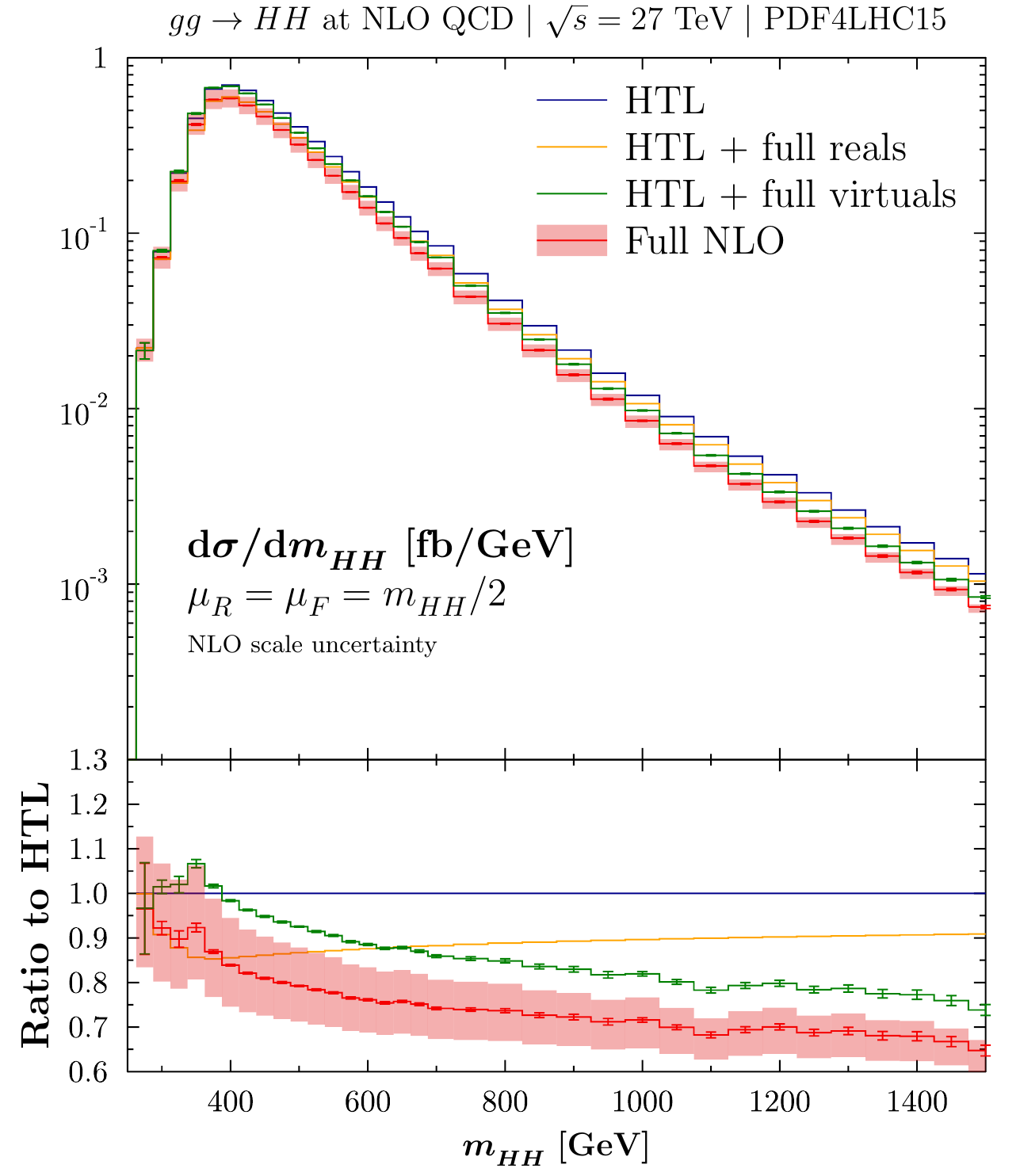}
  \hspace{3mm}
  \caption[]{Same as Fig.~\ref{fig:distrib_14} but for a c.m.~energy
$\sqrt{s}=27$ TeV.}
  \label{fig:distrib_27}
\end{figure}

\begin{figure}[htb!]
  \centering
%  \fbox{
    \includegraphics[width=0.45\textwidth]{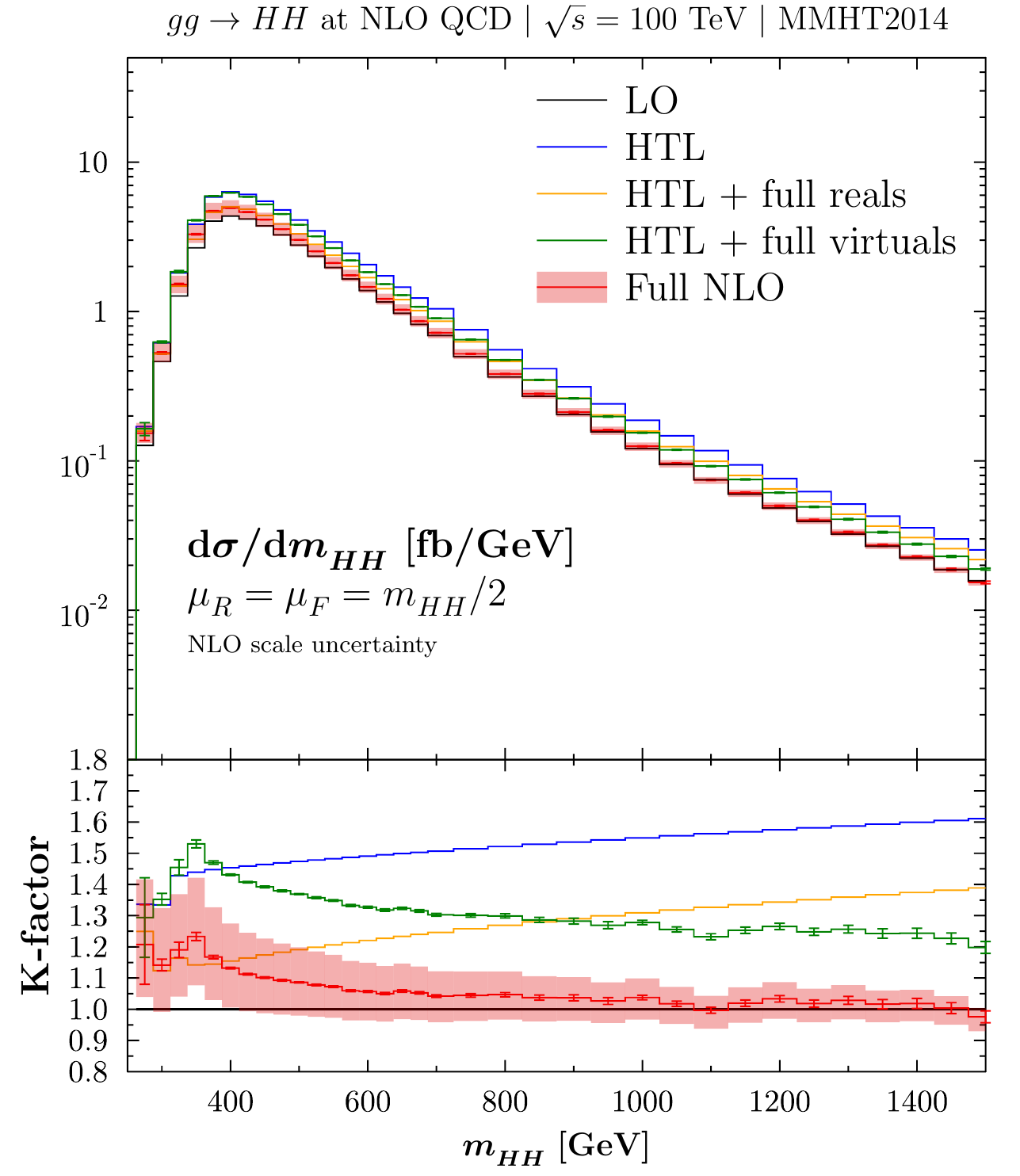}
    %  }
       \hspace{3mm}
%  \fbox{
    \includegraphics[width=0.45\textwidth]{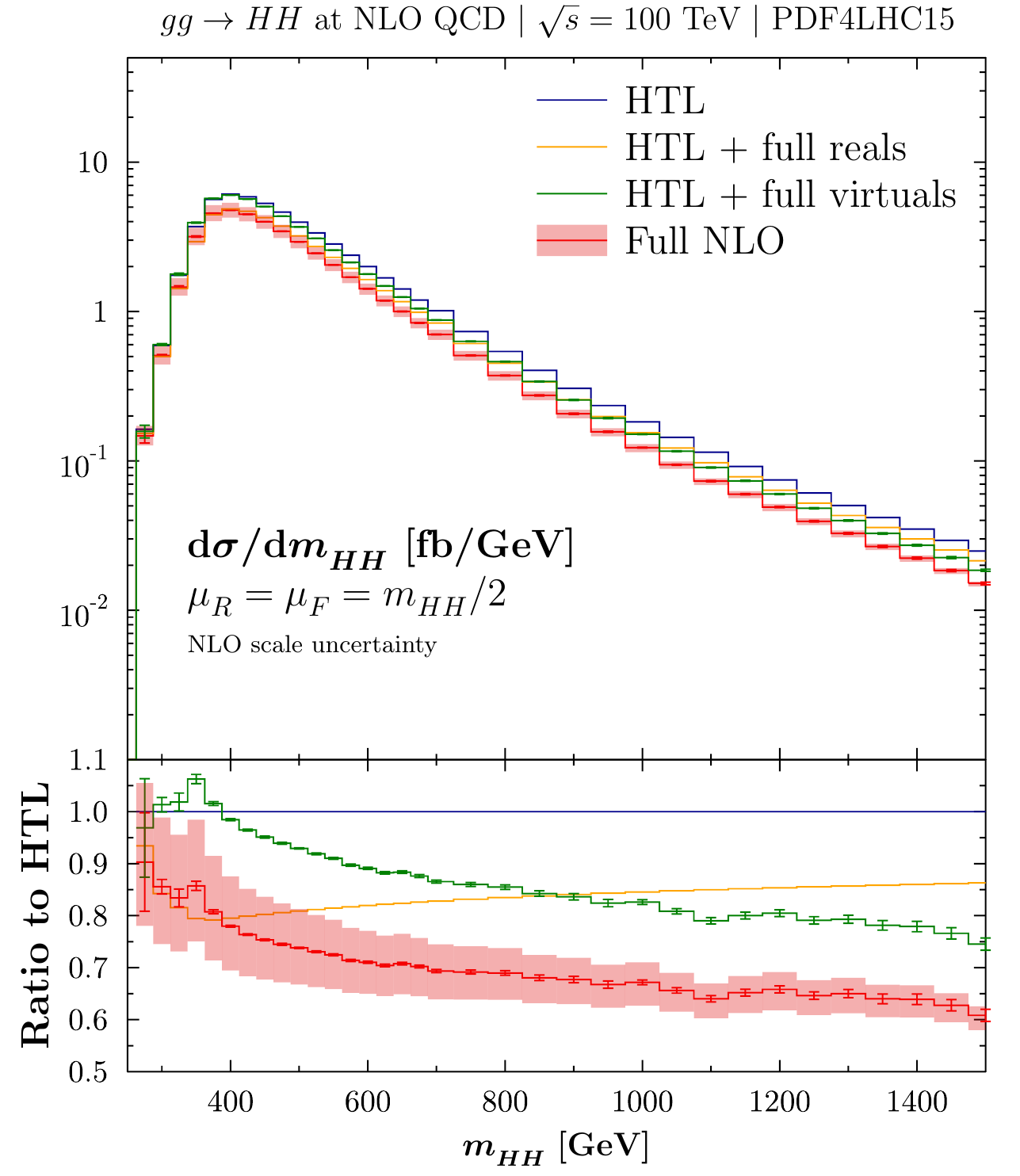}
%  }
  \caption[]{Same as Fig.~\ref{fig:distrib_14} but for a c.m.~energy
$\sqrt{s}=100$ TeV.}
  \label{fig:distrib_100}
\end{figure}

Since we have subtracted the (Born-improved) HTL consistently from the
virtual and real corrections, we are left with the pure top-mass effects
at NLO that are infrared and ultraviolet finite individually after
renormalization. This part has then been added to the results of {\tt
Hpair} \cite{hpair} to derive the full NLO cross section. The final
invariant Higgs-pair-mass distributions are displayed in
Figs.~\ref{fig:distrib_14}--\ref{fig:distrib_100} for the three
c.m.~energies, 14, 27, 100 TeV. The blue curves show the Born-improved
result in the HTL of Ref.~\cite{Dawson:1998py} as implemented in {\tt
Hpair} \cite{hpair}, the yellow ones the Born-improved HTL result plus
the mass effects of the real corrections, the green curves the
Born-improved HTL result plus the mass effects of the virtual
corrections and the red curves the full NLO results. The plots on the
left side of each figure have been obtained by using {\tt MMHT2014} PDFs
\cite{Harland-Lang:2014zoa} and the ones on the right with {\tt PDF4LHC}
PDFs \cite{Butterworth:2015oua}. The lower panel on the left shows the
consistently defined K-factors $K=d\sigma_{NLO}/d\sigma_{LO}$. The lower
panel on the right shows the ratio of the differential NLO cross
section to the one obtained in the Born-improved HTL.

While the Born-improved HTL provides a reasonable approximation for
$Q$-values close to threshold, the real corrections add a negative mass
effect of about $-10\%$ for $\sqrt{s}=14$ TeV (yellow curves) that is
approximately uniform in the entire $Q$ range.  The (negative) mass
effects of the virtual corrections (green curves), however, become large
at large values of $Q$ reaching a level of more than 20\% for $Q$ beyond
about 1 TeV. While the relative mass effects of the virtual corrections
at NLO are independent of the collider energy (see the right plots
showing the ratios to the HTL in the lower panels) in agreement with
Eq.~(\ref{eq:nlodiff}), the NLO mass effects of the real corrections
become larger with rising collider energy, reaching a level of $-20\%$
for $\sqrt{s}=100$ TeV.  Both mass effects of the virtual and real
corrections add up in the same direction and result in a total
modification of the differential cross section of up to $-40\%$ compared
to the Born-improved HTL at large $Q$ values for $\sqrt{s}=100$ TeV.
While (as for the ratios) the full NLO K-factors shown in the left plots
are close to the Born-improved HTL (blue curves) at $Q$ values close to
the production threshold, they deviate significantly at larger values of
$Q$ due to the additional NLO top-mass effects that decrease the total
size of the NLO QCD corrections compared to the HTL as expected from
unitarity arguments.

To estimate the theoretical uncertainties, we have varied the
renormalization and factorization scales for each bin in $Q$ by a factor
of 2 up and down around the central scale $\mu_R=\mu_F=Q/2$ and derived
the envelope of a 7-point variation, i.e.~excluding points where the
renormalization and factorization scales differ by more than a factor of
two. The residual uncertainties are shown by the red band around the
full NLO results (red curves) in
Figs.~\ref{fig:distrib_14}--\ref{fig:distrib_100}. They range at the
level of 10--15\% in total as can be inferred from the explicit numbers
for $\sqrt{s}=14$ TeV (using {\tt PDF4LHC} PDFs),
\begin{eqnarray}
\frac{d\sigma_{NLO}}{dQ}\Big|_{Q=300~{\rm GeV}} & = &
0.02978(7)^{+15.3\%}_{-13.0\%}\
{\rm fb/GeV},\nonumber\\
\frac{d\sigma_{NLO}}{dQ}\Big|_{Q=400~{\rm GeV}} & = &
0.1609(4)^{+14.4\%}_{-12.8\%}\ 
{\rm fb/GeV},\nonumber\\
\frac{d\sigma_{NLO}}{dQ}\Big|_{Q=600~{\rm GeV}} & = &
0.03204(9)^{+10.9\%}_{-11.5\%}\
{\rm fb/GeV},\nonumber\\
\frac{d\sigma_{NLO}}{dQ}\Big|_{Q=1200~{\rm GeV}} & = &
0.000435(4)^{+7.1\%}_{-10.6\%}\
{\rm fb/GeV} \, .
\end{eqnarray}

\begin{figure}[htb!]
  \centering
%  \fbox{
    \includegraphics[width=0.45\textwidth]{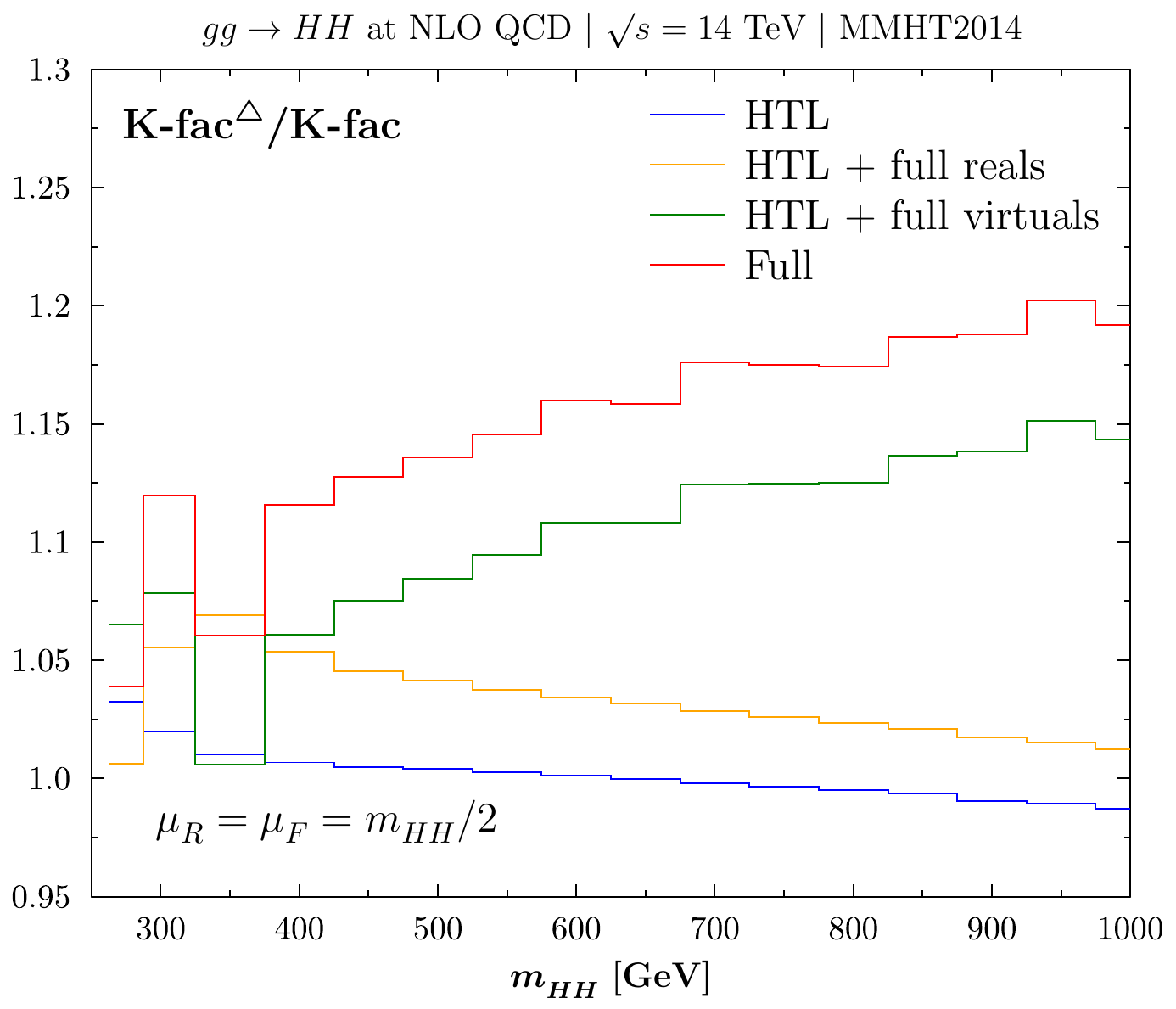}
    %  }
       \hspace{3mm}
%  \fbox{
    \includegraphics[width=0.45\textwidth]{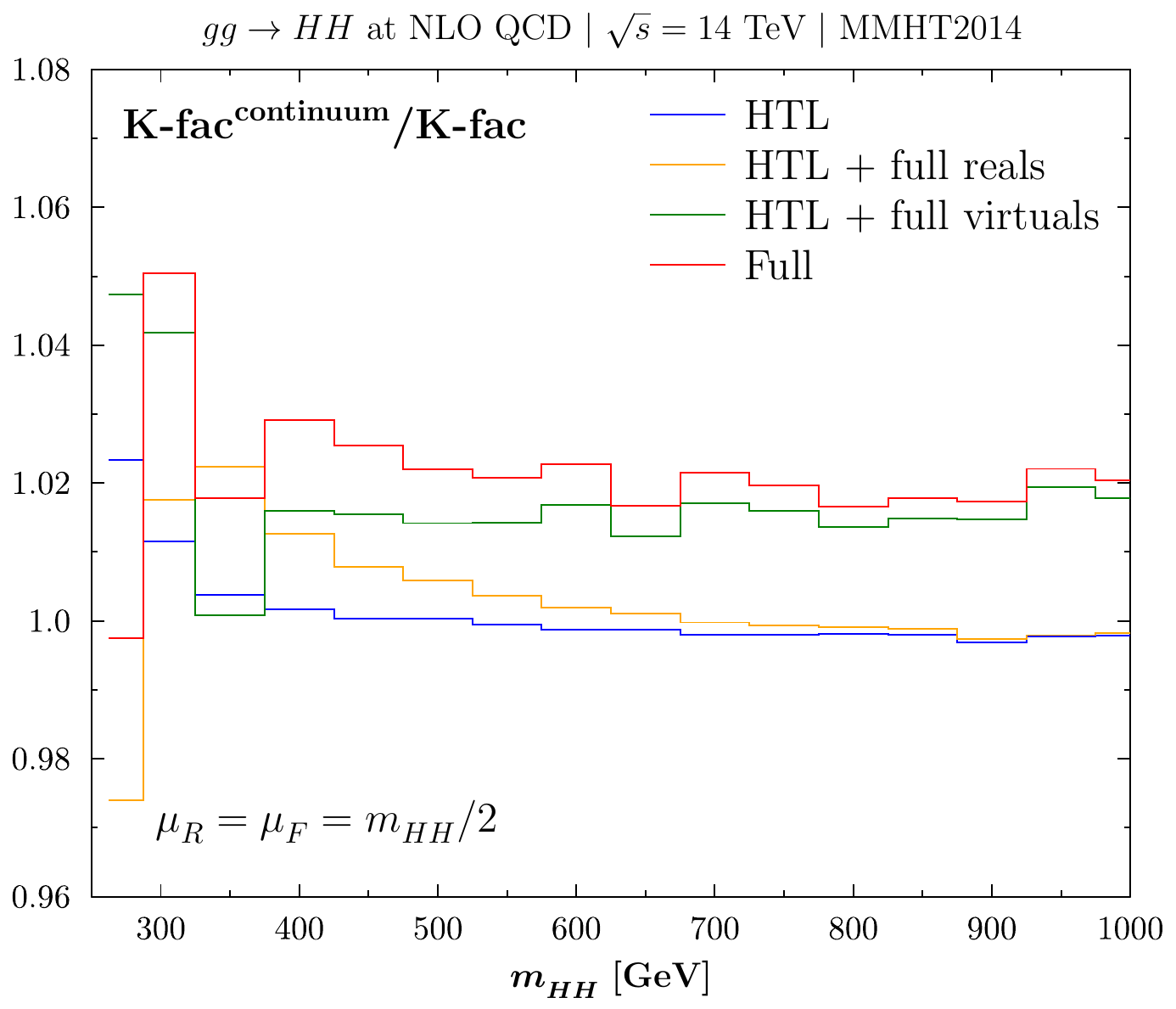}
%  }
  \caption[]{Ratios of the K-factor including (left) only triangle
diagrams and (right) only continuum diagrams to the full K-factor of
Higgs-pair production as a function of the invariant Higgs-pair mass
$Q=m_{HH}$ for the LHC with a c.m.~energy $\sqrt{s}=14$ TeV and using
{\tt MMHT2014} parton densities.}
  \label{fg:ktriafull}
\end{figure}

We have analyzed the structure of the NLO QCD corrections in more detail
by comparing the K-factor with the one of the triangle diagrams alone,
i.e.~with the K-factor of single-Higgs production with mass $M_H=Q$, in
all individual approximations. This will determine the amount of
universal NLO top-mass effects, common in the triangle and box diagrams.
We define the ratio of the NLO triangle-diagram K-factor to the one
including all diagrams as K-fac$^\triangle$/K-fac. This is shown, as a
function of $Q=m_{HH}$, in Fig.~\ref{fg:ktriafull} (left). It is visible
that the triangle-diagram K-factor provides an acceptable approximation
to the full NLO K-factor only for $Q$ values below about 500--600 GeV if
maximal deviations of about 15\% are allowed (red histogram). The break
down into the different mass effects of the virtual (green histogram)
and real (yellow histogram) corrections singles out the origin of
non-universal mass effects in the virtual corrections, while the
non-universal mass effects beyond the single-Higgs case of the real
corrections are limited to less than about 5\% (apart from the virtual
$t\bar t$-threshold region). In comparison to the contribution of the
triangle diagrams alone, we also present the ratio of the K-factor
obtained by including only the continuum diagrams (box diagrams of the
virtual corrections and all box and pentagon diagrams of the real
corrections without trilinear Higgs couplings) to the full K-factor in
Fig.~\ref{fg:ktriafull} (right). The different curves show the results
for the various approximations, i.e.~the blue curves for the
Born-improved HTL, the yellow ones with the inclusion of the NLO mass
effects of the real corrections, the green curves with only the virtual
NLO mass effects and the red curves the full NLO results. The right
figure shows that the full NLO K-factor (red curve) is well-described
(within 5\%) by the one for the continuum diagrams alone which coincides
with the observation that the continuum diagrams play a significant role
for small values of $Q$ (where the K-factor does not deviate much from
the single-Higgs case) and are dominant for large $Q$. This result shows
that the K-factor cannot be approximated well by the one of single-Higgs
production for large values of $Q$ due to the large mass effects of the
virtual corrections.

\subsection{Total cross section}
%              ===================
The total cross section has been obtained from the invariant Higgs-pair
mass distribution by means of a numerical integration of the bins in $Q$
with the trapezoidal method for $Q>300$ GeV. For a reliable result, we
used a Richardson extrapolation \cite{Richardson} in terms of the bin
size in $Q$ also for this step. For $Q<300$ GeV, we have adopted the
extension of Boole's rule to six nodes \cite{Abramowitz}. We obtain the
following values for the total cross section at various c.m.~energies,
\begin{eqnarray}
\sqrt{s} = 13~{\rm TeV}: \quad
\sigma_{tot} & = & 27.73(7)^{+13.8\%}_{-12.8\%}~{\rm fb}, \nonumber \\
\sqrt{s} = 14~{\rm TeV}: \quad
\sigma_{tot} & = & 32.81(7)^{+13.5\%}_{-12.5\%}~{\rm fb}, \nonumber \\
\sqrt{s} = 27~{\rm TeV}: \quad
\sigma_{tot} & = & 127.0(2)^{+11.7\%}_{-10.7\%}~{\rm fb}, \nonumber \\
\sqrt{s} = 100~{\rm TeV}: \quad
\sigma_{tot} & = & 1140(2)^{+10.7\%}_{-10.0\%}~{\rm fb},
\label{eq:signlo}
\end{eqnarray}
where we have used the {\tt PDF4LHC} parton densities with
$\alpha_s(M_Z)=0.118$ and added for completeness also the value for a
c.m.~energy of 13 TeV. The numbers in brackets show the numerical
errors, while the upper and lower per-centage entries determine the
(asymmetric) renormalization and factorization scale dependences. The
corresponding results in the Born-improved HTL with {\tt PDF4LHC} PDFs,
obtained with the program {\tt Hpair} \cite{hpair}, read
\begin{eqnarray}
\sqrt{s} = 13~{\rm TeV}: \quad
\sigma_{HTL} & = & 32.51^{+18\%}_{-15\%}~{\rm fb}, \nonumber \\
\sqrt{s} = 14~{\rm TeV}: \quad
\sigma_{HTL} & = & 38.65^{+18\%}_{-15\%}~{\rm fb}, \nonumber \\
\sqrt{s} = 27~{\rm TeV}: \quad
\sigma_{HTL} & = & 156.2^{+17\%}_{-13\%}~{\rm fb}, \nonumber \\
\sqrt{s} = 100~{\rm TeV}: \quad
\sigma_{HTL} & = & 1521^{+16\%}_{-13\%}~{\rm fb}.
\label{eq:sightl}
\end{eqnarray}
Comparing the results of Eqs.~(\ref{eq:signlo}) and (\ref{eq:sightl}), we
observe a reduction of the total cross section by about 15\% due to the
top-mass effects at NLO and a reduction of the scale uncertainty. These
numbers, as well as the differential distributions presented in
Section~\ref{sec:diffxs}, agree with the results of Refs.~\cite{Borowka:2016ehy,
Borowka:2016ypz}\footnote{The small differences of the total cross
sections at the few-per-mille level between the results originate from
the slightly different values of the top mass ($m_t=172.5$ GeV in our
analysis, $m_t=173$ GeV in Refs.~\cite{Borowka:2016ehy,
Borowka:2016ypz}).}. It should be noted that a comparison of the full
virtual corrections with the analytical large top-mass expansion
presented in Ref.~\cite{Grigo:2015dia} was performed in
Refs.~\cite{Borowka:2016ehy,Borowka:2016ypz} and shows a convergence
to the full result below the $t\bar t$-threshold, as expected.

\subsection{Uncertainties originating from the top-mass definition}
%              ======================================================
An uncertainty that has been neglected or underestimated often
previously is the intrinsic uncertainty due to the scheme and scale
choice of the virtual top mass. This does not play a large role for
single on-shell Higgs-boson production via gluon fusion, $gg\to H$,
since the Higgs mass is small and thus the HTL works well, i.e.~top-mass
effects are suppressed. This uncertainty, however, plays a significant
role for the larger values of $Q$ in Higgs-pair production. Top-mass
effects are already sizeable at LO, but the NLO corrections add
additional relevant top-mass dependences on top of the LO result as we
have discussed in the previous subsection. The top mass is a scheme and
scale dependent quantity so that the related uncertainties need to be
estimated for a reliable determination of the total theoretical
uncertainties. For this analysis, we have evaluated the differential
cross section for the top mass defined in the on-shell scheme (default)
and in the $\overline{\rm MS}$-scheme at the scale $\mu_t$,
i.e.~adjusting the counterterms and input parameters to the choices
$\overline{m}_t(\overline{m}_t)$ and $\overline{m}_t(\mu_t)$ with
$\mu_t$ in the range between $Q/4$ and $Q$ according to Section
\ref{sc:renorm}\footnote{We do not separate the treatment of the
  top-Yukawa couplings and the propagator-top mass, since both are
  linked by the sum rule emerging from the electroweak $SU(2)\times
  U(1)$ symmetry, $y_t - \sqrt{2} m_t/v = 0$, which is
  needed for the cancellation of divergences in electroweak
  corrections.}. Since the scale dependence on $\mu_t$ is a
monotonously falling function, we evaluated the differential cross
section for four choices of the top mass, $m_t$,
$\overline{m}_t(\overline{m}_t)$, $\overline{m}_t(Q/4)$ and
$\overline{m}_t(Q)$, for each bin in $Q$.

\begin{figure}[htb!]
  \centering
  \includegraphics[width=0.6\textwidth]{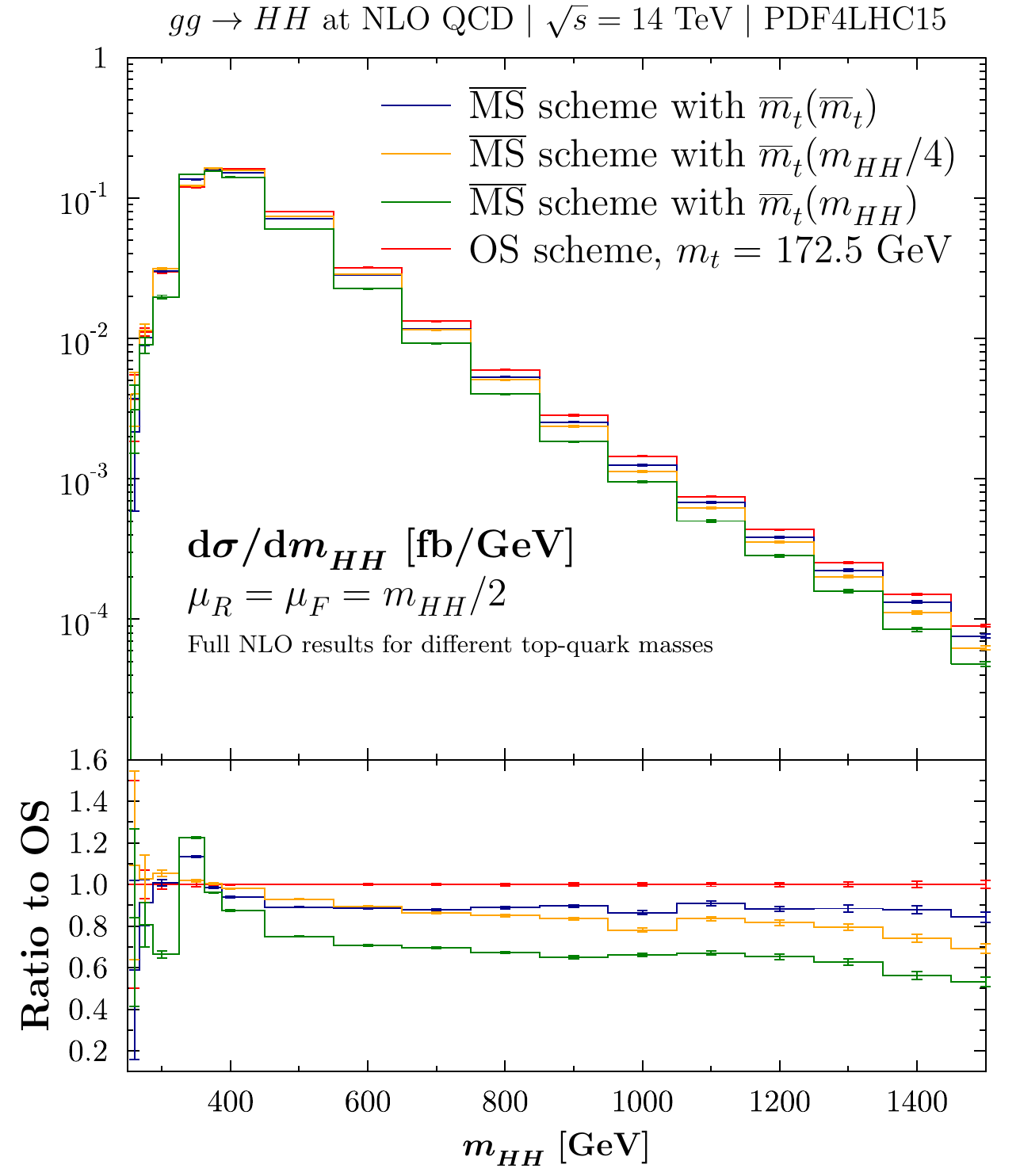}
  %                  \hspace{3mm}
  \caption[]{The differential Higgs-pair production cross section at NLO
as a function of the invariant Higgs-pair mass for a c.m.~energy of 14
TeV for four different choices of the scheme and scale of the top mass.
The lower panel shows the ratio of all results to the default results
with the top pole mass (OS scheme). {\tt PDF4LHC} PDFs have been used
and the renormalization and factorization scales of $\alpha_s$ and the
PDFs have been fixed at our central scale choice $\mu_R=\mu_F=Q/2$.}
  \label{fig:distribmt_1}
\end{figure}

\begin{figure}[htb!]
  \centering
 \includegraphics[width=0.45\textwidth]{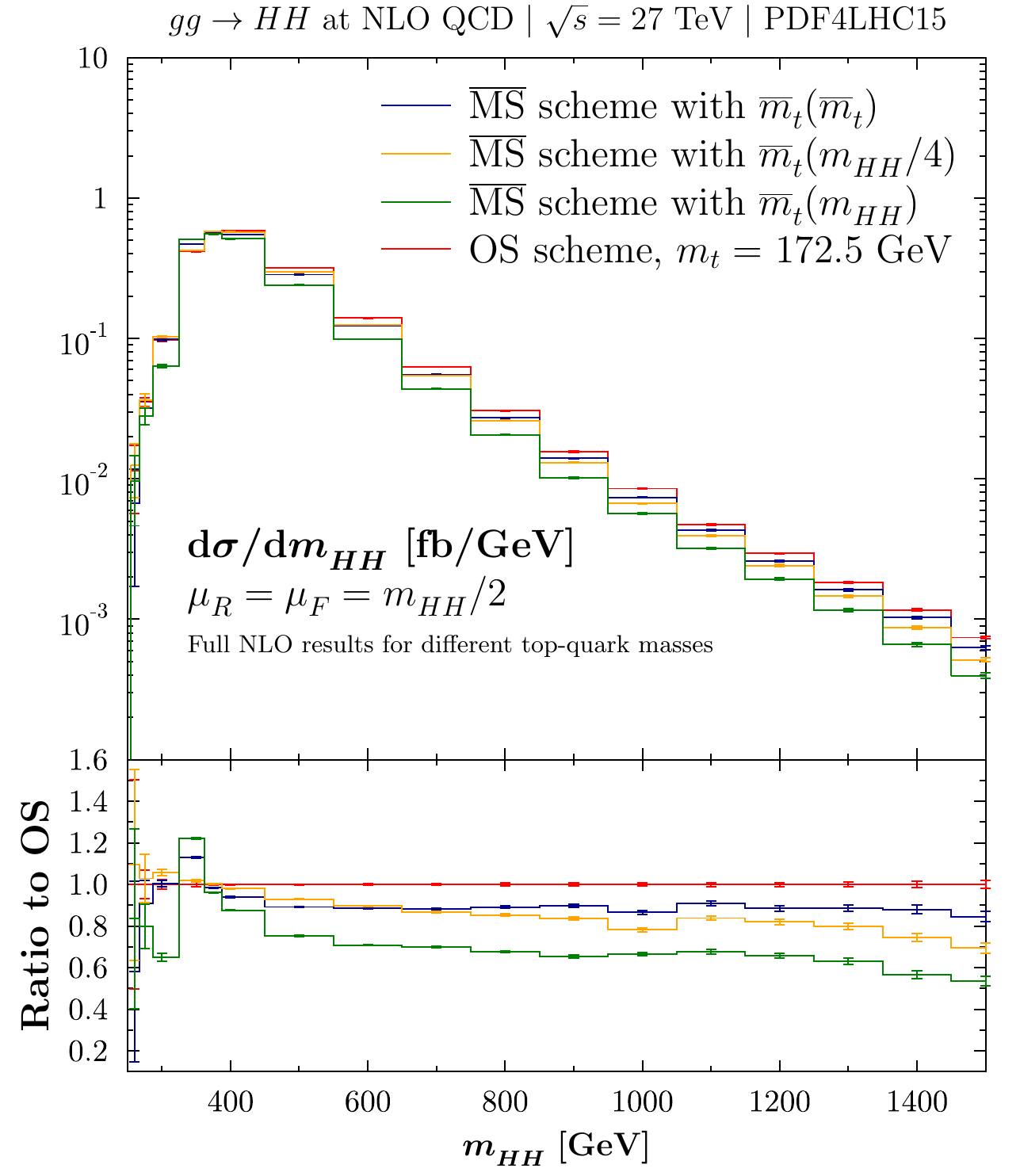}
  \hspace{3mm}
  \includegraphics[width=0.45\textwidth]{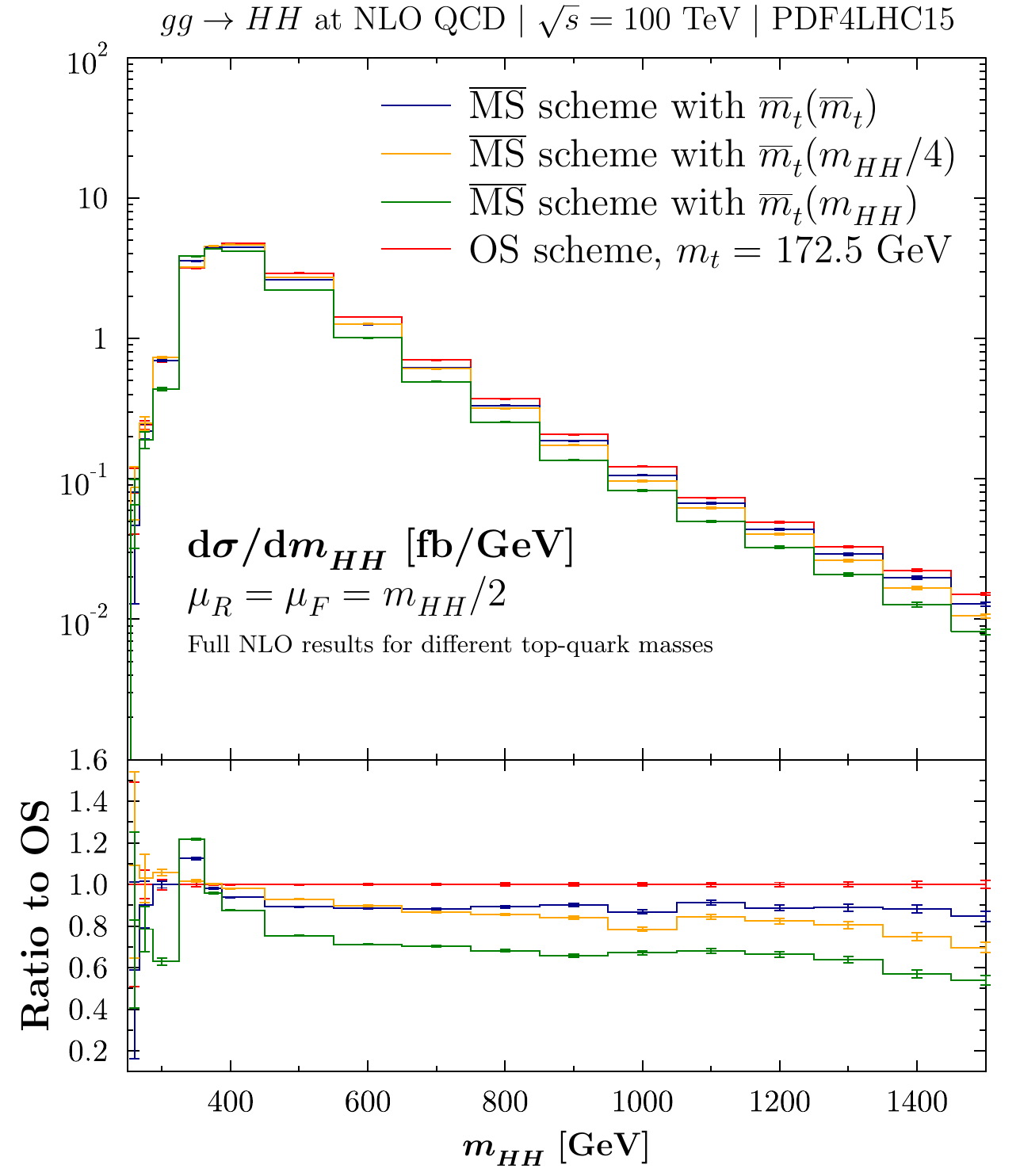}
  \caption[]{Same as Fig.~\ref{fig:distribmt_1} but for c.m.~energies of
27 (left) and 100 (right) TeV.}
  \label{fig:distribmt_2}
\end{figure}

For the three c.m.~energies of 14, 27 and 100 TeV the differential cross
sections are presented in Figs.~\ref{fig:distribmt_1},
\ref{fig:distribmt_2} as a function of $Q=m_{HH}$ for the various
definitions of the top mass.  The lower panels exhibit the ratios of
the differential cross sections to the ones in terms of the top pole
mass (OS scheme).%
It is clearly visible that the scale and scheme dependence of the top
mass induces sizeable variations of the NLO Higgs-pair production cross
section and thus contributes to the theoretical uncertainties. For
small $Q$ values, the size pattern of the differential cross section due
to the different scale and scheme choices is varying. For large values
of $Q$, the maximum is always given by the on-shell scheme and the
minimum in terms of the $\overline{\rm MS}$-top mass $\overline{m}_t(Q)$
with sizeable differences to the on-shell scheme. Adopting the related
uncertainties as the envelope of the cross sections for our four
choices, we arrive at the following uncertainties of the differential
cross section for a c.m.~energy $\sqrt{s}=14$ TeV,
\begin{eqnarray}
\frac{d\sigma_{NLO}}{dQ}\Big|_{Q=300~{\rm GeV}} & = &
0.02978(7)^{+6\%}_{-34\%}\
{\rm fb/GeV},\nonumber\\
\frac{d\sigma_{NLO}}{dQ}\Big|_{Q=400~{\rm GeV}} & = &
0.1609(4)^{+0\%}_{-13\%}\ 
{\rm fb/GeV},\nonumber\\
\frac{d\sigma_{NLO}}{dQ}\Big|_{Q=600~{\rm GeV}} & = &
0.03204(9)^{+0\%}_{-30\%}\
{\rm fb/GeV},\nonumber\\
\frac{d\sigma_{NLO}}{dQ}\Big|_{Q=1200~{\rm GeV}} & = &
0.000435(4)^{+0\%}_{-35\%}\
{\rm fb/GeV} \, .
\end{eqnarray}
Since these uncertainties are given relative to the on-shell results,
the upper uncertainty vanishes for $Q\geq 400$ GeV, because the on-shell
results provide the maximal values. These uncertainties turn out to be
significant and at a similar level as the usual renormalization and
factorization scale uncertainties. Thus, they constitute an additional
contribution to the total theoretical uncertainties that has to be taken
into account. The uncertainties due to the top-mass scheme and scale are
about a factor of two smaller than at LO,
\begin{eqnarray}
\frac{d\sigma_{LO}}{dQ}\Big|_{Q=300~{\rm GeV}} & = &
0.01656^{+62\%}_{-2.4\%}\, \mathrm{fb/GeV},\nonumber\\
\frac{d\sigma_{LO}}{dQ}\Big|_{Q=400~{\rm GeV}} & = &
0.09391^{+0\%}_{-20\%}\, \mathrm{fb/GeV},\nonumber\\
\frac{d\sigma_{LO}}{dQ}\Big|_{Q=600~{\rm GeV}} & = &
0.02132^{+0\%}_{-48\%}\, \mathrm{fb/GeV},\nonumber\\
\frac{d\sigma_{LO}}{dQ}\Big|_{Q=1200~{\rm GeV}} & = &
0.0003223^{+0\%}_{-56\%}\, \mathrm{fb/GeV}
\end{eqnarray}
that have been obtained for a c.m.~energy of 14 TeV and using PDF4LHC15
NLO parton densities with a NLO strong coupling normalized to
$\alpha_s(M_Z)=0.118$\footnote{Note that these choices are incompatible
with a consistent LO prediction, but the relative uncertainties related
to the scheme and scale choice of the top mass will be hardly affected
by this inconsistency. These uncertainties are just parametric at LO.}.
Their reduction from LO to NLO underlines that the NLO QCD corrections
stabilize the theoretical prediction for the Higgs-pair production cross
section. The large size of the residual uncertainties is just a
consequence of the large NLO QCD corrections as is the case for the
renormalization and factorization scale dependences, too. Adopting the
envelope for each $Q$-bin individually and integrating over $Q$, we
arrive at the impact of these uncertainties on the total cross section
for various c.m.~energies,
\begin{eqnarray}
\sqrt{s} = 13~{\rm TeV}: \quad
\sigma_{tot} & = & 27.73(7)^{+4\%}_{-18\%}~{\rm fb}, \nonumber \\
\sqrt{s} = 14~{\rm TeV}: \quad
\sigma_{tot} & = & 32.81(7)^{+4\%}_{-18\%}~{\rm fb}, \nonumber \\
\sqrt{s} = 27~{\rm TeV}: \quad
\sigma_{tot} & = & 127.0(2)^{+4\%}_{-18\%}~{\rm fb}, \nonumber \\
\sqrt{s} = 100~{\rm TeV}: \quad
\sigma_{tot} & = & 1140(2)^{+3\%}_{-18\%}~{\rm fb}
\end{eqnarray}
using {\tt PDF4LHC} PDFs. A further reduction of these uncertainties can
only be achieved by the determination or reliable estimate of the full
mass effects at NNLO.

Since these uncertainties are sizeable, one may wonder why this has not
been observed already for single-Higgs boson production $gg\to H$. The
measured value of the Higgs mass $M_H=125$ GeV is small compared to the
top mass so that for single on-shell Higgs production we are close to
the HTL, i.e.~finite top-mass effects are small and thus the related
uncertainties, too. However, going to larger virtualities $Q$ for
off-shell Higgs production $gg\to H^*$ (or larger Higgs masses for
on-shell Higgs production), we arrive at similar uncertainties for
$\sqrt{s}=14$ TeV,
\begin{eqnarray}
\sigma_{NLO}\Big|_{Q=125~{\rm GeV}} & = 42.17^{+0.4\%}_{-0.5\%}\,
\mathrm{pb}, \qquad
\sigma_{NLO}\Big|_{Q=300~{\rm GeV}} & = 9.85^{+7.5\%}_{-0.3\%}\,
\mathrm{pb},\nonumber \\[0.5cm]
\sigma_{NLO}\Big|_{Q=400~{\rm GeV}} & = 9.43^{+0.1\%}_{-0.9\%}\,
\mathrm{pb}, \qquad
\sigma_{NLO}\Big|_{Q=600~{\rm GeV}} & = 1.97^{+0.0\%}_{-15.9\%}\,
\mathrm{pb},\nonumber \\[0.5cm]
\sigma_{NLO}\Big|_{Q=900~{\rm GeV}} & = 0.230^{+0.0\%}_{-22.3\%}\,
\mathrm{pb}, \quad
\sigma_{NLO}\Big|_{Q=1200~{\rm GeV}} & = 0.0402^{+0.0\%}_{-26.0\%}\,
\mathrm{pb}
\end{eqnarray}
using PDF4LHC PDFs. This has been known for a long time since there are
sizeable effects on the virtual corrections due to the scale choice of
the top mass for larger values of $Q$ or the Higgs mass (see Fig. 7a of
Ref.~\cite{Spira:1995rr}). For the single off-shell Higgs case, a
reduction of the top-mass scale dependence by roughly a factor of two by
going from LO to NLO has been observed, too, as can be inferred from the
comparison with the explicit LO numbers for $\sqrt{s}=14$ TeV,
\begin{eqnarray}
\sigma_{LO}\Big|_{Q=125~{\rm GeV}} & = 18.43^{+0.8\%}_{-1.1\%}\,
\mathrm{pb}, \qquad
\sigma_{LO}\Big|_{Q=300~{\rm GeV}} & = 4.88^{+23.1\%}_{-1.1\%}\,
\mathrm{pb}, \nonumber \\[0.5cm]
\sigma_{LO}\Big|_{Q=400~{\rm GeV}} & = 4.94^{+1.2\%}_{-1.8\%}\,
\mathrm{pb}, \qquad
\sigma_{LO}\Big|_{Q=600~{\rm GeV}} & = 1.13^{+0.0\%}_{-26.2\%}\,
\mathrm{pb}, \nonumber \\[0.5cm]
\sigma_{LO}\Big|_{Q=900~{\rm GeV}} & = 0.139^{+0.0\%}_{-36.0\%}\,
\mathrm{pb}, \quad
\sigma_{LO}\Big|_{Q=1200~{\rm GeV}} & = 0.0249^{+0.0\%}_{-41.1\%}\,
\mathrm{pb}
\end{eqnarray}
that have been obtained with PDF4LHC PDFs as in the Higgs-pair case. On
the other hand, the uncertainties for $Q=125$ GeV confirm that they are
small for on-shell Higgs production via gluon fusion (already at LO) in
agreement with the analysis of the LHC Higgs Cross Section Working
Group \cite{deFlorian:2016spz, Anastasiou:2016cez}.

A relevant issue is the theoretical background of the different scale
choices for the top mass. For small values of $Q$, the matrix element
will be closer to the HTL such that the NLO corrections get closer to
the HTL calculation. The HTL on the other hand can be treated by
starting from the effective Lagrangian of Eq.~(\ref{eq:leff}) which is
the residual effective coupling of Higgs bosons to gluons after
integrating out the top quark. Thus, the corresponding Wilson
coefficients $C_1$ and $C_2$ are determined by matching the full SM with
the top quark to the effective theory without the top quark. The
matching scale is naturally given by the top mass. Performing the proper
matching at the scale of the top mass, i.e.~using either the top pole
mass or the top $\overline{\rm MS}$ mass at the scale of the top mass
itself leads to non-logarithmic (in the top mass) matching contributions
[see Eq.(\ref{eq:leffcoeff}) for $\mu_R=m_t$] also for higher powers in
$1/m_t^2$, i.e.~higher-dimensional operators contributing to the gluonic
Higgs couplings at the subleading level. This implies that the top mass
is the preferred scale choice for small values of $Q$. This is confirmed
by the heavy top expansion of the form factors of
Refs.~\cite{Grigo:2013rya, Grigo:2015dia, Davies:2018qvx}.

At large $Q$ values, on the other hand, we can use the results for the
high-energy expansion of Ref.~\cite{Davies:2018qvx}. In the regime of
large $Q$, the triangle-diagram contributions are suppressed by the
$s$-channel Higgs propagator so that the box diagrams provide the
dominant contributions. In our normalization, the explicit results of the
virtual box-form factors in the high-energy limit ($Q\gg m_t, M_H$) in
terms of the top pole mass $m_t$ are given by\footnote{The NLO form
factors of Eq.~(\ref{eq:ffhe}) correspond to the infrared-subtracted
ones according to Ref.~\cite{Davies:2018qvx} plus the additional
subtraction of the HTL. The piece related to the latter is absorbed in
the functions $G_{1,2}$.}
\begin{eqnarray}
F_i & = & F_{i,LO} + \Delta F_i \, , \nonumber \\
\Delta F_i & = & \Delta F_{i,HTL} + \Delta F_{i,mass} \, , \nonumber \\
F_{1,LO} & \to & 4\frac{m_t^2}{\hat s} \, , \nonumber \\
F_{2,LO} & \to & -\frac{m_t^2}{\hat s \hat t (\hat s+\hat t)} \Big\{
(\hat s+\hat t)^2 L_{1ts}^2 + \hat t^2 L_{ts}^2 + \pi^2 [(\hat s+\hat
t)^2+\hat t^2] \Big\} \, , \nonumber \\
\Delta F_{1,mass} & \to & \frac{\alpha_s}{\pi}\left\{ 2 F_{1,LO}
\log\frac{m_t^2}{\hat s} + \frac{m_t^2}{\hat s} G_1(\hat s, \hat t)
\right\} \, , \nonumber \\
\Delta F_{2,mass} & \to & \frac{\alpha_s}{\pi}\left\{ 2 F_{2,LO}
\log\frac{m_t^2}{\hat s} + \frac{m_t^2}{\hat s} G_2(\hat s, \hat t)
\right\} \, ,
\label{eq:ffhe}
\end{eqnarray}
where $G_{1,2}(\hat s, \hat t)$ denote explicit and lengthy functions
of the kinematical variables $\hat s$ and $\hat t$ that do {\it not}
depend on the top mass \cite{Davies:2018qvx}. The logarithms $L_{ts},
L_{1ts}$ are defined as
\begin{equation}
L_{ts} = \log \left(-\frac{\hat t}{\hat s}\right) + i\pi \, , \qquad
L_{1ts} = \log \left(1+\frac{\hat t}{\hat s}\right) + i\pi \, .
\end{equation}
Transforming the top pole mass $m_t$ into the $\overline{\rm MS}$ mass
$\overline{m}_t(\mu_t)$, we arrive at the LO expressions for
$F_{1/2,LO}$ with $m_t$ replaced by $\overline{m}_t(\mu_t)$ and the
appropriately transformed NLO coefficients
\begin{eqnarray}
F_{1,LO} & \to & 4\frac{\overline{m}_t^2(\mu_t)}{\hat s} \, , \nonumber \\
F_{2,LO} & \to & -\frac{\overline{m}_t^2(\mu_t)}{\hat s \hat t (\hat
s+\hat t)} \Big\{ (\hat s+\hat t)^2 L_{1ts}^2 + \hat t^2 L_{ts}^2 +
\pi^2 [(\hat s+\hat t)^2+\hat t^2] \Big\} \, , \nonumber \\
\Delta F_{1,mass} & \to & \frac{\alpha_s}{\pi}\left\{ 2 F_{1,LO} \left[
\log\frac{\mu_t^2}{\hat s} + \frac{4}{3} \right] +
\frac{\overline{m}_t^2(\mu_t)}{\hat s} G_1(\hat s, \hat t) \right\} \, ,
\nonumber \\
\Delta F_{2,mass} & \to & \frac{\alpha_s}{\pi}\left\{ 2 F_{2,LO} \left[
\log\frac{\mu_t^2}{\hat s} + \frac{4}{3} \right] +
\frac{\overline{m}_t^2(\mu_t)}{\hat s} G_2(\hat s, \hat t) \right\} \, .
\end{eqnarray}
To minimize the logarithms of $\mu_t$, a dynamical scale of the order of
$\sqrt{\hat s}=Q$ has to be chosen, but {\it not} the top mass. A
coefficient $\kappa$ in front of the dynamical scale choice $\mu_t =
\kappa Q$ is still arbitrary (but should not be large) since additional
finite parts of the functions $G_{1,2}(\hat s, \hat t)$ may be absorbed
in the scale choice. Thus, the dynamical scale $Q$ can be identified as
the preferred central scale choice of the Yukawa couplings for large $Q$
values.

The uncertainties originating from the scheme and scale dependence of
the top mass can be reduced by calculating the NNLO mass effects. Such a
three-loop calculation is beyond everything that has been performed so
far with current methods, but for $Q$ values close to threshold a
large-mass expansion at NNLO could be used to reach an approximate
estimate of the finite top-mass effects at NNLO. As a first step,
partial results of the NNLO top-mass effects are known in the
soft+virtual approximation \cite{Grigo:2015dia}. For $Q$ values around
the virtual $t\bar t$ threshold $Q\sim 2m_t$, non-relativistic Green's
functions could be used that allow the introduction of higher-order
corrections to the QCD potential \cite{Fadin:1987wz, Fadin:1988fn,
Fadin:1990wx, Strassler:1990nw, Melnikov:1994jb}. This may lead to an
improved description of the threshold region. However, for the triangle
diagrams, the threshold behaviour is determined by $P$-wave
contributions, since the $t\bar t$-ground state appears as a ${\cal
CP}$-odd configuration that does not mix with the virtual ${\cal
CP}$-even threshold state of the triangle diagrams. For the box
diagrams, the $P$--wave contributions have to be considered, too.
Moreover, it is unclear how large the impact of top-mass effects of the
remainder beyond the non-relativistic Green's functions will be.
Finally, for the high-energy tail, the approximate calculation of
Ref.~\cite{Davies:2018qvx} could be extended to NNLO.

\subsection{Variation of the cross section with $\lambda_{H^3}$}
%              ===================================================
Higgs-pair production at the LHC is directly sensitive to the trilinear
Higgs coupling. The dependence of the total and differential cross
sections on the trilinear coupling $\lambda_{H^3}$ is modified by the
NLO QCD corrections and in particular by the finite mass effects at LO
and NLO. Finite top-mass effects result in a non-vanishing matrix
element at threshold, while in the HTL the matrix element of
Eq.~(\ref{eq:lomat}) vanishes exactly \cite{Glover:1987nx, Plehn:1996wb,
Li:2013rra},
\begin{eqnarray}
{\cal A}^{\mu\nu} & \to & F_1 T_1^{\mu\nu} \, , \nonumber \\
F_1 & \to & \frac{2}{3}(C_\triangle - 1)
\to \frac{2}{3}~\left(\frac{3M_H^2}{4 M_H^2-M_H^2} - 1\right) = 0 \qquad
\mbox{for $Q^2\to (2M_H)^2$} \, ,
\end{eqnarray}
where we have used that the second form factor $G_\Box$ vanishes in the
HTL [see Eq.~(\ref{eq:ffhtl})]. The cancellation is induced by the
destructive interference between the triangle and box diagrams at LO.
This property is modified by finite subleading ${\cal O}(1/m_t^2)$ terms
but explains why the matrix element itself is suppressed at the
production threshold. As a function of $\lambda_{H^3}$, the cross section
develops a minimum at $\lambda_{H^3}$-values around 2.4 times the
SM-value in the Born-improved HTL \cite{Dawson:1998py, Baglio:2012np}
since the phase-space integration adds contributions from above the
production threshold. The NLO QCD corrections will shift the minimum of
the cross section as a function of $\lambda_{H^3}$ and finite top-mass
effects play a prominent role in the amount of these cancellations. For
the determination of the trilinear coupling, the variation of the cross
section with $\lambda_{H^3}$ is of interest. As mentioned in the
introduction, the total cross section behaves approximately as
$\Delta\sigma/\sigma \sim -\Delta\lambda_{H^3}/\lambda_{H^3}$ for
$\lambda_{H^3}$ close to the SM value.

In the following, we will analyze the NLO results, where only the
trilinear coupling has been varied. In general, however, several
coupling modifications contribute to the Higgs-pair production cross
section. This could be treated consistently by extending the SM
Lagrangian by all contributing dimension-6 operators as has been studied
in Ref.~\cite{Grober:2015cwa} in the HTL at NLO and in
Ref.~\cite{deFlorian:2017qfk} at NNLO. Recently the HTL analysis has
been extended to the inclusion of finite top-mass effects at NLO
\cite{Buchalla:2018yce}. However, we will neglect all dimension-6
operators but the one modifying the Higgs self-interactions.  A proper
and consistent effective model of this type has been discussed in
Ref.~\cite{Degrassi:2017ucl} that adds higher-dimension operators to the
scalar Higgs sector only. Thus, a sole variation of the Higgs
self-interactions could be realized within Higgs portal models with
additional heavy scalar states that couple only to the SM-like Higgs
field and are integrated out.
 
\begin{figure}[htb!]
  \centering
  \includegraphics[width=0.6\textwidth]{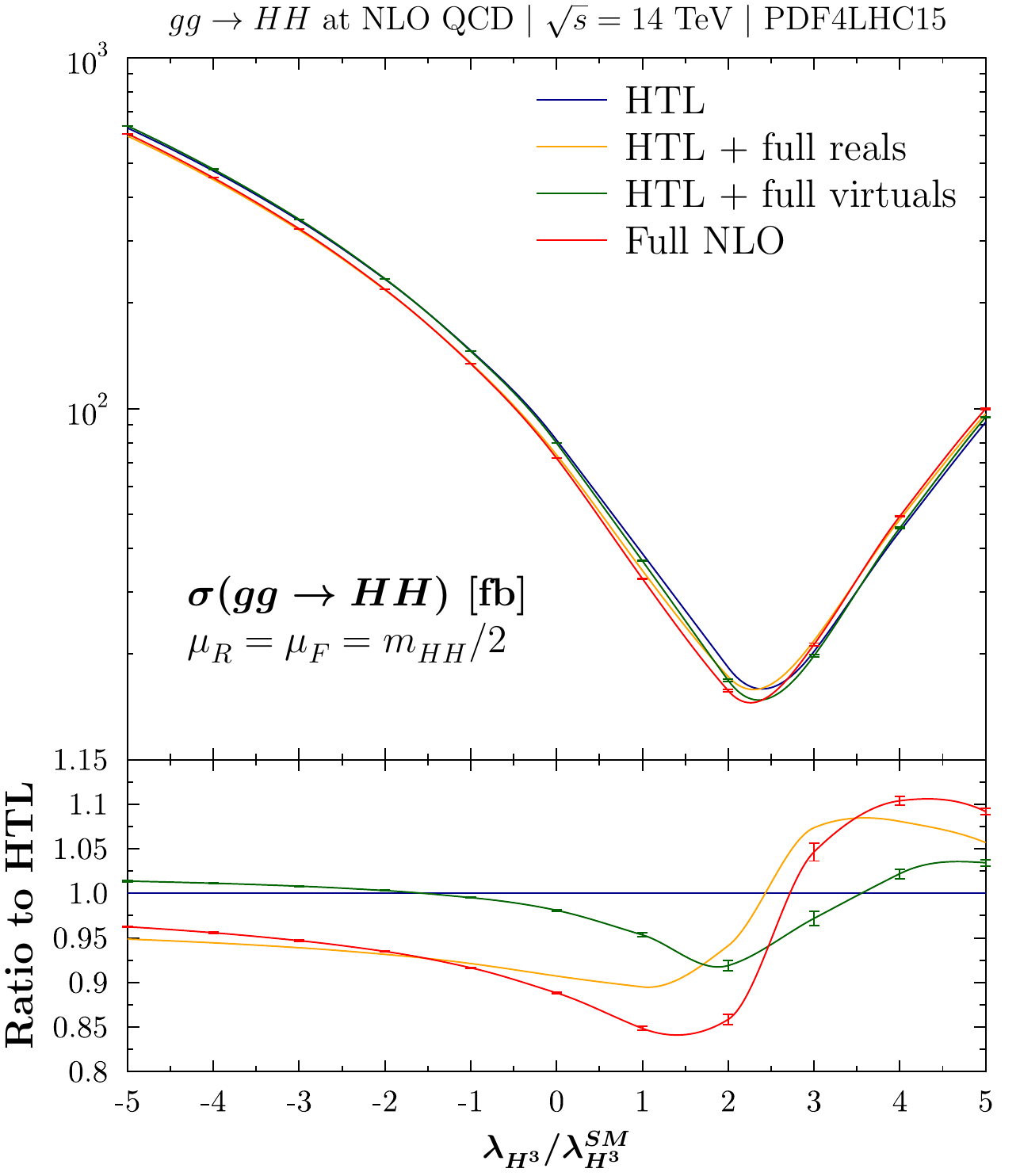}
  %                  \hspace{3mm}
  \caption[]{The total Higgs-pair production cross section at NLO as a
function of the trilinear self-coupling $\lambda_{H^3}$ in units of the
SM value for a c.m.~energy of 14 TeV. The blue curve shows the
Born-improved HTL, the yellow includes the NLO mass effects of the real
corrections in addition and the green curve those of the virtual
corrections in addition. The full NLO result is presented by the red
curve. The lower panel shows the ratio of all results to the
Born-improved HTL. {\tt PDF4LHC} PDFs have been used and the
renormalization and factorization scales of $\alpha_s$ and the PDFs
have been fixed at our central scale choice $\mu_R=\mu_F=Q/2=m_{HH}/2$.}
  \label{fig:lambdavar_1}
\end{figure}
\begin{figure}[htb!]
  \centering
  \includegraphics[width=0.45\textwidth]{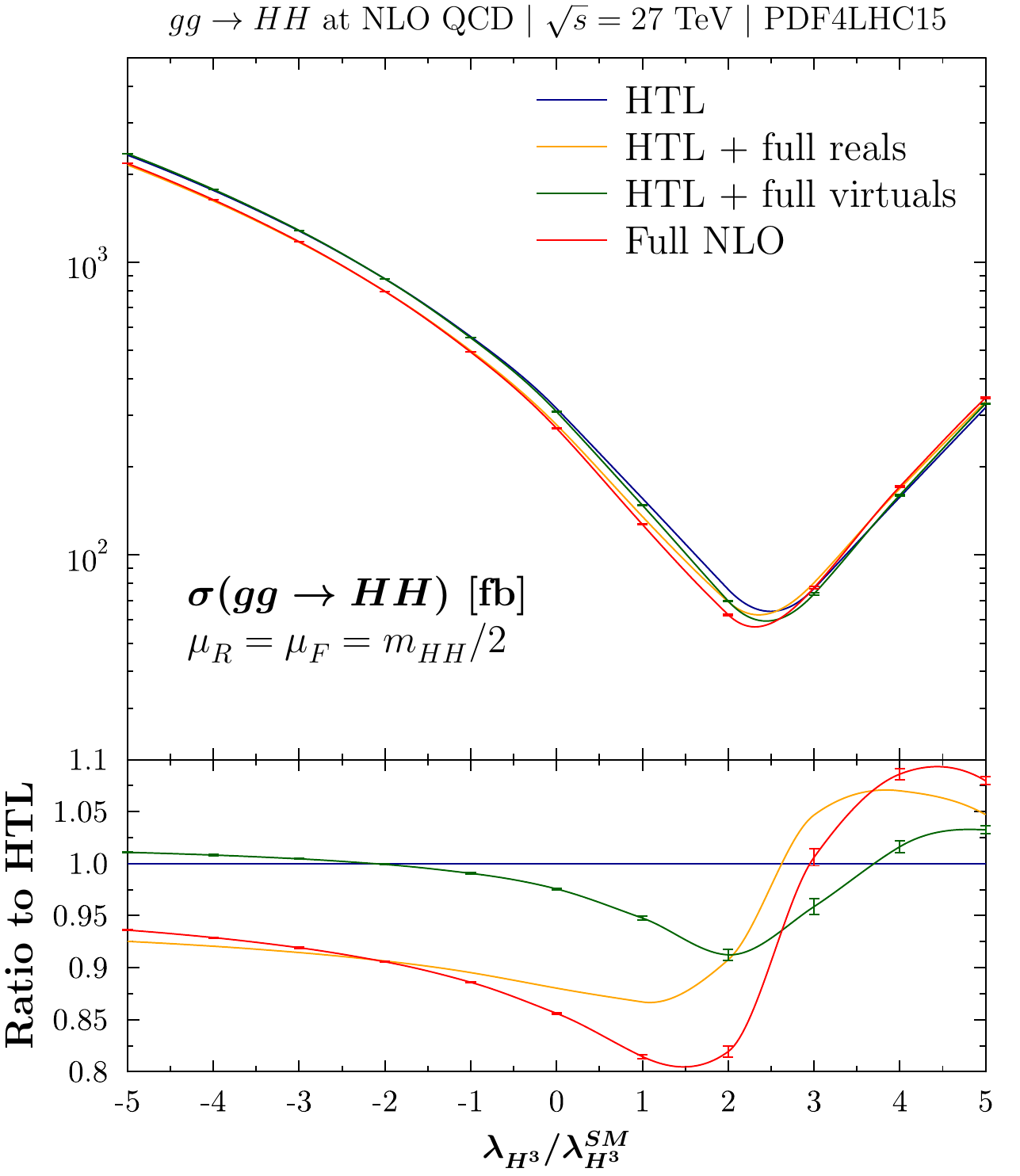}
  \hspace{3mm}
  \includegraphics[width=0.45\textwidth]{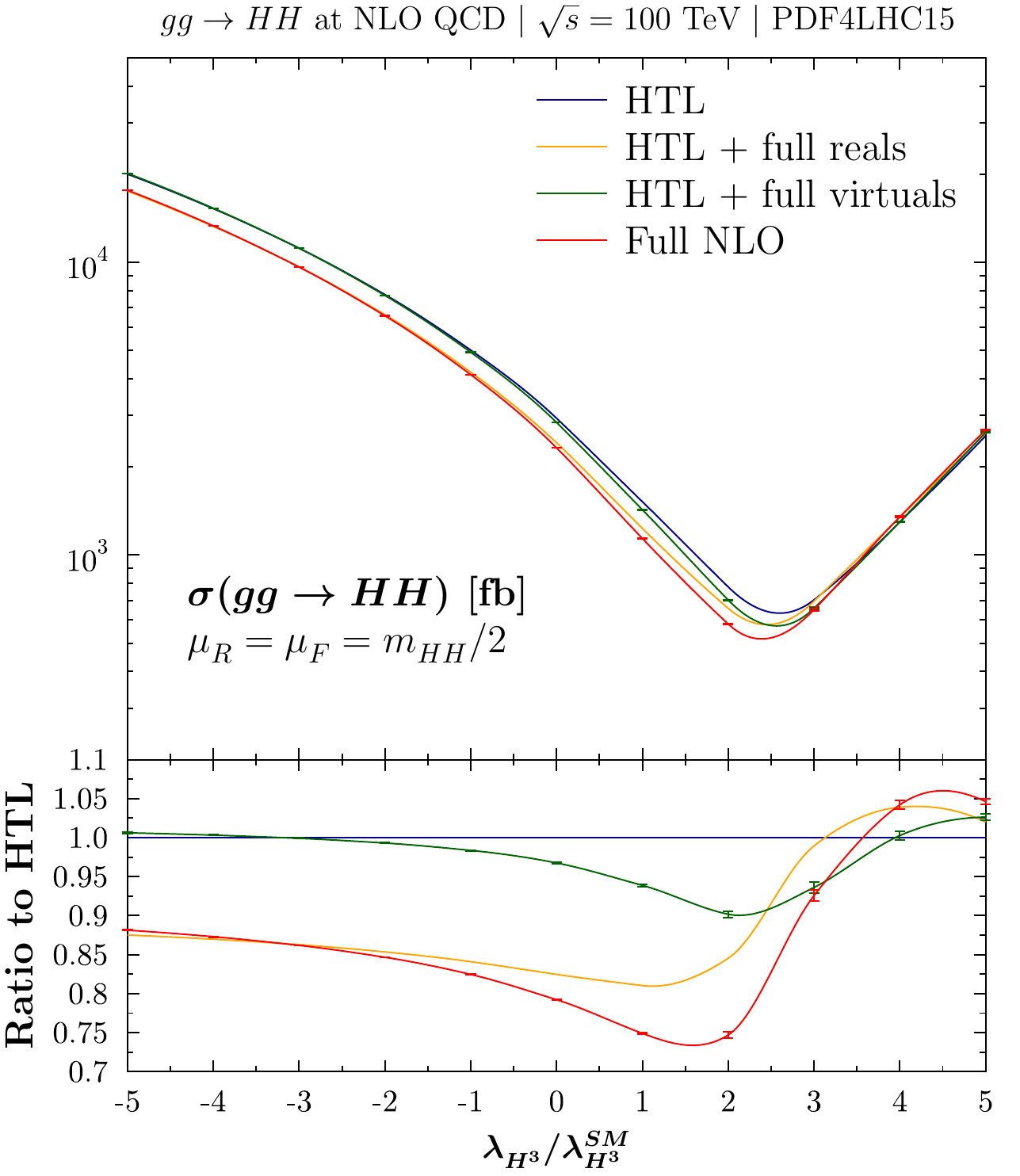}
  \caption[]{Same as Fig.~\ref{fig:lambdavar_1} but for c.m.~energies of
27 (left) and 100 (right) TeV.}
  \label{fig:lambdavar_2}
\end{figure}

In Figs.~\ref{fig:lambdavar_1} and \ref{fig:lambdavar_2}, the dependence
of the total Higgs-pair production cross section is shown as a function
of the trilinear Higgs coupling $\lambda_{H^3}$ in units of the SM
coupling for three c.m.~energies, 14, 27 and 100 TeV. The blue curves
display the results in the Born-improved HTL, the yellow curves include
the mass effects of the real corrections and the green curves the mass
effects of virtual corrections in addition. The red curves exhibit the
complete NLO results. The comparison of the blue and red curves
indicates that the minimum of the $\lambda_{H^3}$-variation is shifted
from about 2.4 times the SM value to about 2.3 times the SM value due to
the NLO mass effects. The yellow and green curves imply that the main
origin of this shift emerges from the mass effects of the real
corrections. The lower panels of Figs.~\ref{fig:lambdavar_1} and
\ref{fig:lambdavar_2} present the ratios of the individual contributions
to the Born-improved HTL.  While the NLO mass effects are of moderate
size for negative values of $\lambda_{H^3}$, where the triangle and box
diagrams interfere constructively, they turn out to be more relevant in
the region of destructive interference, in particular around the minima
of the cross sections. The significantly varying NLO mass effects have
to be taken into account when determining the value of $\lambda_{H^3}$
from the experimental data at the HL-LHC. This agrees with the findings
of Ref.~\cite{Buchalla:2018yce}. The NLO mass effects on the variation
of the total cross section with $\lambda_{H^3}$ become larger with
rising c.m.~energy of the hadron collider.

\begin{figure}[htb!]
  \centering
  \includegraphics[width=0.6\textwidth]{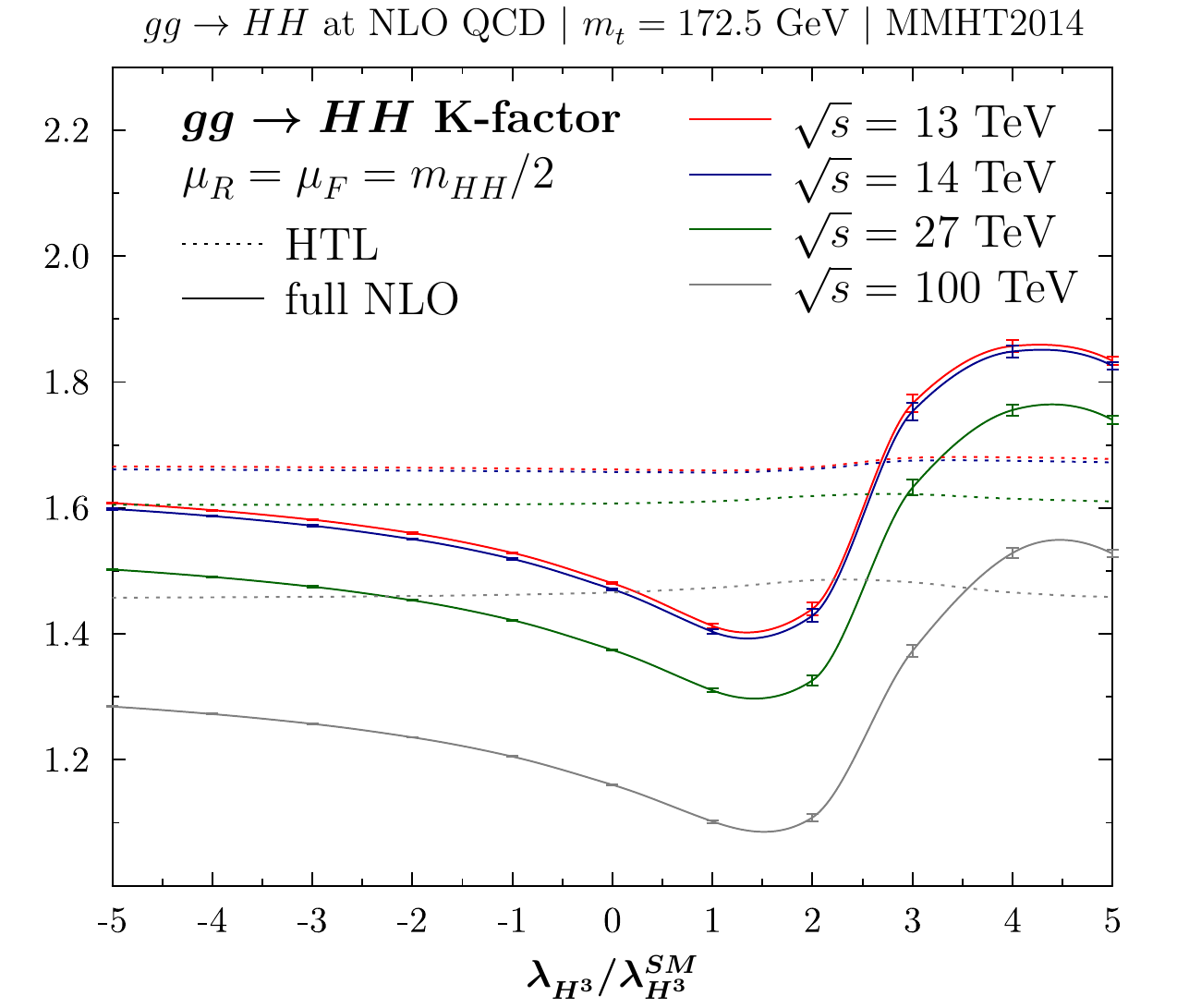}
  %                  \hspace{3mm}
  \caption[]{K-factors of Higgs-pair production at NLO as functions of
the trilinear self-coupling $\lambda_{H^3}$ in units of the SM value
$\lambda_{H^3}^{SM}$ for various c.m.~energies of 13 TeV (red curves),
14 TeV (blue curves), 27 TeV (green curves) and 100 TeV (grey curves).
The full NLO result is presented by the full curves with the error bars
indicating our numerical errors. The dotted curves show the
corresponding K-factors of the Born-improved HTL.  {\tt MMHT2014} PDFs
have been used and the renormalization and factorization scales of
$\alpha_s$ and the PDFs have been fixed at our central scale choice
$\mu_R=\mu_F=Q/2=m_{HH}/2$.}
  \label{fig:lambdakfac}
\end{figure}

In Fig.~\ref{fig:lambdakfac}, we display the consistently defined
K-factors $K=\sigma_{NLO}/\sigma_{LO}$ as a function of $\lambda_{H^3}$
in units of the SM coupling. The full curves
show the NLO K-factors including the NLO top-mass effects for various
c.m.~energies. The dotted curves exhibit the corresponding K-factors in
the Born-improved HTL as computed in Refs.~\cite{Dawson:1998py,
Grober:2015cwa}. The impact of the NLO mass effects on the K-factors
ranges at the level of 10--15\% for negative $\lambda_{H^3}$ values,
where the triangle and box diagrams interfere constructively. For
positive values of $\lambda_{H^3}$ (destructive interference), the size
and sign of the NLO mass effects is changing considerably as can be
inferred from the comparison to the dotted curves. The full K-factors
develop a larger dependence on $\lambda_{H^3}$ than the Born-improved
HTL due to the NLO top-mass effects. This confirms the findings of
Ref.~\cite{Buchalla:2018yce}. The NLO top-mass effects of the total
cross section increase with rising collider energy in general except for
the regions of destructive interference between the triangle and box
diagrams (positive $\lambda_{H^3}$).

The full NLO cross section as a function of $\lambda_{H^3}$ can be
parametrized as
\begin{equation}
\sigma_{NLO} = \sigma_1 + \sigma_2
\frac{\lambda_{H^3}}{\lambda^{SM}_{H^3}} + \sigma_3 \left(
\frac{\lambda_{H^3}}{\lambda^{SM}_{H^3}} \right)^2 \, .
\end{equation}
The coefficients $\sigma_{1\ldots 3}$ depend on the c.m.~energy of the
hadron collider and on the PDFs used in their evaluation. For the various
c.m.~energies, we obtain the following NLO values for {\tt PDF4LHC} PDFs
and our central scale choices $\mu_R=\mu_F=Q/2$,
\begin{eqnarray}
\sqrt{s} = 13~{\rm TeV}: \quad \sigma_1 & = & 61.35(6)~{\rm fb}\, ,
\quad \sigma_2 = -43.26(5)~{\rm fb}\, , \quad \sigma_3 = 9.62(8)~{\rm
fb} \, , \nonumber \\
\sqrt{s} = 14~{\rm TeV}: \quad \sigma_1 & = & 72.27(7)~{\rm fb}\, ,
\quad \sigma_2 = -50.70(6)~{\rm fb}\, , \quad \sigma_3 = 11.23(9)~{\rm
fb} \, , \nonumber \\
\sqrt{s} = 27~{\rm TeV}: \quad \sigma_1 & = & 270.9(3)~{\rm fb}\, ,
\quad \sigma_2 = -183.1(2)~{\rm fb}\, , \quad \sigma_3 = 39.5(4)~{\rm
fb} \, , \nonumber \\
\sqrt{s} = 100~{\rm TeV}: \quad \sigma_1 & = & 2323(2)~{\rm fb}\, ,
\quad\; \sigma_2 = -1496(2)~{\rm fb}\, , \quad\; \sigma_3 = 313(3)~{\rm
fb} \, ,
\label{eq:siglam1}
\end{eqnarray}
where the numbers in brackets denote our numerical errors. The
corresponding coefficients with {\tt MMHT2014} PDFs read
\begin{eqnarray}
\sqrt{s} = 13~{\rm TeV}: \quad \sigma_1 & = & 62.45(7)~{\rm fb}\, ,
\quad \sigma_2 = -44.13(5)~{\rm fb}\, , \quad \sigma_3 = 9.83(9)~{\rm
fb} \, , \nonumber \\
\sqrt{s} = 14~{\rm TeV}: \quad \sigma_1 & = & 73.60(8)~{\rm fb}\, ,
\quad \sigma_2 = -51.75(6)~{\rm fb}\, , \quad \sigma_3 = 11.5(1)~{\rm
fb} \, , \nonumber \\
\sqrt{s} = 27~{\rm TeV}: \quad \sigma_1 & = & 277.4(3)~{\rm fb}\, ,
\quad \sigma_2 = -187.9(2)~{\rm fb}\, , \quad \sigma_3 = 40.6(4)~{\rm
fb} \, , \nonumber \\
\sqrt{s} = 100~{\rm TeV}: \quad \sigma_1 & = & 2401(2)~{\rm fb}\, ,
\quad\; \sigma_2 = -1550(2)~{\rm fb}\, , \quad\; \sigma_3 = 325(3)~{\rm
fb} \, .
\label{eq:siglam2}
\end{eqnarray}
It should be noted that the final numerical errors of the cross sections
as shown in Figs.~\ref{fig:lambdavar_1}--\ref{fig:lambdakfac} are
smaller than the ones emerging from using the coefficients of
Eqs.~(\ref{eq:siglam1}, \ref{eq:siglam2}) since the combinations of
each bin in $Q$ {\it before} integration reduces them.

\section{Conclusions \label{sc:conclusions}}
%        ===========
In this work, we have discussed the full QCD corrections to Higgs-pair
production at NLO. We have explained the details of our numerical
approach to solve the multi-scale two-loop integrals involving
ultraviolet and infrared singularities. The ultraviolet singularities
could be extracted from the finite parts by suitable end-point
subtractions, while the infrared singularities have been isolated by
means of dedicated subtraction terms. The ultraviolet singularities have
been absorbed by the proper renormalization of the strong coupling and
the top mass, while the infrared ones cancel against the one-loop real
corrections involving an additional gluon or quark in the final state of
the Higgs-boson pair. We have performed the evaluation of the virtual
corrections diagram by diagram without tensor reduction.

The emerging integrals develop thresholds if the virtual $t\bar
t$-threshold is crossed, but also at small virtualities due to the
presence of purely gluonic intermediate states. The numerical
stabilization of the virtual two-loop integrals has been achieved
through integrations by parts of the integrands such that the power of
the threshold-singular denominators is reduced. The narrow-width limit
of the virtual top quarks has been obtained by a Richardson
extrapolation of the results for different sizes of an auxiliarly
introduced width parameter. This has allowed a numerical integration of
the virtual two-loop corrections with an accuracy of less than one per
cent.

The matrix elements for the real corrections have been generated with
{\tt FeynArts} and {\tt FormCalc} and integrated using the library {\tt
Collier}.  The collinear region of the phase-space integration has been
regularized numerically by a technical cut.

We have subtracted the Born-improved HTL from the virtual and real
corrections individually so that we have been left with the pure NLO
top-mass effects beyond the Born-improved HTL that is implemented in the
public tool {\tt Hpair}. Thus, the final NLO results have been obtained
by adding back the numbers from {\tt Hpair}.

The final results have been analyzed in detail for the differential
cross section in the invariant Higgs-pair mass and the total cross
section. Finite top-mass effects beyond the Born-improved HTL decrease
the total cross section by about 15\% at the LHC. However, the negative
mass effects are larger for the differential cross section reaching a
level of $-30\%$ or $-40\%$ for large invariant Higgs-pair masses. This
implies that the inclusion of the NLO top-mass effects is crucial for a
reliable analysis at the LHC and future proton colliders. We have
discussed the usual renormalization and factorization scale
uncertainties that are in agreement with previous calculations.
However, we have identified an additional scale and scheme uncertainty
due to the virtual top mass. This uncertainty reaches a level of 15\%
for the total cross section but can be larger (up to 35\%) for the
differential cross section. Based on the heavy-top and high-energy
expansions, we have discussed the preferred scale choices of the running
top mass and identified a large dynamical scale as the proper choice for
large invariant Higgs-pair masses. This additional uncertainty has to be
combined with the usual renormalization and factorization scale
uncertainties. Since the (relative) scheme and scale uncertainties
originating from the top mass only mildly depend on the renormalization
and factorization scale choice, the addition of this uncertainty may lead
to about a {\it linear} addition to the other uncertainties, if the
total uncertainty is defined as the envelope. This, however, has to be
analyzed in more detail which is left for future work.

We have investigated the total cross section as a function of the
trilinear coupling varied from its SM value. We have found significant
NLO mass effects beyond the Born-improved HTL that result in a shift of
the minimum of the cross section at various present and future
c.m.~energies of the hadron colliders. While the main effect of shifting
the minimum originates from the NLO top-mass effects of the real
corrections, the more symmetric virtual mass effects mainly affect the
size of the total cross section as a function of $\lambda_{H^3}$. The
full K-factors develop a larger dependence on $\lambda_{H^3}$ than those
of the Born-improved HTL due to the NLO top-mass effects. \\

\clearpage

\noindent
{\large \bf Acknowledgements} \\
%           ================
We are indebted to S.~Dittmaier for providing us with a copy of his {\tt
mathematica} program for the QCD corrections in the HTL as constructed
for the work of Ref.~\cite{Dawson:1998py} and to R.~Gr\"ober for
useful discussions. The work of S.~G.~is
supported by the Swiss National Science Foundation (SNF). The
work of S.~G.~and M.~M.~is supported by the DFG Collaborative
Research Center TRR 257 ``Particle Physics Phenomenology after the Higgs
Discovery''. F.~C.~and J.~R.~acknowledge financial support by the
Generalitat Valenciana, Spanish Government and ERDF funds from the
European Commission (Grants No.  RYC-2014-16061, SEJI-2017/2017/019,
FPA2017-84543- P,FPA2017-84445-P, and SEV-2014-0398). We acknowledge
support by the state of Baden-W\"urttemberg through bwHPC and the
German Research Foundation (DFG) through grant no INST 39/963-1 FUGG
(bwForCluster NEMO).

\clearpage
\appendix

\section{Two-loop box diagrams of the virtual corrections}
%        ===========
Here we present the two-loop box diagrams (omitting the ones
with reversed fermion flow):
\begin{figure}[ht!]
\begin{center}
 \includegraphics[width=0.75\textwidth]{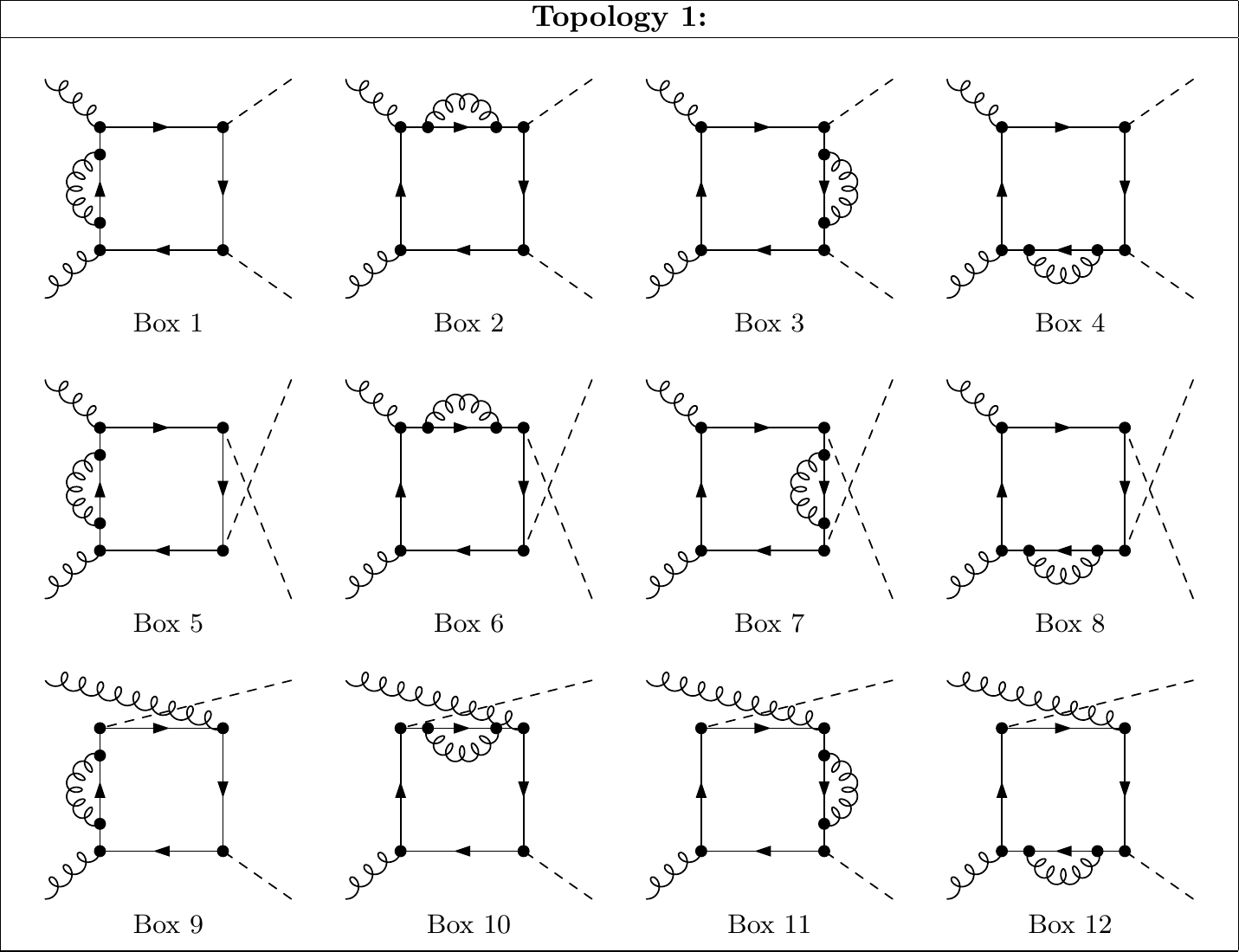} 
%\caption{\label{fg:boxdia1} \it Two-loop box diagrams: topology 1.}
~\\[0.2cm]
%\end{center}
%\end{figure}
%\begin{figure}[hbtp]
%\begin{center}
\hspace*{-0.3cm}
 \includegraphics[width=0.75\textwidth]{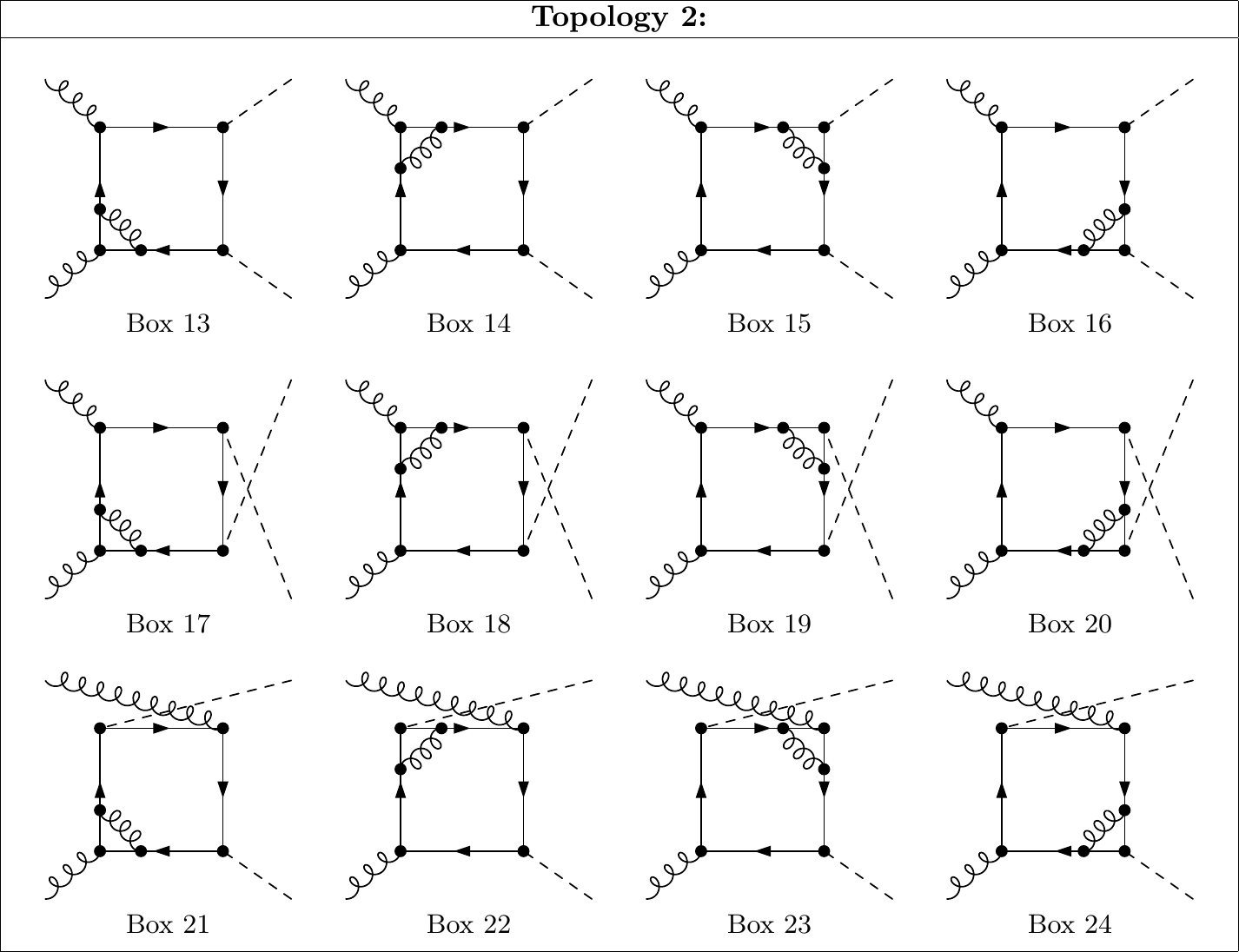} 
\caption{\label{fg:boxdia1} \it Two-loop box diagrams: topologies 1 and 2.}
\vspace*{-2.5cm}

\end{center}
\end{figure}
\begin{figure}[ht!]
\begin{center}
 \includegraphics[width=0.80\textwidth]{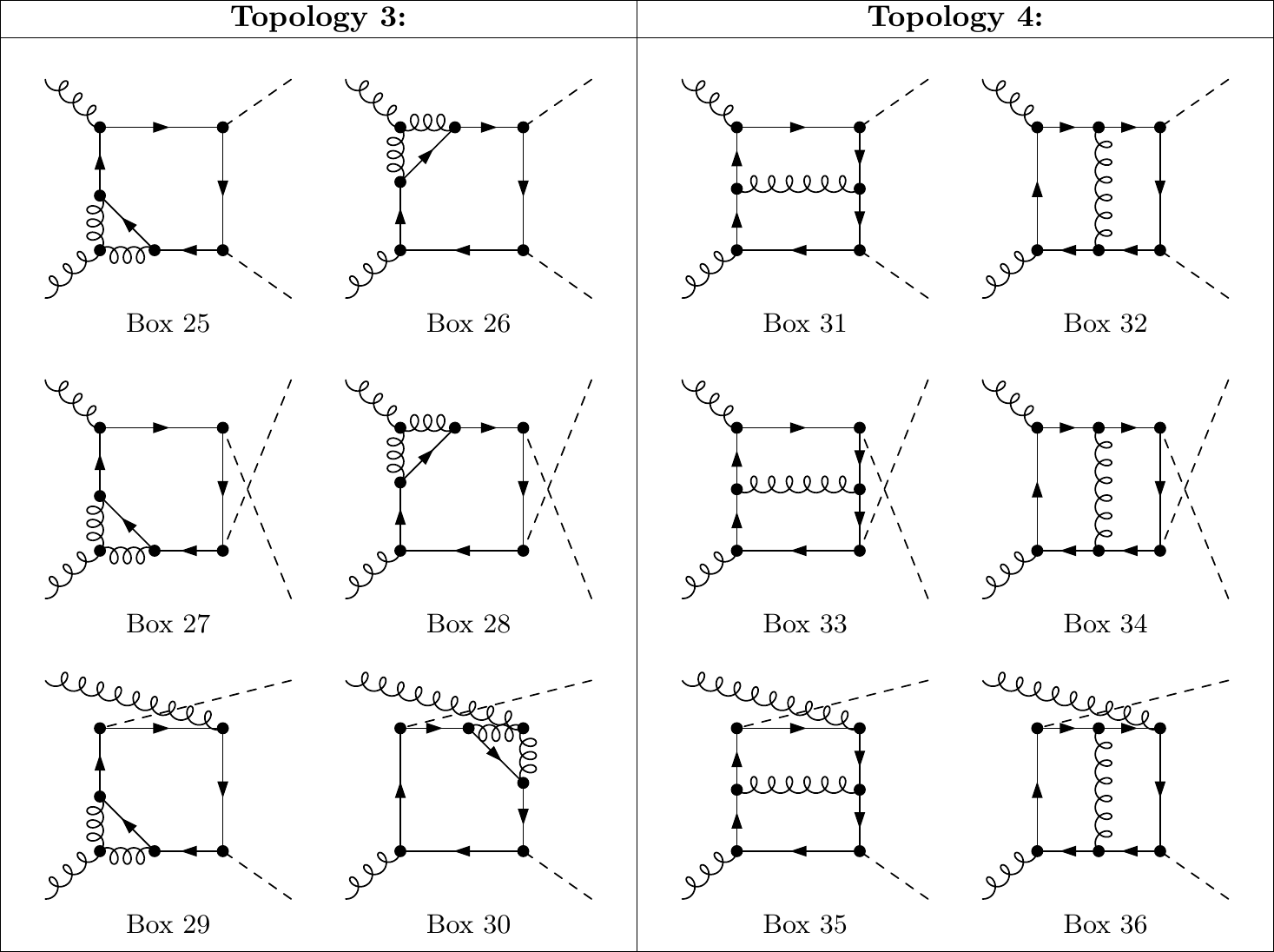} 
\caption{\label{fg:boxdia3} \it Two-loop box diagrams: topologies 3 and 4.}
~\\
%\end{center}
%\end{figure}
%\begin{figure}[hbtp]
%\begin{center}
 \includegraphics[width=0.80\textwidth]{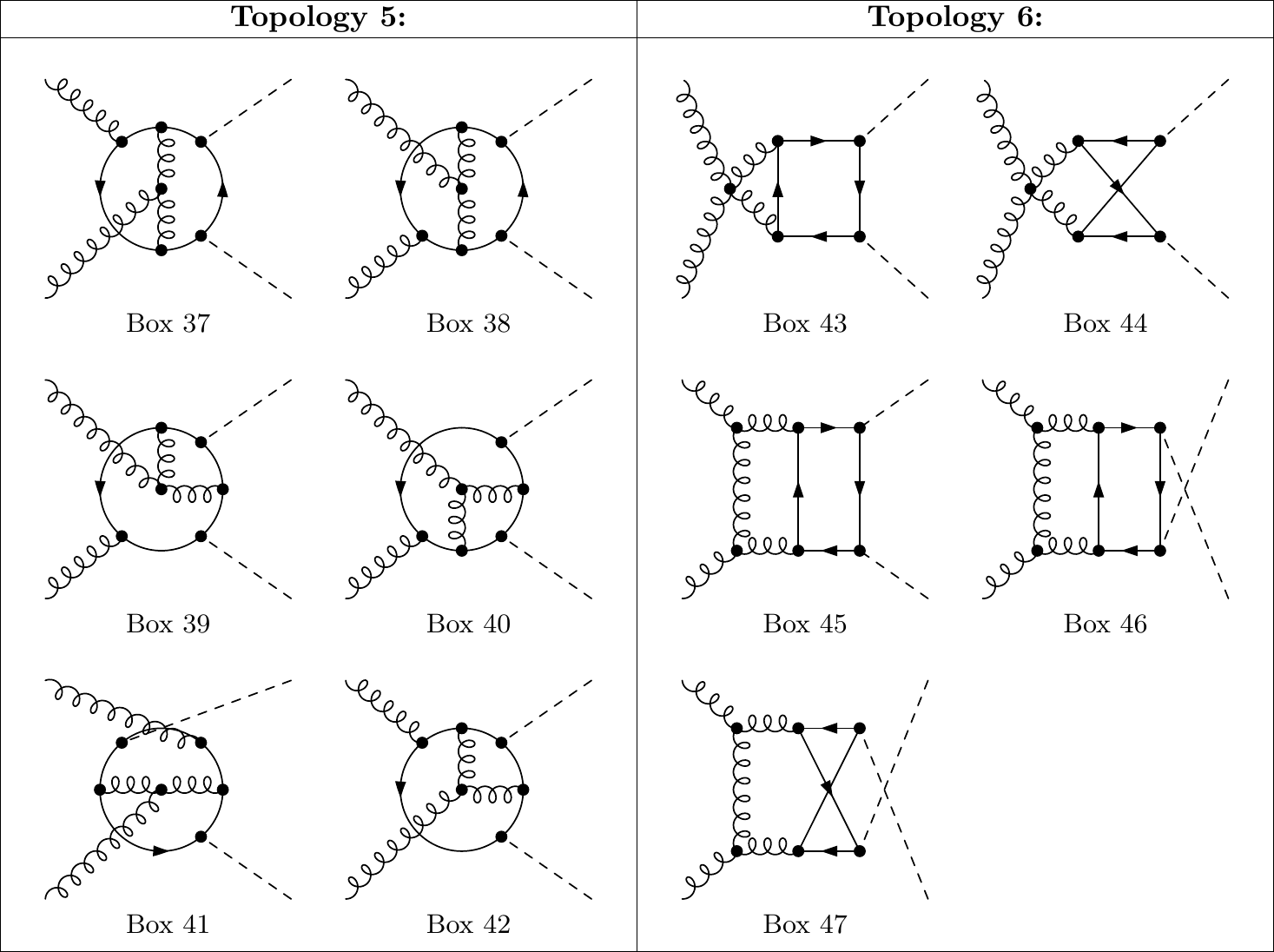} 
\caption{\label{fg:boxdia4} \it Two-loop box diagrams: topologies 5 and 6.}
\end{center}
\end{figure}
\vspace*{1cm}

\clearpage

\bibliographystyle{JHEP}

\bibliography{paper}

\providecommand{\href}[2]{#2}\begingroup\raggedright\begin{thebibliography}{100}

\bibitem{Aad:2012tfa}
{\scshape ATLAS} collaboration, \emph{{Observation of a new particle in the
  search for the Standard Model Higgs boson with the ATLAS detector at the
  LHC}}, \href{https://doi.org/10.1016/j.physletb.2012.08.020}{\emph{Phys.
  Lett.} {\bfseries B716} (2012) 1}
  [\href{https://arxiv.org/abs/1207.7214}{{\ttfamily 1207.7214}}].

\bibitem{Chatrchyan:2012xdj}
{\scshape CMS} collaboration, \emph{{Observation of a new boson at a mass of
  125 GeV with the CMS experiment at the LHC}},
  \href{https://doi.org/10.1016/j.physletb.2012.08.021}{\emph{Phys. Lett.}
  {\bfseries B716} (2012) 30}
  [\href{https://arxiv.org/abs/1207.7235}{{\ttfamily 1207.7235}}].

\bibitem{Khachatryan:2016vau}
{\scshape ATLAS, CMS} collaboration, \emph{{Measurements of the Higgs boson
  production and decay rates and constraints on its couplings from a combined
  ATLAS and CMS analysis of the LHC pp collision data at $ \sqrt{s}=7 $ and 8
  TeV}}, \href{https://doi.org/10.1007/JHEP08(2016)045}{\emph{JHEP} {\bfseries
  08} (2016) 045} [\href{https://arxiv.org/abs/1606.02266}{{\ttfamily
  1606.02266}}].

\bibitem{Higgs:1964ia}
P.~W. Higgs, \emph{{Broken symmetries, massless particles and gauge fields}},
  \href{https://doi.org/10.1016/0031-9163(64)91136-9}{\emph{Phys. Lett.}
  {\bfseries 12} (1964) 132}.

\bibitem{Higgs:1964pj}
P.~W. Higgs, \emph{{Broken Symmetries and the Masses of Gauge Bosons}},
  \href{https://doi.org/10.1103/PhysRevLett.13.508}{\emph{Phys. Rev. Lett.}
  {\bfseries 13} (1964) 508}.

\bibitem{Higgs:1966ev}
P.~W. Higgs, \emph{{Spontaneous Symmetry Breakdown without Massless Bosons}},
  \href{https://doi.org/10.1103/PhysRev.145.1156}{\emph{Phys. Rev.} {\bfseries
  145} (1966) 1156}.

\bibitem{Englert:1964et}
F.~Englert and R.~Brout, \emph{{Broken Symmetry and the Mass of Gauge Vector
  Mesons}}, \href{https://doi.org/10.1103/PhysRevLett.13.321}{\emph{Phys. Rev.
  Lett.} {\bfseries 13} (1964) 321}.

\bibitem{Guralnik:1964eu}
G.~S. Guralnik, C.~R. Hagen and T.~W.~B. Kibble, \emph{{Global Conservation
  Laws and Massless Particles}},
  \href{https://doi.org/10.1103/PhysRevLett.13.585}{\emph{Phys. Rev. Lett.}
  {\bfseries 13} (1964) 585}.

\bibitem{Kibble:1967sv}
T.~W.~B. Kibble, \emph{{Symmetry breaking in nonAbelian gauge theories}},
  \href{https://doi.org/10.1103/PhysRev.155.1554}{\emph{Phys. Rev.} {\bfseries
  155} (1967) 1554}.

\bibitem{Glover:1987nx}
E.~N. Glover and J.~van~der Bij, \emph{{Higgs boson pair production via gluon
  fusion}}, \href{https://doi.org/10.1016/0550-3213(88)90083-1}{\emph{Nucl.
  Phys.} {\bfseries B309} (1988) 282}.

\bibitem{Plehn:1996wb}
T.~Plehn, M.~Spira and P.~M. Zerwas, \emph{{Pair production of neutral Higgs
  particles in gluon-gluon collisions}},
  \href{https://doi.org/10.1016/0550-3213(96)00418-X,
  10.1016/S0550-3213(98)00406-4}{\emph{Nucl. Phys.} {\bfseries B479} (1996) 46}
  [\href{https://arxiv.org/abs/hep-ph/9603205}{{\ttfamily hep-ph/9603205}}].
  [Erratum: Nucl. Phys. {\bf B531} (1998) 655].

\bibitem{Dawson:1998py}
S.~Dawson, S.~Dittmaier and M.~Spira, \emph{{Neutral Higgs boson pair
  production at hadron colliders: QCD corrections}},
  \href{https://doi.org/10.1103/PhysRevD.58.115012}{\emph{Phys. Rev.}
  {\bfseries D58} (1998) 115012}
  [\href{https://arxiv.org/abs/hep-ph/9805244}{{\ttfamily hep-ph/9805244}}].

\bibitem{Djouadi:1999rca}
A.~Djouadi, W.~Kilian, M.~M{\"u}hlleitner and P.~M. Zerwas, \emph{{Production
  of neutral Higgs boson pairs at LHC}},
  \href{https://doi.org/10.1007/s100529900083}{\emph{Eur. Phys. J.} {\bfseries
  C10} (1999) 45} [\href{https://arxiv.org/abs/hep-ph/9904287}{{\ttfamily
  hep-ph/9904287}}].

\bibitem{Baglio:2012np}
J.~Baglio, A.~Djouadi, R.~Gr{\"o}ber, M.~M. M{\"u}hlleitner, J.~Quevillon and
  M.~Spira, \emph{{The measurement of the Higgs self-coupling at the LHC:
  theoretical status}},
  \href{https://doi.org/10.1007/JHEP04(2013)151}{\emph{JHEP} {\bfseries 04}
  (2013) 151} [\href{https://arxiv.org/abs/1212.5581}{{\ttfamily 1212.5581}}].

\bibitem{Degrassi:2016wml}
G.~Degrassi, P.~P. Giardino, F.~Maltoni and D.~Pagani, \emph{{Probing the Higgs
  self coupling via single Higgs production at the LHC}},
  \href{https://doi.org/10.1007/JHEP12(2016)080}{\emph{JHEP} {\bfseries 12}
  (2016) 080} [\href{https://arxiv.org/abs/1607.04251}{{\ttfamily
  1607.04251}}].

\bibitem{Degrassi:2017ucl}
G.~Degrassi, M.~Fedele and P.~P. Giardino, \emph{{Constraints on the trilinear
  Higgs self coupling from precision observables}},
  \href{https://doi.org/10.1007/JHEP04(2017)155}{\emph{JHEP} {\bfseries 04}
  (2017) 155} [\href{https://arxiv.org/abs/1702.01737}{{\ttfamily
  1702.01737}}].

\bibitem{Plehn:2005nk}
T.~Plehn and M.~Rauch, \emph{{The quartic higgs coupling at hadron colliders}},
  \href{https://doi.org/10.1103/PhysRevD.72.053008}{\emph{Phys. Rev.}
  {\bfseries D72} (2005) 053008}
  [\href{https://arxiv.org/abs/hep-ph/0507321}{{\ttfamily hep-ph/0507321}}].

\bibitem{Binoth:2006ym}
T.~Binoth, S.~Karg, N.~Kauer and R.~R{\"u}ckl, \emph{{Multi-Higgs boson
  production in the Standard Model and beyond}},
  \href{https://doi.org/10.1103/PhysRevD.74.113008}{\emph{Phys. Rev.}
  {\bfseries D74} (2006) 113008}
  [\href{https://arxiv.org/abs/hep-ph/0608057}{{\ttfamily hep-ph/0608057}}].

\bibitem{Fuks:2015hna}
B.~Fuks, J.~H. Kim and S.~J. Lee, \emph{{Probing Higgs self-interactions in
  proton-proton collisions at a center-of-mass energy of 100 TeV}},
  \href{https://doi.org/10.1103/PhysRevD.93.035026}{\emph{Phys. Rev.}
  {\bfseries D93} (2016) 035026}
  [\href{https://arxiv.org/abs/1510.07697}{{\ttfamily 1510.07697}}].

\bibitem{deFlorian:2016sit}
D.~de~Florian and J.~Mazzitelli, \emph{{Two-loop corrections to the triple
  Higgs boson production cross section}},
  \href{https://doi.org/10.1007/JHEP02(2017)107}{\emph{JHEP} {\bfseries 02}
  (2017) 107} [\href{https://arxiv.org/abs/1610.05012}{{\ttfamily
  1610.05012}}].

\bibitem{deFlorian:2019app}
D.~de~Florian, J.~Mazzitelli and I.~Fabre, \emph{{Triple Higgs production at
  hadron colliders at NNLO in QCD}},
  \href{https://arxiv.org/abs/1912.02760}{{\ttfamily 1912.02760}}.

\bibitem{Liu:2018peg}
T.~Liu, K.-F. Lyu, J.~Ren and H.~X. Zhu, \emph{{Probing the quartic Higgs boson
  self-interaction}},
  \href{https://doi.org/10.1103/PhysRevD.98.093004}{\emph{Phys. Rev.}
  {\bfseries D98} (2018) 093004}
  [\href{https://arxiv.org/abs/1803.04359}{{\ttfamily 1803.04359}}].

\bibitem{Bizon:2018syu}
W.~Bizoń, U.~Haisch and L.~Rottoli, \emph{{Constraints on the quartic Higgs
  self-coupling from double-Higgs production at future hadron colliders}},
  \href{https://doi.org/10.1007/JHEP10(2019)267}{\emph{JHEP} {\bfseries 10}
  (2019) 267} [\href{https://arxiv.org/abs/1810.04665}{{\ttfamily
  1810.04665}}].

\bibitem{Borowka:2018pxx}
S.~Borowka, C.~Duhr, F.~Maltoni, D.~Pagani, A.~Shivaji and X.~Zhao,
  \emph{{Probing the scalar potential via double Higgs boson production at
  hadron colliders}},
  \href{https://doi.org/10.1007/JHEP04(2019)016}{\emph{JHEP} {\bfseries 04}
  (2019) 016} [\href{https://arxiv.org/abs/1811.12366}{{\ttfamily
  1811.12366}}].

\bibitem{Borowka:2016ehy}
S.~Borowka, N.~Greiner, G.~Heinrich, S.~Jones, M.~Kerner, J.~Schlenk et~al.,
  \emph{{Higgs Boson Pair Production in Gluon Fusion at Next-to-Leading Order
  with Full Top-Quark Mass Dependence}},
  \href{https://doi.org/10.1103/PhysRevLett.117.079901,
  10.1103/PhysRevLett.117.012001}{\emph{Phys. Rev. Lett.} {\bfseries 117}
  (2016) 012001} [\href{https://arxiv.org/abs/1604.06447}{{\ttfamily
  1604.06447}}]. [Erratum: Phys. Rev. Lett. {\bf 117} (2016) 079901].

\bibitem{Borowka:2016ypz}
S.~Borowka, N.~Greiner, G.~Heinrich, S.~P. Jones, M.~Kerner, J.~Schlenk et~al.,
  \emph{{Full top quark mass dependence in Higgs boson pair production at
  NLO}}, \href{https://doi.org/10.1007/JHEP10(2016)107}{\emph{JHEP} {\bfseries
  10} (2016) 107} [\href{https://arxiv.org/abs/1608.04798}{{\ttfamily
  1608.04798}}].

\bibitem{Baglio:2018lrj}
J.~Baglio, F.~Campanario, S.~Glaus, M.~M{\"u}hlleitner, M.~Spira and
  J.~Streicher, \emph{{Gluon fusion into Higgs pairs at NLO QCD and the top
  mass scheme}},
  \href{https://doi.org/10.1140/epjc/s10052-019-6973-3}{\emph{Eur. Phys. J.}
  {\bfseries C79} (2019) 459}
  [\href{https://arxiv.org/abs/1811.05692}{{\ttfamily 1811.05692}}].

\bibitem{deFlorian:2013uza}
D.~de~Florian and J.~Mazzitelli, \emph{{Two-loop virtual corrections to Higgs
  pair production}},
  \href{https://doi.org/10.1016/j.physletb.2013.06.046}{\emph{Phys. Lett.}
  {\bfseries B724} (2013) 306}
  [\href{https://arxiv.org/abs/1305.5206}{{\ttfamily 1305.5206}}].

\bibitem{deFlorian:2013jea}
D.~de~Florian and J.~Mazzitelli, \emph{{Higgs Boson Pair Production at
  Next-to-Next-to-Leading Order in QCD}},
  \href{https://doi.org/10.1103/PhysRevLett.111.201801}{\emph{Phys. Rev. Lett.}
  {\bfseries 111} (2013) 201801}
  [\href{https://arxiv.org/abs/1309.6594}{{\ttfamily 1309.6594}}].

\bibitem{Grigo:2014jma}
J.~Grigo, K.~Melnikov and M.~Steinhauser, \emph{{Virtual corrections to Higgs
  boson pair production in the large top quark mass limit}},
  \href{https://doi.org/10.1016/j.nuclphysb.2014.09.003}{\emph{Nucl. Phys.}
  {\bfseries B888} (2014) 17}
  [\href{https://arxiv.org/abs/1408.2422}{{\ttfamily 1408.2422}}].

\bibitem{Banerjee:2018lfq}
P.~Banerjee, S.~Borowka, P.~K. Dhani, T.~Gehrmann and V.~Ravindran,
  \emph{{Two-loop massless QCD corrections to the $g + g \to H + H$ four-point
  amplitude}}, \href{https://doi.org/10.1007/JHEP11(2018)130}{\emph{JHEP}
  {\bfseries 11} (2018) 130}
  [\href{https://arxiv.org/abs/1809.05388}{{\ttfamily 1809.05388}}].

\bibitem{Chen:2019lzz}
L.-B. Chen, H.~T. Li, H.-S. Shao and J.~Wang, \emph{{Higgs boson pair
  production via gluon fusion at N$^3$LO in QCD}},
  \href{https://arxiv.org/abs/1909.06808}{{\ttfamily 1909.06808}}.

\bibitem{Chen:2019fhs}
L.-B. Chen, H.~T. Li, H.-S. Shao and J.~Wang, \emph{{The gluon-fusion
  production of Higgs boson pair: N$^3$LO QCD corrections and top-quark mass
  effects}}, \href{https://doi.org/10.1007/JHEP03(2020)072}{\emph{JHEP}
  {\bfseries 03} (2020) 072}
  [\href{https://arxiv.org/abs/1912.13001}{{\ttfamily 1912.13001}}].

\bibitem{Spira:2016zna}
M.~Spira, \emph{{Effective Multi-Higgs Couplings to Gluons}},
  \href{https://doi.org/10.1007/JHEP10(2016)026}{\emph{JHEP} {\bfseries 10}
  (2016) 026} [\href{https://arxiv.org/abs/1607.05548}{{\ttfamily
  1607.05548}}].

\bibitem{Heinrich:2017kxx}
G.~Heinrich, S.~P. Jones, M.~Kerner, G.~Luisoni and E.~Vryonidou, \emph{{NLO
  predictions for Higgs boson pair production with full top quark mass
  dependence matched to parton showers}},
  \href{https://doi.org/10.1007/JHEP08(2017)088}{\emph{JHEP} {\bfseries 08}
  (2017) 088} [\href{https://arxiv.org/abs/1703.09252}{{\ttfamily
  1703.09252}}].

\bibitem{Jones:2017giv}
S.~Jones and S.~Kuttimalai, \emph{{Parton Shower and NLO-Matching uncertainties
  in Higgs Boson Pair Production}},
  \href{https://doi.org/10.1007/JHEP02(2018)176}{\emph{JHEP} {\bfseries 02}
  (2018) 176} [\href{https://arxiv.org/abs/1711.03319}{{\ttfamily
  1711.03319}}].

\bibitem{Grazzini:2018bsd}
M.~Grazzini, G.~Heinrich, S.~Jones, S.~Kallweit, M.~Kerner, J.~M. Lindert
  et~al., \emph{{Higgs boson pair production at NNLO with top quark mass
  effects}}, \href{https://doi.org/10.1007/JHEP05(2018)059}{\emph{JHEP}
  {\bfseries 05} (2018) 059}
  [\href{https://arxiv.org/abs/1803.02463}{{\ttfamily 1803.02463}}].

\bibitem{Abada:2019ono}
{\scshape FCC} collaboration, \emph{{HE-LHC: The High-Energy Large Hadron
  Collider}}, \href{https://doi.org/10.1140/epjst/e2019-900088-6}{\emph{Eur.
  Phys. J. ST} {\bfseries 228} (2019) 1109}.

\bibitem{Abada:2019lih}
{\scshape FCC} collaboration, \emph{{FCC Physics Opportunities}},
  \href{https://doi.org/10.1140/epjc/s10052-019-6904-3}{\emph{Eur. Phys. J.}
  {\bfseries C79} (2019) 474}.

\bibitem{Benedikt:2018csr}
{\scshape FCC} collaboration, \emph{{FCC-hh: The Hadron Collider}},
  \href{https://doi.org/10.1140/epjst/e2019-900087-0}{\emph{Eur. Phys. J. ST}
  {\bfseries 228} (2019) 755}.

\bibitem{Graudenz:1992pv}
D.~Graudenz, M.~Spira and P.~M. Zerwas, \emph{{QCD corrections to Higgs boson
  production at proton proton colliders}},
  \href{https://doi.org/10.1103/PhysRevLett.70.1372}{\emph{Phys. Rev. Lett.}
  {\bfseries 70} (1993) 1372}.

\bibitem{Spira:1995rr}
M.~Spira, A.~Djouadi, D.~Graudenz and P.~M. Zerwas, \emph{{Higgs boson
  production at the LHC}},
  \href{https://doi.org/10.1016/0550-3213(95)00379-7}{\emph{Nucl. Phys.}
  {\bfseries B453} (1995) 17}
  [\href{https://arxiv.org/abs/hep-ph/9504378}{{\ttfamily hep-ph/9504378}}].

\bibitem{Harlander:2005rq}
R.~Harlander and P.~Kant, \emph{{Higgs production and decay: Analytic results
  at next-to-leading order QCD}},
  \href{https://doi.org/10.1088/1126-6708/2005/12/015}{\emph{JHEP} {\bfseries
  12} (2005) 015} [\href{https://arxiv.org/abs/hep-ph/0509189}{{\ttfamily
  hep-ph/0509189}}].

\bibitem{Anastasiou:2009kn}
C.~Anastasiou, S.~Bucherer and Z.~Kunszt, \emph{{HPro: A NLO Monte-Carlo for
  Higgs production via gluon fusion with finite heavy quark masses}},
  \href{https://doi.org/10.1088/1126-6708/2009/10/068}{\emph{JHEP} {\bfseries
  10} (2009) 068} [\href{https://arxiv.org/abs/0907.2362}{{\ttfamily
  0907.2362}}].

\bibitem{Aglietti:2006tp}
U.~Aglietti, R.~Bonciani, G.~Degrassi and A.~Vicini, \emph{{Analytic Results
  for Virtual QCD Corrections to Higgs Production and Decay}},
  \href{https://doi.org/10.1088/1126-6708/2007/01/021}{\emph{JHEP} {\bfseries
  01} (2007) 021} [\href{https://arxiv.org/abs/hep-ph/0611266}{{\ttfamily
  hep-ph/0611266}}].

\bibitem{Altarelli:1977zs}
G.~Altarelli and G.~Parisi, \emph{{Asymptotic Freedom in Parton Language}},
  \href{https://doi.org/10.1016/0550-3213(77)90384-4}{\emph{Nucl. Phys.}
  {\bfseries B126} (1977) 298}.

\bibitem{Ellis:1975ap}
J.~R. Ellis, M.~K. Gaillard and D.~V. Nanopoulos, \emph{{A Phenomenological
  Profile of the Higgs Boson}},
  \href{https://doi.org/10.1016/0550-3213(76)90382-5}{\emph{Nucl. Phys.}
  {\bfseries B106} (1976) 292}.

\bibitem{Shifman:1979eb}
M.~A. Shifman, A.~I. Vainshtein, M.~B. Voloshin and V.~I. Zakharov,
  \emph{{Low-Energy Theorems for Higgs Boson Couplings to Photons}},
  {\emph{Sov. J. Nucl. Phys.} {\bfseries 30} (1979) 711}. [Yad.
  Fiz.30,1368(1979)].

\bibitem{Inami:1982xt}
T.~Inami, T.~Kubota and Y.~Okada, \emph{{Effective Gauge Theory and the Effect
  of Heavy Quarks in Higgs Boson Decays}},
  \href{https://doi.org/10.1007/BF01571710}{\emph{Z. Phys.} {\bfseries C18}
  (1983) 69}.

\bibitem{Kniehl:1995tn}
B.~A. Kniehl and M.~Spira, \emph{{Low-energy theorems in Higgs physics}},
  \href{https://doi.org/10.1007/s002880050007}{\emph{Z. Phys.} {\bfseries C69}
  (1995) 77} [\href{https://arxiv.org/abs/hep-ph/9505225}{{\ttfamily
  hep-ph/9505225}}].

\bibitem{Chetyrkin:1997iv}
K.~G. Chetyrkin, B.~A. Kniehl and M.~Steinhauser, \emph{{Hadronic Higgs decay
  to order $\alpha_s^4$}},
  \href{https://doi.org/10.1103/PhysRevLett.79.353}{\emph{Phys. Rev. Lett.}
  {\bfseries 79} (1997) 353}
  [\href{https://arxiv.org/abs/hep-ph/9705240}{{\ttfamily hep-ph/9705240}}].

\bibitem{Kramer:1996iq}
M.~Kr{\"a}mer, E.~Laenen and M.~Spira, \emph{{Soft gluon radiation in Higgs
  boson production at the LHC}},
  \href{https://doi.org/10.1016/S0550-3213(97)00679-2}{\emph{Nucl. Phys.}
  {\bfseries B511} (1998) 523}
  [\href{https://arxiv.org/abs/hep-ph/9611272}{{\ttfamily hep-ph/9611272}}].

\bibitem{Schroder:2005hy}
Y.~Schr{\"o}der and M.~Steinhauser, \emph{{Four-loop decoupling relations for
  the strong coupling}},
  \href{https://doi.org/10.1088/1126-6708/2006/01/051}{\emph{JHEP} {\bfseries
  01} (2006) 051} [\href{https://arxiv.org/abs/hep-ph/0512058}{{\ttfamily
  hep-ph/0512058}}].

\bibitem{Baikov:2016tgj}
P.~A. Baikov, K.~G. Chetyrkin and J.~H. K{\"u}hn, \emph{{Five-Loop Running of
  the QCD coupling constant}},
  \href{https://doi.org/10.1103/PhysRevLett.118.082002}{\emph{Phys. Rev. Lett.}
  {\bfseries 118} (2017) 082002}
  [\href{https://arxiv.org/abs/1606.08659}{{\ttfamily 1606.08659}}].

\bibitem{Gerlach:2018hen}
M.~Gerlach, F.~Herren and M.~Steinhauser, \emph{{Wilson coefficients for Higgs
  boson production and decoupling relations to $
  \mathcal{O}\left({\alpha}_s^4\right) $}},
  \href{https://doi.org/10.1007/JHEP11(2018)141}{\emph{JHEP} {\bfseries 11}
  (2018) 141} [\href{https://arxiv.org/abs/1809.06787}{{\ttfamily
  1809.06787}}].

\bibitem{Shao:2013bz}
D.~Y. Shao, C.~S. Li, H.~T. Li and J.~Wang, \emph{{Threshold resummation
  effects in Higgs boson pair production at the LHC}},
  \href{https://doi.org/10.1007/JHEP07(2013)169}{\emph{JHEP} {\bfseries 07}
  (2013) 169} [\href{https://arxiv.org/abs/1301.1245}{{\ttfamily 1301.1245}}].

\bibitem{deFlorian:2015moa}
D.~de~Florian and J.~Mazzitelli, \emph{{Higgs pair production at
  next-to-next-to-leading logarithmic accuracy at the LHC}},
  \href{https://doi.org/10.1007/JHEP09(2015)053}{\emph{JHEP} {\bfseries 09}
  (2015) 053} [\href{https://arxiv.org/abs/1505.07122}{{\ttfamily
  1505.07122}}].

\bibitem{Frederix:2014hta}
R.~Frederix, S.~Frixione, V.~Hirschi, F.~Maltoni, O.~Mattelaer et~al.,
  \emph{{Higgs pair production at the LHC with NLO and parton-shower effects}},
  \href{https://doi.org/10.1016/j.physletb.2014.03.026}{\emph{Phys. Lett.}
  {\bfseries B732} (2014) 142}
  [\href{https://arxiv.org/abs/1401.7340}{{\ttfamily 1401.7340}}].

\bibitem{Maltoni:2014eza}
F.~Maltoni, E.~Vryonidou and M.~Zaro, \emph{{Top-quark mass effects in double
  and triple Higgs production in gluon-gluon fusion at NLO}},
  \href{https://doi.org/10.1007/JHEP11(2014)079}{\emph{JHEP} {\bfseries 1411}
  (2014) 079} [\href{https://arxiv.org/abs/1408.6542}{{\ttfamily 1408.6542}}].

\bibitem{Alwall:2014hca}
J.~Alwall, R.~Frederix, S.~Frixione, V.~Hirschi, F.~Maltoni et~al., \emph{{The
  automated computation of tree-level and next-to-leading order differential
  cross sections, and their matching to parton shower simulations}},
  \href{https://doi.org/10.1007/JHEP07(2014)079}{\emph{JHEP} {\bfseries 1407}
  (2014) 079} [\href{https://arxiv.org/abs/1405.0301}{{\ttfamily 1405.0301}}].

\bibitem{Hirschi:2015iia}
V.~Hirschi and O.~Mattelaer, \emph{{Automated event generation for loop-induced
  processes}}, \href{https://doi.org/10.1007/JHEP10(2015)146}{\emph{JHEP}
  {\bfseries 10} (2015) 146}
  [\href{https://arxiv.org/abs/1507.00020}{{\ttfamily 1507.00020}}].

\bibitem{Grigo:2013rya}
J.~Grigo, J.~Hoff, K.~Melnikov and M.~Steinhauser, \emph{{On the Higgs boson
  pair production at the LHC}},
  \href{https://doi.org/10.1016/j.nuclphysb.2013.06.024}{\emph{Nucl. Phys.}
  {\bfseries B875} (2013) 1} [\href{https://arxiv.org/abs/1305.7340}{{\ttfamily
  1305.7340}}].

\bibitem{Grigo:2015dia}
J.~Grigo, J.~Hoff and M.~Steinhauser, \emph{{Higgs boson pair production: top
  quark mass effects at NLO and NNLO}},
  \href{https://doi.org/10.1016/j.nuclphysb.2015.09.012}{\emph{Nucl. Phys.}
  {\bfseries B900} (2015) 412}
  [\href{https://arxiv.org/abs/1508.00909}{{\ttfamily 1508.00909}}].

\bibitem{Grober:2017uho}
R.~Gr{\"o}ber, A.~Maier and T.~Rauh, \emph{{Reconstruction of top-quark mass
  effects in Higgs pair production and other gluon-fusion processes}},
  \href{https://doi.org/10.1007/JHEP03(2018)020}{\emph{JHEP} {\bfseries 03}
  (2018) 020} [\href{https://arxiv.org/abs/1709.07799}{{\ttfamily
  1709.07799}}].

\bibitem{Bonciani:2018omm}
R.~Bonciani, G.~Degrassi, P.~P. Giardino and R.~Gr{\"o}ber, \emph{{Analytical
  Method for Next-to-Leading-Order QCD Corrections to Double-Higgs
  Production}},
  \href{https://doi.org/10.1103/PhysRevLett.121.162003}{\emph{Phys. Rev. Lett.}
  {\bfseries 121} (2018) 162003}
  [\href{https://arxiv.org/abs/1806.11564}{{\ttfamily 1806.11564}}].

\bibitem{Davies:2018qvx}
J.~Davies, G.~Mishima, M.~Steinhauser and D.~Wellmann, \emph{{Double Higgs
  boson production at NLO in the high-energy limit: complete analytic
  results}}, \href{https://doi.org/10.1007/JHEP01(2019)176}{\emph{JHEP}
  {\bfseries 01} (2019) 176}
  [\href{https://arxiv.org/abs/1811.05489}{{\ttfamily 1811.05489}}].

\bibitem{Davies:2019dfy}
J.~Davies, G.~Heinrich, S.~P. Jones, M.~Kerner, G.~Mishima, M.~Steinhauser
  et~al., \emph{{Double Higgs boson production at NLO: combining the exact
  numerical result and high-energy expansion}},
  \href{https://doi.org/10.1007/JHEP11(2019)024}{\emph{JHEP} {\bfseries 11}
  (2019) 024} [\href{https://arxiv.org/abs/1907.06408}{{\ttfamily
  1907.06408}}].

\bibitem{hpair}
{\emph{{\rm The program {\tt Hpair} can be obtained at the URL:} {\tt
  http://tiger.web.psi.ch/hpair/}} }.

\bibitem{Cahn:1978nz}
R.~N. Cahn, M.~S. Chanowitz and N.~Fleishon, \emph{{Higgs Particle Production
  by $Z \to H \gamma$}},
  \href{https://doi.org/10.1016/0370-2693(79)90438-6}{\emph{Phys. Lett.}
  {\bfseries 82B} (1979) 113}.

\bibitem{Bergstrom:1985hp}
L.~Bergstr{\"o}m and G.~Hulth, \emph{{Induced Higgs Couplings to Neutral Bosons
  in $e^+ e^-$ Collisions}},
  \href{https://doi.org/10.1016/0550-3213(86)90074-X,
  10.1016/0550-3213(85)90302-5}{\emph{Nucl. Phys.} {\bfseries B259} (1985)
  137}. [Erratum: Nucl. Phys.B276,744(1986)].

\bibitem{Gunion:1989we}
J.~F. Gunion, H.~E. Haber, G.~L. Kane and S.~Dawson, \emph{{The Higgs Hunter's
  Guide}}, {\emph{Front. Phys.} {\bfseries 80} (2000) 1}.

\bibitem{Degrassi:2016vss}
G.~Degrassi, P.~P. Giardino and R.~{Gr{\"o}ber}, \emph{{On the two-loop virtual
  QCD corrections to Higgs boson pair production in the Standard Model}},
  \href{https://doi.org/10.1140/epjc/s10052-016-4256-9}{\emph{Eur. Phys. J.}
  {\bfseries C76} (2016) 411}
  [\href{https://arxiv.org/abs/1603.00385}{{\ttfamily 1603.00385}}].

\bibitem{deFlorian:2017qfk}
D.~de~Florian, I.~Fabre and J.~Mazzitelli, \emph{{Higgs boson pair production
  at NNLO in QCD including dimension 6 operators}},
  \href{https://doi.org/10.1007/JHEP10(2017)215}{\emph{JHEP} {\bfseries 10}
  (2017) 215} [\href{https://arxiv.org/abs/1704.05700}{{\ttfamily
  1704.05700}}].

\bibitem{Vermaseren:2000nd}
J.~A.~M. Vermaseren, \emph{{New features of FORM}},
  \href{https://arxiv.org/abs/math-ph/0010025}{{\ttfamily math-ph/0010025}}.

\bibitem{Kuipers:2012rf}
J.~Kuipers, T.~Ueda, J.~A.~M. Vermaseren and J.~Vollinga, \emph{{FORM version
  4.0}}, \href{https://doi.org/10.1016/j.cpc.2012.12.028}{\emph{Comput. Phys.
  Commun.} {\bfseries 184} (2013) 1453}
  [\href{https://arxiv.org/abs/1203.6543}{{\ttfamily 1203.6543}}].

\bibitem{Hearn:1971zza}
A.~C. Hearn, \emph{{Reduce 2: A System and Language For Algebraic
  Manipulation}}, {\emph{PRINT-71-1192} (1971) }.

\bibitem{Mathematica}
{\emph{{\rm Wolfram Research{,} Inc., Mathematica, Version 9.0, Champaign, IL
  (2012)}} }.

\bibitem{Djouadi:1990aj}
A.~Djouadi, M.~Spira, J.~J. van~der Bij and P.~M. Zerwas, \emph{{QCD
  corrections to gamma gamma decays of Higgs particles in the intermediate mass
  range}}, \href{https://doi.org/10.1016/0370-2693(91)90879-U}{\emph{Phys.
  Lett.} {\bfseries B257} (1991) 187}.

\bibitem{Spira:1991tj}
M.~Spira, A.~Djouadi and P.~M. Zerwas, \emph{{QCD corrections to the H Z gamma
  coupling}}, \href{https://doi.org/10.1016/0370-2693(92)90331-W}{\emph{Phys.
  Lett.} {\bfseries B276} (1992) 350}.

\bibitem{Muhlleitner:2006wx}
M.~M{\"u}hlleitner and M.~Spira, \emph{{Higgs Boson Production via Gluon
  Fusion: Squark Loops at NLO QCD}},
  \href{https://doi.org/10.1016/j.nuclphysb.2007.08.011}{\emph{Nucl. Phys.}
  {\bfseries B790} (2008) 1}
  [\href{https://arxiv.org/abs/hep-ph/0612254}{{\ttfamily hep-ph/0612254}}].

\bibitem{Spira:1995es}
M.~Spira, \emph{{Radiative QCD corrections to decay and production of Higgs
  bosons at $e^+ e^-$ and $p p$ accelerators. (In German)}}, Ph.D. thesis,
  Aachen, Tech. Hochsch., 1993.

\bibitem{Muhlleitner:2010nm}
M.~M{\"u}hlleitner, H.~Rzehak and M.~Spira, \emph{{SUSY-QCD Corrections to MSSM
  Higgs Boson Production via Gluon fusion}},
  \href{https://doi.org/10.22323/1.092.0043}{\emph{PoS} {\bfseries RADCOR2009}
  (2010) 043} [\href{https://arxiv.org/abs/1001.3214}{{\ttfamily 1001.3214}}].

\bibitem{Muhlleitner:2010zz}
M.~M{\"u}hlleitner, H.~Rzehak and M.~Spira, \emph{{MSSM Higgs boson production
  via gluon fusion}},  in \emph{{Physics at the LHC2010. Proceedings, 5th
  Conference, PLHC2010, Hamburg, Germany, June 7-12, 2010}}, pp.~415--417,
  2010, \href{https://doi.org/10.3204/DESY-PROC-2010-01/rzehak}{DOI}.

\bibitem{Richardson}
L.~F. Richardson, \emph{{The approximate arithmetical solution by finite
  differences of physical problems including differential equations, with an
  application to the stresses in a masonry dam}},
  \href{https://doi.org/10.1098/rsta.1911.0009}{\emph{Philosophical
  Transactions of the Royal Society} {\bfseries A210} (1911) 307}.

\bibitem{Gray:1990yh}
N.~Gray, D.~J. Broadhurst, W.~Grafe and K.~Schilcher, \emph{{Three Loop
  Relation of Quark (Modified) Ms and Pole Masses}},
  \href{https://doi.org/10.1007/BF01614703}{\emph{Z. Phys.} {\bfseries C48}
  (1990) 673}.

\bibitem{Chetyrkin:1999ys}
K.~G. Chetyrkin and M.~Steinhauser, \emph{{Short distance mass of a heavy quark
  at order $\alpha_s^3$}},
  \href{https://doi.org/10.1103/PhysRevLett.83.4001}{\emph{Phys. Rev. Lett.}
  {\bfseries 83} (1999) 4001}
  [\href{https://arxiv.org/abs/hep-ph/9907509}{{\ttfamily hep-ph/9907509}}].

\bibitem{Chetyrkin:1999qi}
K.~G. Chetyrkin and M.~Steinhauser, \emph{{The Relation between the MS-bar and
  the on-shell quark mass at order $\alpha_s^3$}},
  \href{https://doi.org/10.1016/S0550-3213(99)00784-1}{\emph{Nucl. Phys.}
  {\bfseries B573} (2000) 617}
  [\href{https://arxiv.org/abs/hep-ph/9911434}{{\ttfamily hep-ph/9911434}}].

\bibitem{Melnikov:2000qh}
K.~Melnikov and T.~v. Ritbergen, \emph{{The Three loop relation between the
  MS-bar and the pole quark masses}},
  \href{https://doi.org/10.1016/S0370-2693(00)00507-4}{\emph{Phys. Lett.}
  {\bfseries B482} (2000) 99}
  [\href{https://arxiv.org/abs/hep-ph/9912391}{{\ttfamily hep-ph/9912391}}].

\bibitem{Tarasov:1982gk}
O.~V. Tarasov, \emph{{Anomalous Dimensions of Quark Masses in three Loop
  Approximation}}, {\emph{JINR-P2-82-900} (1982) }.

\bibitem{Chetyrkin:1997dh}
K.~G. Chetyrkin, \emph{{Quark mass anomalous dimension to O($\alpha_s^4$)}},
  \href{https://doi.org/10.1016/S0370-2693(97)00535-2}{\emph{Phys. Lett.}
  {\bfseries B404} (1997) 161}
  [\href{https://arxiv.org/abs/hep-ph/9703278}{{\ttfamily hep-ph/9703278}}].

\bibitem{Lepage:1980dq}
G.~P. Lepage, \emph{{{\tt Vegas}: An adaptive multidimensional Integration
  Program}}, {\emph{CLNS-80/447} (1980) }.

\bibitem{Catani:1996vz}
S.~Catani and M.~H. Seymour, \emph{{A General algorithm for calculating jet
  cross-sections in NLO QCD}},
  \href{https://doi.org/10.1016/S0550-3213(96)00589-5,
  10.1016/S0550-3213(98)81022-5}{\emph{Nucl. Phys.} {\bfseries B485} (1997)
  291} [\href{https://arxiv.org/abs/hep-ph/9605323}{{\ttfamily
  hep-ph/9605323}}]. [Erratum: Nucl. Phys.B510,503(1998)].

\bibitem{Hahn:2000kx}
T.~Hahn, \emph{{Generating Feynman diagrams and amplitudes with FeynArts 3}},
  \href{https://doi.org/10.1016/S0010-4655(01)00290-9}{\emph{Comput. Phys.
  Commun.} {\bfseries 140} (2001) 418}
  [\href{https://arxiv.org/abs/hep-ph/0012260}{{\ttfamily hep-ph/0012260}}].

\bibitem{Hahn:1998yk}
T.~Hahn and M.~Perez-Victoria, \emph{{Automatized one loop calculations in
  four-dimensions and D-dimensions}},
  \href{https://doi.org/10.1016/S0010-4655(98)00173-8}{\emph{Comput. Phys.
  Commun.} {\bfseries 118} (1999) 153}
  [\href{https://arxiv.org/abs/hep-ph/9807565}{{\ttfamily hep-ph/9807565}}].

\bibitem{tHooft:1978jhc}
G.~'t~Hooft and M.~J.~G. Veltman, \emph{{Scalar One Loop Integrals}},
  \href{https://doi.org/10.1016/0550-3213(79)90605-9}{\emph{Nucl. Phys.}
  {\bfseries B153} (1979) 365}.

\bibitem{vanOldenborgh:1990yc}
G.~J. van Oldenborgh, \emph{{FF: A Package to evaluate one loop Feynman
  diagrams}}, \href{https://doi.org/10.1016/0010-4655(91)90002-3}{\emph{Comput.
  Phys. Commun.} {\bfseries 66} (1991) 1}.

\bibitem{Denner:2002ii}
A.~Denner and S.~Dittmaier, \emph{{Reduction of one loop tensor five point
  integrals}}, \href{https://doi.org/10.1016/S0550-3213(03)00184-6}{\emph{Nucl.
  Phys.} {\bfseries B658} (2003) 175}
  [\href{https://arxiv.org/abs/hep-ph/0212259}{{\ttfamily hep-ph/0212259}}].

\bibitem{Denner:2005nn}
A.~Denner and S.~Dittmaier, \emph{{Reduction schemes for one-loop tensor
  integrals}},
  \href{https://doi.org/10.1016/j.nuclphysb.2005.11.007}{\emph{Nucl. Phys.}
  {\bfseries B734} (2006) 62}
  [\href{https://arxiv.org/abs/hep-ph/0509141}{{\ttfamily hep-ph/0509141}}].

\bibitem{Denner:2010tr}
A.~Denner and S.~Dittmaier, \emph{{Scalar one-loop 4-point integrals}},
  \href{https://doi.org/10.1016/j.nuclphysb.2010.11.002}{\emph{Nucl. Phys.}
  {\bfseries B844} (2011) 199}
  [\href{https://arxiv.org/abs/1005.2076}{{\ttfamily 1005.2076}}].

\bibitem{Denner:2016kdg}
A.~Denner, S.~Dittmaier and L.~Hofer, \emph{{Collier: a fortran-based Complex
  One-Loop LIbrary in Extended Regularizations}},
  \href{https://doi.org/10.1016/j.cpc.2016.10.013}{\emph{Comput. Phys. Commun.}
  {\bfseries 212} (2017) 220}
  [\href{https://arxiv.org/abs/1604.06792}{{\ttfamily 1604.06792}}].

\bibitem{Harland-Lang:2014zoa}
L.~A. Harland-Lang, A.~D. Martin, P.~Motylinski and R.~S. Thorne, \emph{{Parton
  distributions in the LHC era: MMHT 2014 PDFs}},
  \href{https://doi.org/10.1140/epjc/s10052-015-3397-6}{\emph{Eur. Phys. J.}
  {\bfseries C75} (2015) 204}
  [\href{https://arxiv.org/abs/1412.3989}{{\ttfamily 1412.3989}}].

\bibitem{Butterworth:2015oua}
J.~Butterworth et~al., \emph{{PDF4LHC recommendations for LHC Run II}},
  \href{https://doi.org/10.1088/0954-3899/43/2/023001}{\emph{J. Phys.}
  {\bfseries G43} (2016) 023001}
  [\href{https://arxiv.org/abs/1510.03865}{{\ttfamily 1510.03865}}].

\bibitem{Buckley:2014ana}
A.~Buckley, J.~Ferrando, S.~Lloyd, K.~{Nordstr\"om}, B.~Page, M.~{R\"ufenacht}
  et~al., \emph{{LHAPDF6: parton density access in the LHC precision era}},
  \href{https://doi.org/10.1140/epjc/s10052-015-3318-8}{\emph{Eur. Phys. J.}
  {\bfseries C75} (2015) 132}
  [\href{https://arxiv.org/abs/1412.7420}{{\ttfamily 1412.7420}}].

\bibitem{Abramowitz}
M.~Abramowitz and I.~Stegun, \emph{{Handbook of Mathematical Functions with
  Formulas, Graphs, and Mathematical Tables}}, {\emph{Applied Mathematics
  Series} {\bfseries 55} (1964) 307}.

\bibitem{deFlorian:2016spz}
{\scshape LHC Higgs Cross Section Working Group} collaboration, \emph{{Handbook
  of LHC Higgs Cross Sections: 4. Deciphering the Nature of the Higgs Sector}},
   \href{https://arxiv.org/abs/1610.07922}{{\ttfamily 1610.07922}}.

\bibitem{Anastasiou:2016cez}
C.~Anastasiou, C.~Duhr, F.~Dulat, E.~Furlan, T.~Gehrmann, F.~Herzog et~al.,
  \emph{{High precision determination of the gluon fusion Higgs boson
  cross-section at the LHC}},
  \href{https://doi.org/10.1007/JHEP05(2016)058}{\emph{JHEP} {\bfseries 05}
  (2016) 058} [\href{https://arxiv.org/abs/1602.00695}{{\ttfamily
  1602.00695}}].

\bibitem{Fadin:1987wz}
V.~S. Fadin and V.~A. Khoze, \emph{{Threshold Behavior of Heavy Top Production
  in $e^+e^-$ Collisions}}, {\emph{JETP Lett.} {\bfseries 46} (1987) 525}.
  [Pisma Zh. Eksp. Teor. Fiz.46,417(1987)].

\bibitem{Fadin:1988fn}
V.~S. Fadin and V.~A. Khoze, \emph{{Production of a pair of heavy quarks in
  $e^+e^-$ annihilation in the threshold region}}, {\emph{Sov. J. Nucl. Phys.}
  {\bfseries 48} (1988) 309}. [Yad. Fiz.48,487(1988)].

\bibitem{Fadin:1990wx}
V.~S. Fadin, V.~A. Khoze and T.~Sjostrand, \emph{{On the Threshold Behavior of
  Heavy Top Production}}, \href{https://doi.org/10.1007/BF01614696}{\emph{Z.
  Phys.} {\bfseries C48} (1990) 613}.

\bibitem{Strassler:1990nw}
M.~J. Strassler and M.~E. Peskin, \emph{{The Heavy top quark threshold: QCD and
  the Higgs}}, \href{https://doi.org/10.1103/PhysRevD.43.1500}{\emph{Phys.
  Rev.} {\bfseries D43} (1991) 1500}.

\bibitem{Melnikov:1994jb}
K.~Melnikov, M.~Spira and O.~I. Yakovlev, \emph{{Threshold effects in two
  photon decays of Higgs particles}},
  \href{https://doi.org/10.1007/BF01560100}{\emph{Z. Phys.} {\bfseries C64}
  (1994) 401} [\href{https://arxiv.org/abs/hep-ph/9405301}{{\ttfamily
  hep-ph/9405301}}].

\bibitem{Li:2013rra}
X.~Li and M.~B. Voloshin, \emph{{Remarks on double Higgs boson production by
  gluon fusion at threshold}},
  \href{https://doi.org/10.1103/PhysRevD.89.013012}{\emph{Phys. Rev.}
  {\bfseries D89} (2014) 013012}
  [\href{https://arxiv.org/abs/1311.5156}{{\ttfamily 1311.5156}}].

\bibitem{Grober:2015cwa}
R.~Gr{\"o}ber, M.~M{\"u}hlleitner, M.~Spira and J.~Streicher, \emph{{NLO QCD
  Corrections to Higgs Pair Production including Dimension-6 Operators}},
  \href{https://doi.org/10.1007/JHEP09(2015)092}{\emph{JHEP} {\bfseries 09}
  (2015) 092} [\href{https://arxiv.org/abs/1504.06577}{{\ttfamily
  1504.06577}}].

\bibitem{Buchalla:2018yce}
G.~Buchalla, M.~Capozi, A.~Celis, G.~Heinrich and L.~Scyboz, \emph{{Higgs boson
  pair production in non-linear Effective Field Theory with full
  $m_t$-dependence at NLO QCD}},
  \href{https://doi.org/10.1007/JHEP09(2018)057}{\emph{JHEP} {\bfseries 09}
  (2018) 057} [\href{https://arxiv.org/abs/1806.05162}{{\ttfamily
  1806.05162}}].

\end{thebibliography}\endgroup

\end{document}